\documentclass[useAMS,usenatbib]{mnras}
\usepackage{graphicx,natbib,amssymb,amsmath,stmaryrd,makecell}
\usepackage{txfonts}
\usepackage{xcolor}
\usepackage{array}
\usepackage{multirow}
\usepackage{fancyhdr}
\newcommand{\eV}{{\rm eV}}
\newcommand{\keV}{{\rm keV}}
\newcommand{\MeV}{{\rm MeV}}
\newcommand{\Kel}{{\rm K}}

\newcommand{\secs}{{\rm s}}

\newcommand{\GHz}{{\rm\, GHz}}

\newcommand{\expf}[1]{{{\rm e}^{#1}}}

\renewcommand{\ion}[2]{{\text{{#1}\,{#2}}}\xspace}
\newcommand{\HI}{\ion{H}{I}}
\newcommand{\HeI}{\ion{He}{I}}
\newcommand{\HeII}{\ion{He}{II}}

\newcommand{\TCMB}{T_{\rm CMB}}

\newcommand{\zX}{{z_{\rm X}}}

\newcommand{\zmu}{{z_{\mu}}}

\newcommand{\zi}{{z_{\rm i}}}

\newcommand{\nbb}{{n^{\rm pl}}}

\newcommand{\xe}{x_{\rm e}}

\newcommand{\xc}{x_{\rm c}}

\newcommand{\id}{{\,\rm d}}

\newcommand{\beq}{\begin{equation}}   %

\newcommand{\eeq}{\end{equation}}   %

\newcommand{\beqa}{\begin{eqnarray}}   %

\newcommand{\eeqa}{\end{eqnarray}}   %

\newcommand{\beal}{\begin{align}}
\newcommand{\enal}{\end{align}}

\newcommand{\bspl}{\begin{split}}

\newcommand{\espl}{\end{split}}

\newcommand{\bsub}{\begin{subequations}}

\newcommand{\esub}{\end{subequations}}

\newcommand{\bmulti}{\begin{multline}}   %

\newcommand{\beqm}{\begin{mathletters}}   %

\newcommand{\eeqm}{\end{mathletters}}   %

\newcommand{\me}{m_{\rm e}}

\newcommand{\Ne}{N_{\rm e}}

\newcommand{\Te}{T_{\rm e}}

\newcommand{\Tg}{T_{\gamma}}

\newcommand{\sigT}{\sigma_{\rm T}}

\newcommand{\vek} [1]{\mbox{\boldmath${#1}$\unboldmath}}

\newcommand{\pot}[2]{#1 \times 10^{#2}}

\newcommand{\zeq}{z_{\rm eq}}
\usepackage{xspace}
\newcommand{\FIRAS}{{\it COBE/FIRAS}\xspace}
\newcommand{\EDGES}{{EDGES}\xspace}
\newcommand{\ARCADE}{{ARCADE}\xspace}
\newcommand{\PIXIE}{{\it PIXIE}\xspace}

\newcommand{\Planck}{{\it Planck}\xspace}

\newcommand{\mXmax}{{20\,\keV}}
\newcommand{\xinj}{{x_{\rm inj}}}
\newcommand{\xinjc}{{x_{\rm inj, 0}}}
\newcommand{\nuinj}{{\nu_{\rm inj}}}
\newcommand{\Einj}{{E_{\rm inj}}}
\newcommand{\Einjc}{{E_{\rm inj, 0}}}
\newcommand{\Ginj}{\Gamma_X}
\newcommand{\Ginjeq}{{\Ginj^{\rm eq}}}
\newcommand{\Ginjeqv}{{\pot{2}{-13}\,{\rm s}^{-1}}}
\newcommand{\ratesec}{{{\rm s}^{-1}}}
\newcommand{\tinj}{{t_{\rm X}}}
\newcommand{\finj}{{f_{\rm inj}}}
\newcommand{\meV}{{{\rm meV}}}
\newcommand{\GeV}{{{\rm GeV}}}
\newcommand{\TeV}{{{\rm TeV}}}
\newcommand{\fdm}{f_{\rm dm}}
\newcommand{\fdms}{f^*_{\rm dm}}
\newcommand{\efdm}{\epsilon \fdm}

\renewcommand{\zi}{{z_{\rm inj}}}
\newcommand{\LyC}{{\rm Ly{\textendash}c}}

\date{\vspace{-3mm}{Accepted XXX --. Received December 2020}}

\begin{document}
\title[Spectral distortion from photon injection]{Spectral distortion constraints on photon injection from low-mass decaying particles}

\author[Bolliet et al.]{Boris Bolliet$^{1,2}$\thanks{boris.bolliet@gmail.com}, 
Jens Chluba$^{1}$\thanks{jens.chluba@manchester.ac.uk} and 
Richard Battye$^{1}$\thanks{richard.battye@manchester.ac.uk}
\\
$^1$ Jodrell Bank Centre for Astrophysics, School of Physics and Astronomy,
The University of Manchester, Manchester, M13 9PL, U.K.\\
$^2$ Columbia Astrophysics Laboratory, Columbia University, 550 West 120th Street, New York, NY, 10027, USA
\vspace{-3mm}
}

\maketitle	

\begin{abstract}
Spectral distortions (SDs) of the cosmic microwave background (CMB) provide a powerful tool for studying particle physics. Here we compute the distortion signals from decaying particles that convert directly into photons at different epochs during cosmic history, focusing on injection energies $E_\mathrm{inj}\lesssim 20\,\mathrm{keV}$. We deliver a comprehensive library of SD solutions, using {\tt CosmoTherm} to compute the SD signals, including effects on the ionization history and opacities of the Universe, and blackbody-induced stimulated decay. Then, we use data from \textit{COBE/FIRAS} and EDGES to constrain the properties of the decaying particles. We explore scenarios where these provide a dark matter (DM) candidate or constitute only a small fraction of DM.   We complement the SD constraints with CMB anisotropy constraints, highlighting new effects from injections at very-low photon energies ($h\nu\lesssim 10^{-4}\,\eV$).  Our model-independent constraints exhibit rich structures in the lifetime-energy domain, covering injection energies $E_\mathrm{inj}\simeq 10^{-10}\mathrm{eV}-10\mathrm{keV}$ and lifetimes $\tau_X\simeq 10^5\,\mathrm{s}-10^{33}\,\mathrm{s}$. We discuss the constraints on axions and axion-like particles, revising existing SD constraints in the literature. Our limits are competitive with other constraints for axion masses $m_a c^2\gtrsim 27\,\eV$ and we find that simple estimates based on the overall energetics are generally inaccurate. Future CMB spectrometers could significantly improve the obtained constraints, thus providing an important complementary probe of early-universe particle physics.
\end{abstract}

\begin{keywords}
Cosmology: cosmic microwave background -- theory -- observations.
\vspace{-4mm}
\end{keywords}

%-------------------------------------------------------------------
\section{Introduction}
\label{sec:IN}
%-------------------------------------------------------------------
The average energy spectrum of the cosmic microwave background (CMB) is known to be extremely close to that of a perfect blackbody at a temperature $\TCMB=2.7255\,\Kel$ \citep{Mather1994, Fixsen1996, Fixsen2009}. However, out-of-equilibrium processes lead to departures from the Planckian spectrum \citep{Zeldovich1969, Sunyaev1970mu, Illarionov1974, Danese1977}, causing so-called CMB spectral distortions (SDs).
The presence of SDs has been tightly constrained using \FIRAS, ruling out significant episodes of energy release after thermalization becomes inefficient at redshift $z\lesssim \pot{2}{6}$  \citep{Mather1994, Wright1994, Fixsen1996, Fixsen2009}. 
This provides strong limits on various early-energy release scenarios \citep[e.g.,][]{Burigana1991, Hu1993, Chluba2013PCA}, and in the future these limits are expected to be significantly improved with novel spectrometer concepts such as \PIXIE \citep{Kogut2011PIXIE, Kogut2016SPIE} and its enhanced versions \citep{PRISM2013WPII, Kogut2019BAAS}, including a possible mission concept for the ESA Voyage 2050 program \citep[see][]{Chluba2019Voyage}. This promising perspective has spurred significant interest in SD science in the last decade \citep[e.g.,][]{Chluba2011therm, Sunyaev2013, deZotti2015, Chluba2019BAAS, Chluba2019Voyage, Lucca2020}. We refer to \citet{Chluba2018Varenna} for a broad introduction to the science of CMB spectral distortions.

Moreover, the results from ARCADE2 \citep{Fixsen2011excess} and EDGES \citep{Bowman:2018vm} have proven difficult to interpret with standard astrophysical assumptions \citep[e.g.,][]{arcade2, Feng2018, Hardcastle2020}, and may be consistent with a brightness temperature in the Rayleigh-Jeans tail of the CMB that is significantly larger than the one corresponding to the \FIRAS measurement (see Sect.~\ref{ss:edges} for discussion). These low-frequency measurements of the background radiation with unexpected feature are another motivation for the study of SDs.

There are two main ways of creating distortions. The most commonly considered is due to the {\it injection of energy}, which leads to the heating of electrons, generally causing the classical $\mu$- and $y$-type distortions through Comptonization, i.e., the repeated scattering of photons by free electrons \citep{Zeldovich1969, Sunyaev1970mu}. 
The other is related to directly {\it adding or removing photons} from the CMB \citep{Hu1995PhD, Chluba2015GreensII}. One concrete example is the SD caused by the cosmological recombination process \citep[see][for an  overview]{Sunyaev2009},  primarily due to the photons created in uncompensated atomic transitions of hydrogen and helium.  The corresponding signal is small  but extremely rich  \citep[see][for recent calculations and forecasts]{Hart2020CRR}. For the cosmological recombination radiation, Comptonization only plays a minor role because around recombination (and after) the Compton scattering time scale $t_\mathrm{C}$ is already much longer than the expansion time scale $t_{H_0}$ \citep{Jose2008, Chluba2016CosmoSpec}; however, decaying or annihilating particle scenarios can in principle cause direct photon injection at earlier times, requiring a careful thermalization treatment, which is expected to result in signals that differ significantly from the classical Comptonization distortions, in particular when occurring at redshift $z\lesssim 10^5$ (i.e., when  $t_\mathrm{C}\gtrsim t_{H_0}$) and for photon energies $h\nu\lesssim 10\times k\TCMB(z)$ \citep[see][]{Chluba2015GreensII}. 

While the data from \FIRAS has been extensively used to constrain energy release processes, no comprehensive analysis of decaying particle scenarios with photon injection has been carried out.
The main goal of this paper is to derive SD constraints on decaying particle scenarios that directly lead to the production of photons at energies $h\nu \lesssim \mXmax$. This upper limit in energy is chosen for two reasons. First, to avoid complications from high-energy processes   associated with production of non-thermal electrons (see discussion Sect.~\ref{sec:domains_ana}). Second, because at even higher energies most of the effects on the CMB spectrum are indeed captured by treating the transfer of energy to the baryons \citep[e.g.,][]{Chluba2015GreensII}, thus essentially mimicking pure energy release \citep[e.g.,][for classical references on the topic]{Ellis1992, Sarkar1984, Hu1993b, McDonald2001}. 
For our computations, the particles are assumed to be non-relativistic and cold, with a constant lifetime parameter, injection energy and free abundance, which we express relative to the dark matter (DM) density. Both decays in vacuum and within the ambient CMB field, which gives rise to stimulated decay, are considered.
We furthermore take the effect of photon injection on the ionization history into account, carefully accounting for the extra heating, ionizations and collisional processes by modifying {\tt CosmoTherm} \citep{Chluba2011therm} and {\tt Cosmorec/Recfast++} \citep{Chluba2010b}. We use these computations to create a comprehensive library of spectral distortion solutions that are then translated into limits on the particle properties.
This extends the computations carried out by \citet{Chluba2015GreensII}, which focused on single injections in time and energy, during the pre-recombination era, bringing us one step closer to treating more general scenarios.

To derive constraints, we use the spectral distortion data from \FIRAS, marginalizing over a  galactic foreground in the same way as \cite{Fixsen1996}. For injections at energies well below the maximum of the CMB blackbody, we also consider the observations of \EDGES \citep{Bowman:2018vm} as an additional upper limit on the radio background, showing that this significantly tightens the obtained limits for injection energies $10^{-6}\,\eV\lesssim h\nu \lesssim \pot{\rm few}{-4}\,\eV$ (see Fig.~\ref{fig:fdm_with_edges_contours}).

Further improvements on our limits can be achieved by considering the effects of photon injection on the ionization history,  since these can be constrained by CMB anisotropy data, as  measured by \Planck \citep{Planck2018params}. Here, in order to avoid computing the CMB anisotropy spectra and comparing them with \Planck for each ionisation history, we use the projection method introduced by \citet{Hart2020PCA} to estimate the constraints.  We find that  for injection at very low ($h\nu\lesssim 10^{-4}\,\eV$) and high ($h\nu\gtrsim 13.6\,\eV$) energies the addition of \Planck data tightens the bounds on decays in the post-recombination era. See Sect.~\ref{xeconstraints} for details.

In that last part of this work, we translate our constraints to the parameter space of axions and axion-like particles (ALPS), highlighting how SD measurements can provide a sensitive probe of particle physics.
These particles are being considered as a possible DM candidate \citep[see, for example][]{Marsh_2016}, given that the standard WIMP scenario is seeing increased observational pressure from direct-detection and collider experiments.
Our constraints at masses $m c^2 \gtrsim \keV$ are comparable to those previously published \citep[e.g.][]{Cadamuro2011,Millea_2015}, but unlike these previous works we use the full SD spectra, including late time evolution during reionization and Lyman absorption to place our constraints, rather than basing them on the approximate $\mu$ and $y$ estimates based on heating. At lower masses, our constraints are not competitive with CAST \citep{CAST2007}; however, we emphasize that this is the first time SD data is used to place an independent constraints in this part of the parameter space, with data that predates many measurements by decades. 
We close by briefly discussing other particle physics scenarios that can be constrained with the SD library we provide here (see Sect.~\ref{sec:num_sols} and conclusions), as a more detailed analysis is left to future work.

Our fiducial cosmology is assumed to be a spatially flat Friedmann-Lema\^itre-Robertson-Walker universe with a cosmological constant and  $\TCMB=2.726\,\mathrm{K}$ for the present CMB temperature, $Y_\mathrm{P}=0.24$ for the helium mass fraction, $N_\mathrm{eff} = 3.046$ effective  relativistic species, $\Omega_\mathrm{m}=0.31$, $\Omega_\mathrm{b}=0.049$ for the present density parameters of matter and baryons respectively, and Hubble constant $H_0 = 67.5\,\mathrm{km}\,\mathrm{s}^{-1}\,\mathrm{Mpc}$. Our results are not strongly dependent on the specific value of the cosmological parameters. 
%

%-------------------------------------------------------------------
\vspace{-3mm}
%\section{Photon injection distortions from decay}
\section{Computing photon injection distortions}
\label{sec:PHO}
%-------------------------------------------------------------------
In this section, we explain how the distortions created by photon injection from decay processes are computed. The starting point is the standard thermalization problem, which accounts for Compton (C), double Compton (DC) and Bremsstrahlung (BR) interactions to evolve the CMB spectrum across time. Schematically, the thermalization equation for the photon occupation number $n_\nu$, with respect to time $\mathrm{d}\tau=\mathrm{d}t/t_\mathrm{C}$ where $t_\mathrm{C}$ is the Thomson scattering time $t_\mathrm{C}=1/\sigT \Ne c$, reads as

\begin{equation}
    \frac{\partial n_\nu}{\partial \tau}-H t_\mathrm{C}\nu \frac{\partial n_\nu}{\partial \nu} = \left.\frac{\mathrm{d} n_\nu}{\mathrm{d} \tau}\right|_\mathrm{C} + \left.\frac{\mathrm{d} n_\nu}{\mathrm{d} \tau}\right|_\mathrm{DC} + \left.\frac{\mathrm{d} n_\nu}{\mathrm{d} \tau}\right|_\mathrm{BR}.\label{eq:pde}
\end{equation}
This problem is solved numerically using {\tt CosmoTherm} \citep{Chluba2011therm}, which we modify by adding an explicit time-dependent source term on the right-hand-side of Eq.~\eqref{eq:pde}, corresponding to photon injection. 
We also account for extra ionizations of hydrogen and helium by hard photons as well as atomic collision and recombination processes.

We consider the decay of a massive particle or an excited state of matter that both lead to the injection of photons at frequency $\nuinj$. For the decay of a particle with mass $m_X$ we assume that two photons are produced, $X\rightarrow \gamma + \gamma$, such that $\Einj=h\nuinj=m_X c^2/2$. In contrast, for the decay of an excited state we have $X^*\rightarrow X+\gamma$ and $\Einj=h\nuinj=E^{\rm ex}_X$, where $E^{\rm ex}_X$ is the particle's excitation energy. The efficiency of photon production per particle thus differs by a factor of two. For simplicity we just refer to both cases as {\it decaying particle scenarios}.\\
%\newpage
We assume that the production of the decaying particle has happened before the distortion era ($z\gtrsim 2\times 10^{6}$). 
The number density of the decaying particle then evolves according to the exponential law
%-------------------------------------------------------------------
\begin{align}
\label{eq:NX_ev_def}
\frac{\id \ln a^3 N_X}{\id t}=-\Ginj,
\end{align}
%-------------------------------------------------------------------
governed by the decay rate $\Ginj$, or the lifetime $\tinj=1/\Ginj$, of the particle. 
For the main discussion below we consider vacuum decay, while Sect.~\ref{sec:stim} is dedicated to  blackbody-stimulated decay for which the time-dependence of the photon injection process in the low-energy limit, i.e., $\Einj \lesssim 0.1\,k\TCMB(z)$, is modified. Note that in Eq.~\eqref{eq:NX_ev_def}, $a$ is the scale factor and $t$ the cosmic time of the FLRW model.

Without decays, from Eq.~\eqref{eq:NX_ev_def} one has $N_X=N_{X, 0} (1+z)^3$, where $N_{X, 0}$ is the would-be number density today. The solution of the equation above then reads $N_X(t)=N_{X, 0} (1+z)^3 \, \expf{-\Ginj t}$. This allows us to define the photon source term for the occupation number $n_x=n(t, x)$ at frequency $x=h\nu/k\TCMB(z)$ in the form
%-------------------------------------------------------------------
\begin{align}
\label{eq:dnx_dt_inj}
\left.\frac{\id n_{x}}{\id t}\right|_{\mathrm{inj}} 
&=\mathcal{G}_{2}\,\finj\Ginj \exp\left(-\Ginj t\right)\times\frac{G\left(x,\xinj,\sigma_{x}\right)}{x^{2}}.
\end{align}
%-------------------------------------------------------------------
Here, we introduced the injection efficiency $\finj$ and the integrals $\mathcal{G}_k=\int x^k/(\expf{x}-1)\id x$ ($\mathcal{G}_2\approx 2.404$ and $\mathcal{G}_3\approx 6.494$). We assume that the photons are injected in a narrow Gaussian with width $\sigma_x$ centered around $\xinj(z)=\Einj/k\TCMB(z)=\xinjc/(1+z)$, where $\xinjc=\Einj/k\TCMB= 4.257 \,[\Einj/\meV]$ with $\TCMB=2.726\,\Kel$. We usually set $\sigma_x=0.05\xinj$ in our computations, but the results do not crucially depend on this choice. Our calculations are thus applicable to cold non-relativistic relic particles such as axions or excited internal states of cold dark matter. 

Equation~\eqref{eq:dnx_dt_inj} is normalized such that the total number of photons injected at any moment is given by
%-------------------------------------------------------------------
\begin{align}
\label{eq:dNg_dt_inj}
\left.\frac{\id \ln a^3 N_{\gamma}}{\id t}\right|_{\mathrm{inj}} 
&=\finj\Ginj \exp\left(-\Ginj t\right)\equiv f_\gamma \Ginj \frac{N_X}{N_{\gamma}},
\end{align}
%-------------------------------------------------------------------
where $f_\gamma=2$ for decaying massive particles and $f_\gamma=1$ for decaying excited states. This then implies 
%-------------------------------------------------------------------
\begin{align}
\label{eq:f_inj}
\finj&=f_\gamma\,\frac{N_{X, 0}}{N_{\gamma, 0}}=\frac{\mathcal{G}_3}{\mathcal{G}_2}\,\frac{\epsilon}{\xinjc}\,\frac{\rho_{X, 0}}{\rho_{\gamma, 0}}
=\frac{\mathcal{G}_3}{\mathcal{G}_2}\,\frac{\epsilon\fdm}{\xinjc} \frac{\rho_{\rm cdm, 0}}{\rho_{\gamma, 0}}
\nonumber
\\
&\approx \pot{1.31}{4}\,\,\frac{\epsilon\fdm}{\xinjc} 
\left[\frac{\Omega_{\rm cdm}h^2}{0.12}\right],%\,\left[\frac{\TCMB}{2.726\,\Kel}\right]^{-4},
\end{align}
%-------------------------------------------------------------------
where in the last steps we give the expression in terms of the dark matter energy density, $\rho_{\rm cdm, 0}$. Furthermore, we have $\epsilon=1$ for decaying particles and $\epsilon=E^{\rm ex}_X/m_X c^2<  1$ for excited states. The excited state scenario is suppressed by a factor $\epsilon<1$ to account for the reduction of the particle number density, which in both cases is given by $N_{X, 0}=\rho_{X, 0}/m_X c^2$. Model-independent constraints are then obtained for the effective dark matter fraction, $\fdms=\efdm$.

The injected photons can i) Compton scatter with electrons; ii) be absorbed in a BR or DC event; iii) interact with the atoms in the Universe and iv) simply remain in the CMB spectrum as a {\it direct distortion}. The processes i)-iii) all lead to heating/cooling of the matter. This in turn causes $\mu$ and $y$-type distortion contributions and indirectly changes to the ionization history, depending on the epoch at which the injection happens. In addition, iii) directly changes the ionization history, as we explain below.

\vspace{-3mm}
\subsection{Energy release histories}
%----------------------------------------------------------
When studying decaying particle scenarios, it is instructive to first understand the time-dependence of the injection process. The parameters $\finj$  and $\Einj$ determine how much energy is added and at which frequency. These do not directly affect the time-dependence of the injection process, solely controlled by $\Ginj$. 
The energy release history is directly obtained from Eq.~\eqref{eq:dnx_dt_inj} as
%-------------------------------------------------------------------
\begin{align}
\label{eq:drho_dt_inj}
\frac{\id \ln a^4 \rho_{\gamma}}{\id t} 
&=\frac{\mathcal{G}_{2}}{\mathcal{G}_{3}}\,\finj \xinj\,\Ginj \exp\left(-\Ginj t\right)
\nonumber
\\
&=
\epsilon \fdm \frac{\rho_{\rm cdm, 0}}{\rho_{\gamma, 0}}\, \frac{\Ginj \exp\left(-\Ginj t\right)}{1+z}, 
\end{align}
%-------------------------------------------------------------------
where $\xinj=\xinjc/(1+z)$ gives rise to a factor $1/(1+z)$. Note that we also have $\rho_{\rm cdm, 0}/\rho_{\gamma, 0}\simeq \pot{4.85}{3}\,[\Omega_{\rm cdm}h^2/0.12]$. 

%----------------------------------------------------------
\begin{figure}
\begin{centering}
\vspace{-0mm}
\includegraphics[trim={2mm 3mm 0 0},clip, width=\columnwidth]{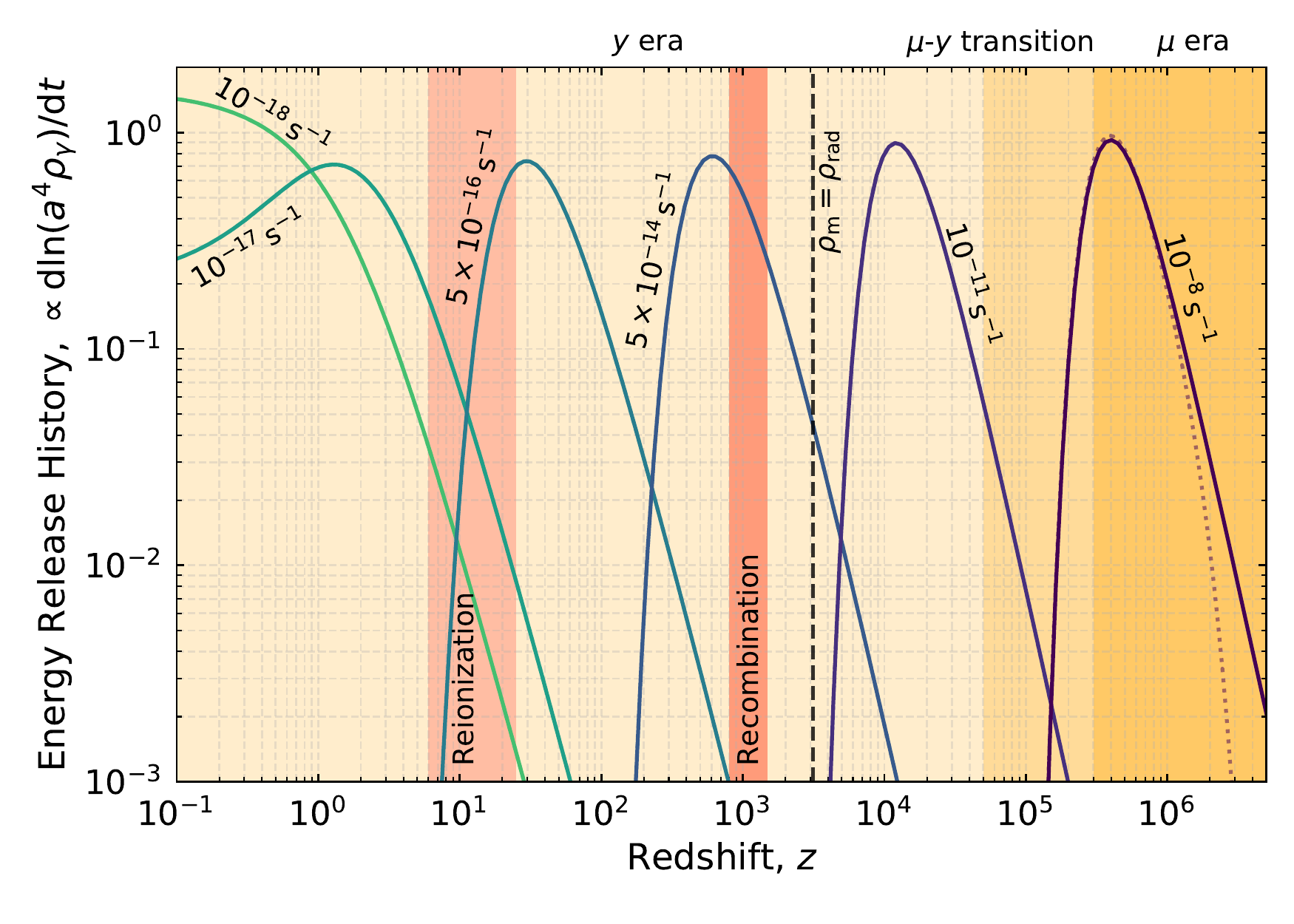}
\end{centering}
 \vspace{-0mm}
 \caption{Energy release history for several lifetimes. For distortions created by heating, the function already determines the form of the spectral distortion, while for photon injection distortions a more complicated interplay of injection time and energy is found. The dotted line illustrates the effective energy release history when accounting for thermalization (see Sect.~\ref{sec:distortion_results})
} 
\label{fig:drho_rho}
\vspace{-0mm}
\end{figure}
%----------------------------------------------------------

%----------------------------------------------------------
\begin{figure}
\begin{centering}
\includegraphics[trim={3mm 3mm 4mm 0},clip, width=\columnwidth]{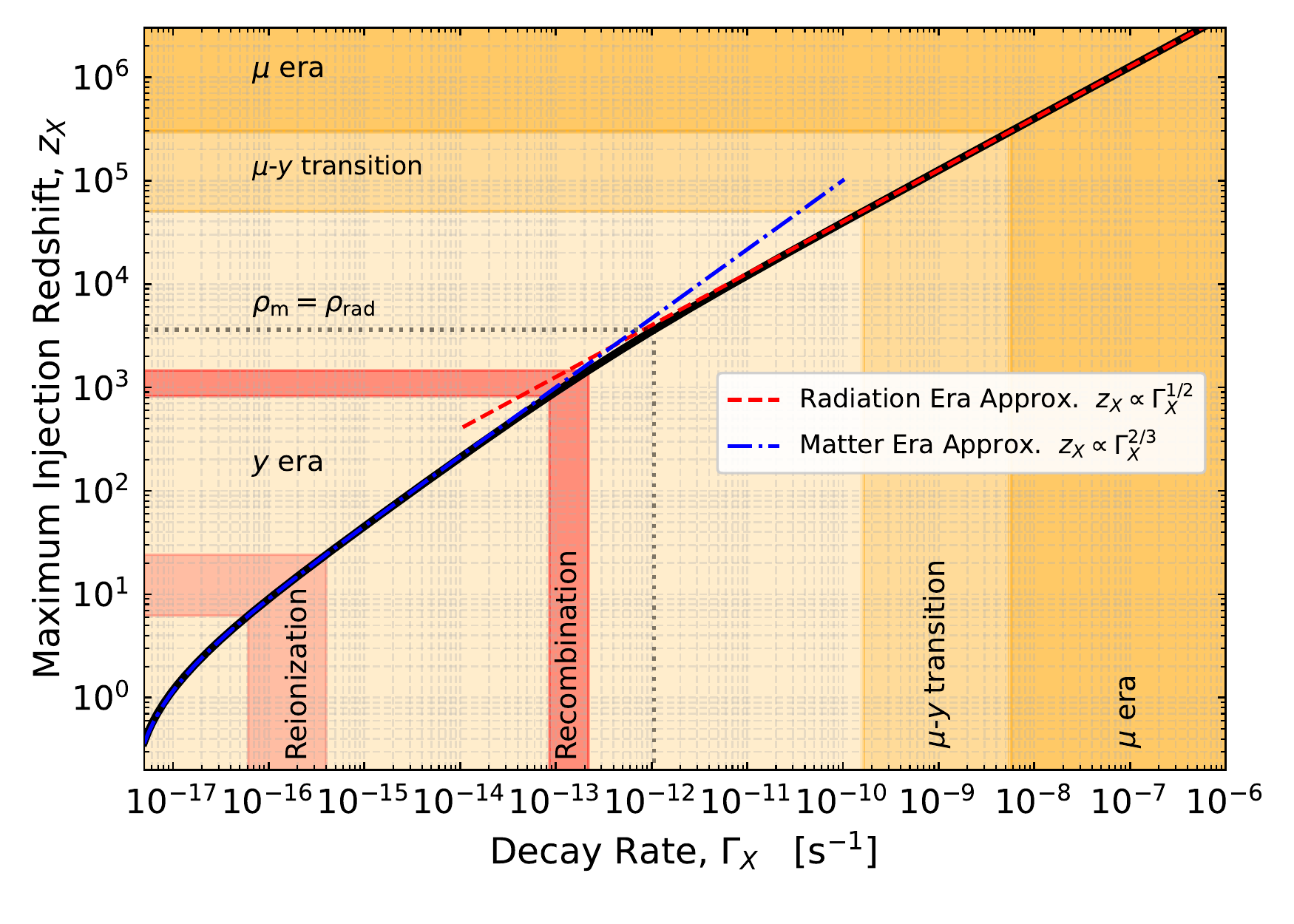}
\end{centering}
\vspace{-0mm}
\caption{The relationship between $\Ginj$ and the maximal injection redshift $\zX$. The solid black line shows the exact numerical result, while the two other coloured curves are the approximations in Eq.~\eqref{eq:zx_max}.} 
\label{fig:zx_Ginj}
\vspace{2mm}
\end{figure}
%----------------------------------------------------------

In Fig.~\ref{fig:drho_rho}, we show the energy release histories for several lifetimes. We normalized all of them to a total energy release of $\Delta\rho/\rho=\pot{3}{-5}$, which corresponds to the 68\% CL limit from \FIRAS  (see Section \ref{sec:FIRAS_setup}). Using the analytical approximations for the relationship between time and redshift in the matter and radiation dominated era, we find that the maximum of energy release occurs at redshift $z_X$ such that
%-------------------------------------------------------------------
\begin{align}
\label{eq:zx_max}
1+z_X &\simeq
\begin{cases}
 1.26\times10^{4}\left[\frac{\Ginj}{10^{-11}\mathrm{s^{-1}}}\right]^{1/2} 
 & \,\Ginj \gtrsim \Ginjeq
 \\[1mm]
 2.17\times10^{2}\left[\frac{\Omega_{\mathrm{m}}h^{2}}{0.15}\right]^{-1/3}\left[\frac{\Ginj}{10^{-14} \,{\rm s}^{-1}}\right]^{2/3} 
 &\,\Ginj \lesssim \Ginjeq.
\end{cases}
\end{align}
%-------------------------------------------------------------------
A comparison of these approximations with the exact result is shown in Fig.~\ref{fig:zx_Ginj}.
 We note that the redshift of matter-radiation equality is given by $1+\zeq=\Omega_\mathrm{m}/\Omega_\mathrm{rel}\simeq \pot{3.58}{3}\,[\Omega_{\mathrm{m}}h^2/0.15]$, and the decay rate at $\zeq$ is $\Ginjeq\simeq \Ginjeqv$. For $\Ginj\lesssim \pot{2.3}{-18}\,\ratesec$ (i.e., lifetime longer than the age of the Universe), no maximum in the injection history is present at $z>0$.

The total number of injected photon and energy release are obtained by integrating Eq. \ref{eq:dNg_dt_inj} and \ref{eq:drho_dt_inj} over time. One finds
%----------------------------------------------------------
\begin{subequations}
\begin{align}
\left.\frac{\Delta N_\gamma}{N_\gamma}\right|_{\rm tot}
&= \finj \,\left[1-\exp\left(-\Ginj t_0\right)\right]
\\
\left.\frac{\Delta \rho_\gamma}{\rho_\gamma}\right|_{\rm tot}
&= \frac{\mathcal{G}_2 }{\mathcal{G}_3} \finj\,\xinjc \,
\int_0^{t_0} \frac{\Ginj \, \expf{-\Ginj t}}{(1+z)}\,\id t
\label{eq:norms_b}
\\
&\approx 
\frac{\mathcal{G}_2 }{\mathcal{G}_3} \finj\,\xinjc
\times
\begin{cases}
\frac{\Ginj\,t_0}{2} & \text{for}\,\,\Ginj\,t_0 \ll 1
\\[1mm]
\sqrt{\frac{\pi}{2}}\,\frac{\Omega_{\rm rel}^{1/4}}{\sqrt{\Ginj\,t_0}} & \text{for}\,\,\Ginj\,t_0 \gg 1,
\end{cases}
\label{eq:norms_c}
\end{align}
\end{subequations}
%----------------------------------------------------------
where $t_0= 1/H_0=3.086\times 10^{17}\mathrm{s}/h \approx \pot{4.41}{17}\,{\rm s}$ is the Hubble time. These two relations can be used to set approximate initial conditions for $\finj$ once $\xinjc$ and $\Ginj$ are chosen. (In our computations, we also take into account the thermalization efficiency, as explained in Sect.~\ref{sec:distortion_results}.)

For distortions created by energy release, simple estimates for $\mu$ and $y$ distortions can be obtained by integrating the energy release history multiplied by appropriate distortion visibility functions \citep[e.g.,][]{Chluba2013Green, Chluba2016}.
In particular, for distortions created by photon injection from a decaying particle with short lifetime ($\tau_X\lesssim 10^{10}\,{\rm s}$) it is crucial to consider the number of photons added by the injection process. This can lead to both negative and positive $\mu$-distortions if injection occurs for $z\gtrsim \pot{3}{5}$ \citep{Chluba2015GreensII}, as we will see below.
In addition, for longer lifetime when the injection happens in the post-recombination era, the transparency of the Universe to photons is a strong function of their energy \citep[e.g.,][Fig.~10]{Chluba2015GreensII}, and the simple $\mu$ and $y$ distortions are not appropriate to describe the final spectrum. 

For $\Ginj \ll 1/t_0$, one furthermore finds that the particles are essentially stable and the decay follows a power-law in redshift. This leads to a quasi-universal distortion shape that mainly depends on the injection energy and scales linearly with $\Ginj$ (see Sect.~\ref{LLp}).

\subsection{Interactions with atomic species}
%----------------------------------------------------------
Around the recombination era, significant fractions of neutral hydrogen and helium atoms in the ground-state form \citep{Zeldovich68, Peebles68, Seager2000}. For photons injected at energies above the ionization thresholds of hydrogen and helium, this strongly affects the distortion evolution, as photo-ionization processes not only remove photons from the CMB bands, but also affect the ionization history and lead to heating of the medium.  Accounting for both absorption and emission to the ground state, the terms in the evolution equation of the photon occupation number Eq.~\eqref{eq:pde}, related to hydrogen Lyman-continuum (Ly-c) transfer, can be written as
%----------------------------------------------------------
\begin{align}
\label{eq:em_ab_higha}
\frac{1}{c}\,\frac{\partial n_\nu}{\partial t}\Bigg|_{\LyC}
=  N^{\rm eq}_{\rm 1s}\,\sigma_{\rm 1s}(\nu) \,\expf{-\xe}\,(1+n_\nu) - N_{\rm 1s} \sigma_{\rm 1s}(\nu) \,n_\nu,
\end{align}
%----------------------------------------------------------
where $\sigma_{\rm 1s}(\nu)$ is the photoionization cross section\footnote{We neglect the thermal broadening due to the motion of the atoms when computing the photoionization cross section.} of the ground-state, $N^{\rm eq}_{\rm 1s}$ is the equilibrium population with respect to the continuum, $N^{\rm eq}_{\rm 1s}=\Ne\,N_{\rm p} f_{\rm 1s}(\Te)$, and $\xe=h\nu/k\Te$.  Here, $f_{\rm 1s}(\Te)$ is a temperature-dependent factor that follows from Saha-equilibrium and detailed balance \citep[e.g., see][for explicit expression]{Seager2000,Chluba2007b}. Note that these terms are time dependent, in particular due to the evolution of the electron temperature $\Te$, which is solved simultaneously \citep[see][]{Chluba2011therm}.
Similar terms arise for neutral and singly-ionized helium. Ionizations from excited states remain negligible, since at any stage during recombination the populations of the excited levels remain small \citep[e.g.,][]{Jose2006, Chluba2007}. We also omit the effects of atomic excitations from the ground-state (Lyman-series for hydrogen), as they  do not lead to a large energy transfer \citep[e.g.,][]{Basko1981, Grachev2008, Hirata2009, Chluba2009b} and would only affect the CMB spectrum in the distant Wien tail.

To account for the continuum transfer self-consistently is beyond the scope of this paper; however, to capture the main effects, we approach the problem in the following way: once significant absorption occurs, the re-emission process is heavily suppressed because the Lyman-continuum escape probability is small \citep{Chluba2007b, Chluba2008c}. Therefore, photons escaping from the Lyman-continuum will only lead to a very small distortion in the Wien tail of the CMB blackbody, which we shall neglect here. 
We thus assume that photons injected in excess of the CMB blackbody can be efficiently absorbed, but will not be re-distributed or re-emitted. The kinetic energy of the liberated electron is quickly thermalized by Coulomb interactions and thus leads to heating of the medium. We will we show that for photons below a certain critical energy this is a good approximation. 
We thus approximate Eq.~\eqref{eq:em_ab_higha} by
%----------------------------------------------------------
\begin{align}
\label{eq:em_ab_highb}
\frac{\partial n_x}{\partial t}\Bigg|_{\rm Ly-c}
\approx - N_{\rm 1s} \sigma_{\rm 1s}\,c \,\Delta n_x
\end{align}
%----------------------------------------------------------
only accounting for the absorption of the distortion $\Delta n(x)=n(x)-\nbb(x)$ with respect to the CMB blackbody, $\nbb(x)=[e^x-1]^{-1}$. This leads to a matter heating term
%----------------------------------------------------------
\begin{align}
\label{eq:em_ab_high_heating}
\frac{\partial \rho_{\rm m}}{\partial t}\Bigg|_{\rm Ly-c}
&\approx \frac{8\pi}{c^2} \int^\infty_{\nu_{\rm 1sc}}  N_{\rm 1s}\sigma_{\rm 1s} \,h[\nu-\nu_{\rm 1sc}]\, \nu^2 \Delta n_\nu \id \nu
\nonumber\\
&=\frac{\rho_\gamma^{\rm pl}(\Tg)}{G^{\rm pl}_3}
\int^\infty_{x_{\rm 1sc}} c N_{\rm 1s} \sigma_{\rm 1s} [x-x_{\rm 1sc}]\, x^2 \Delta n_x \id x,
\end{align}
%----------------------------------------------------------
where $x=h\nu/k\Tg$ and $\nu_{\rm 1sc}$ is the ionization frequency (the subscript 1sc denotes 1s to continuum). In particular at low redshifts in the post-recombination era, this term will be crucial, as it leads to a strong thermal coupling of high energy photons to the matter, thereby creating $y$-type distortions that would otherwise not appear.
We also include the extra ionizations into the rate equations for the ionization history calculation using
%----------------------------------------------------------
\begin{align}
\label{eq:em_ab_high_ion}
\frac{\partial N_{\rm 1s}}{\partial t}\Bigg|_{\rm Ly-c}
&\approx-\frac{8\pi}{c^2} \int^\infty_{\nu_{\rm 1sc}}  N_{\rm 1s}\sigma_{\rm 1s} \, \nu^2 \Delta n_\nu \id \nu
\nonumber\\
&=-\frac{N_\gamma^{\rm pl}(\Tg)}{G^{\rm pl}_2}
\int^\infty_{x_{\rm 1sc}} c N_{\rm 1s} \sigma_{\rm 1s} \, x^2 \Delta n_x \id x,
\end{align}
%----------------------------------------------------------
which again can lead to significant changes in the ionization history for decay with energies above $\simeq 13.6\,\eV$.

\subsubsection{Scattering of high energy photons by neutral atoms}
\label{ss:high_energy_photons}
%----------------------------------------------------------
It is well-known that for energies far above the ionization threshold of the atoms, photons do not distinguish between free or bound electrons \citep{Eisenberger1970, Sunyaev1996, Houamer2020}. Therefore, the interaction cross section for scattering and photo-ionization essentially become indistinguishable and approach the Compton scattering cross section for free electrons. Treating this problem in full detail is far beyond the scope of this paper. In particular for photons close to the ionization thresholds, differences in the way energy and momentum are redistributed will affect some of the scattering dynamics that in principle requires an independent Fokker-Planck treatment.

However, one can estimate the photon energy above which the scattering by bound electrons becomes similar to a Compton event by computing the energy that is transferred to the electron. At high photon energies, in each scattering event the energy transfer is dominated by electron recoil $\Delta E\simeq (h\nu)^2/\me c^2$ \citep{Sunyaev1996}. Equating this to the ionization energy, yields the critical photon energies $h\nu_{\rm crit}\simeq 2.6\,\keV$ for hydrogen, $h\nu_{\rm crit}\simeq 3.5\,\keV$ for \HeI and $h\nu_{\rm crit}\simeq 5.3\,\keV$ for \HeII. Below these energies, we can assume that for neutral atoms only photo-ionization matters, while above, we may assume that neutral atoms directly contribute to Compton scattering. In this case, a small correction to the Compton heating should be added for bound electrons, as they lose the ionization energy without actually heating the medium; however, we shall neglect this effect in our computations. This will slightly affect the results for post-recombination injection above these thresholds, however, not by order of magnitude.

\subsection{Including collisions}
\label{sssec:colls}
%----------------------------------------------------------
Collisional processes can directly affect the recombination dynamics. This usually has a minor effect in the standard calculation \citep{Seager2000, Chluba2010, Chluba2016CosmoSpec}, but during the dark ages and reionization in particular collisional ionization can become important \citep{Chluba2015PMF}. Since the photon injection scenarios considered here can lead to significant heating, the contributions from collisions between free electrons and ions are indeed found to become relevant for injections associated with large heating (see Sect.~\ref{sec:collisions_effect}). 

To estimate the collisional ionization rates, $R_{{\rm 1s c}}^{\rm coll}$, due to electron impact, we use the fits given for hydrogen and helium given by \citet{Bell1983}. At low temperatures, these rates are often simplified \citep[e.g., see expressions given in][]{Chluba2015PMF}, but here we include the more broadly applicable representations of \citet{Bell1983}, which correctly capture the decrease of the collisional coefficients at high temperatures. This improvement is found to be noticeable during the dark ages. For collisions connecting to the continuum, one has the net rate
%----------------------------------------------------------------
\begin{align}
\label{eq:net_collisions}
\Delta C_{{\rm1s c}}^{\rm coll}
\approx R_{{\rm 1s c}}^{\rm coll}\,\Ne\,\left[N^{\rm eq}_{\rm 1s}-N_{\rm 1s}\right],
\end{align}
%----------------------------------------------------------------
where we only consider transitions from the ground-state. If requested, the required rate equations are added to {\tt CosmoTherm} to allow following the ionization history. We note that here we use rates that account for thermal population of electron and omit potential non-thermal electron contributions. 

Collisions lead to extra ionizations, while collisional recombination is only relevant in the pre-recombination era, since three-body interaction require high densities. The liberated electron carries some excess kinetic energy, which is quickly thermalized inside the baryonic plasma. However, the energy required for the ionization event causes a net cooling of the baryons and is added as a heat sink to the electron temperature equation. 

\vspace{-3mm}
\section{Main domains for the solution}
\label{sec:domains_ana}
%----------------------------------------------------------
The main properties of the solutions can be understood by considering the fate of photons injected at a single frequency and time. For the pre-recombination Universe, this was already done previously \citep{Chluba2015GreensII}. Here, we extend discussion to lower redshifts and also consider the energy exchange for energetic photons more carefully, including ionizations and thresholds for the production of non-thermal electrons. The resulting domains are summarized in Fig.~\ref{fig:regimes} and will be explained in the proceeding sections.

%----------------------------------------------------------
\begin{figure}
\includegraphics[width=\columnwidth]{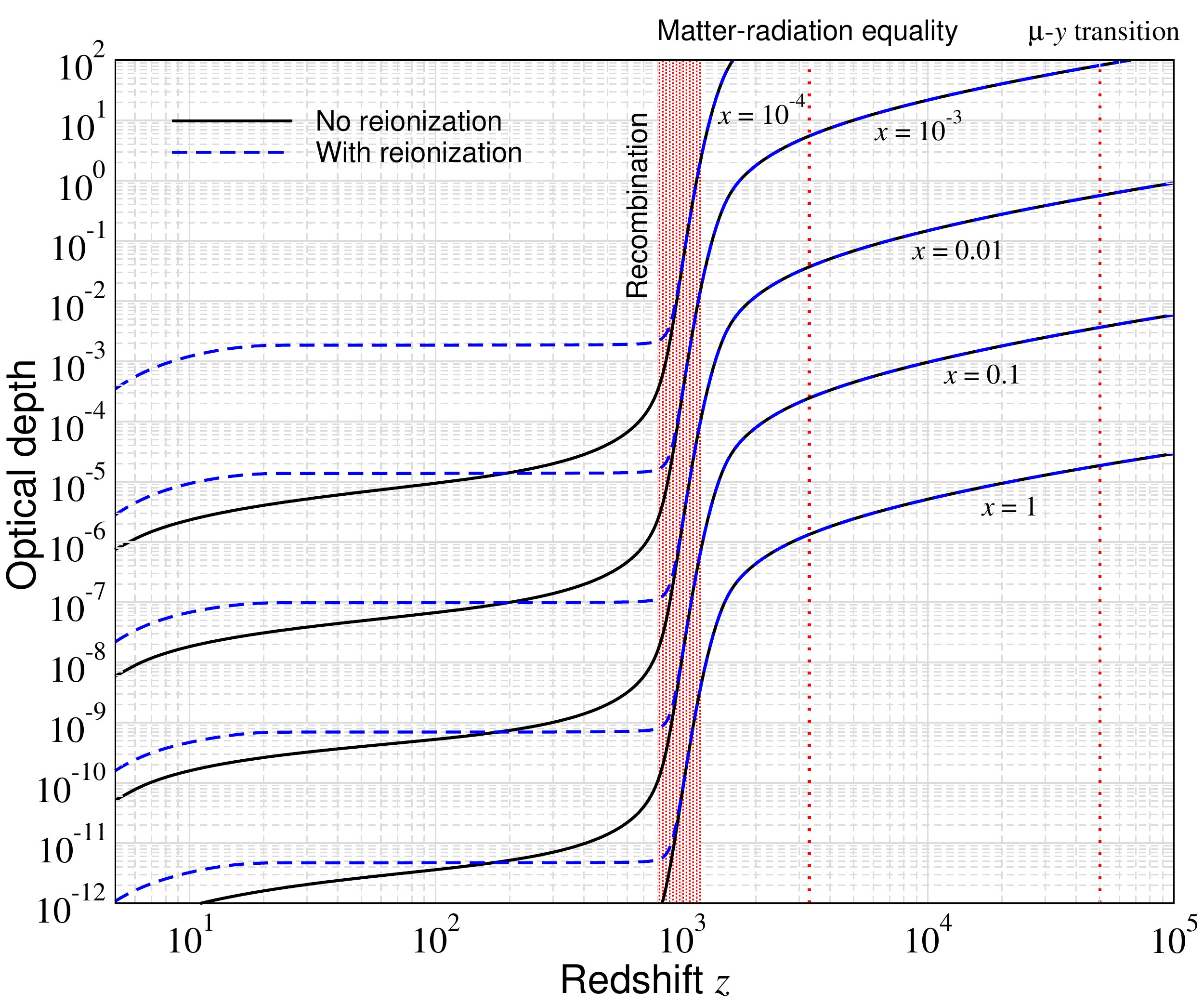}
\vspace{-0mm}
 \caption{Absorption optical depth at low frequencies with and without reionization. The medium becomes optically thin before reionization, explaining the plateaus at $z\simeq 20-1000$. }
\label{fig:tau_abs}
\vspace{-0mm}
\end{figure}
%----------------------------------------------------------

%----------------------------------------------------------
\begin{figure*}
\includegraphics[width=1.7\columnwidth]{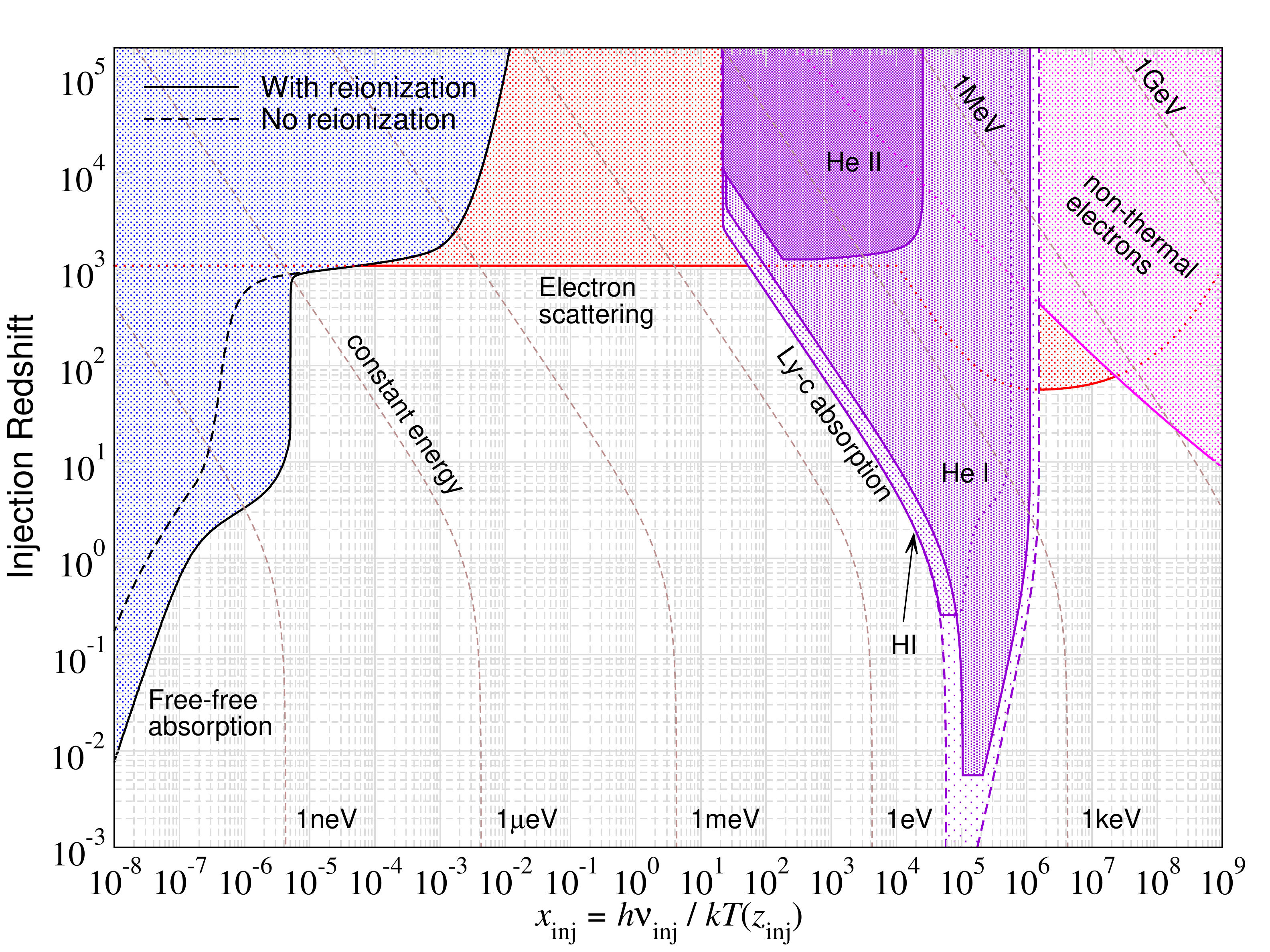}
\vspace{-2mm}
 \caption{Regions of large opacity ($\tau\geq 1$) to free-free absorption (blue; see Sect.~\ref{sec:low_freq_abs}), hydrogen and helium photo-ionization (purple; see Sect.~\ref{sec:high_freq_abs}), electron scattering (bound and free; red; see Sect.~\ref{sec:energy_exchange}) and non-thermal electron production (magenta; see Sect.~\ref{sec:energy_exchange}). Lines of constant energy in the dimensionless variable $\xinj=h\nuinj/k\Tg(\zi)=\xinjc/(1+\zi)$ are shown as brown dashed lines. The pair-production threshold is around the 1~\MeV~line. Non-thermal electron production at $z\lesssim 10^5$ is negligible at photon energies $h\nuinj\lesssim 20\,\keV$ or $\xinjc\lesssim 10^{8}$. The standard ionization history obtained using {\tt CosmoRec/Recfast++} with and without reionization module activated (see Sect.~\ref{sec:low_freq_abs} for details) was used to compute the domains. (HI and HeI denote neutral hydrogen and helium, HeII denotes singly ionized helium.)} 
\label{fig:regimes}
\vspace{-0mm}
\end{figure*}
%----------------------------------------------------------

\vspace{-3mm}
\subsection{Photon absorption at low frequencies}
\label{sec:low_freq_abs}
%----------------------------------------------------------
At low frequencies, DC and BR become the dominant processes controlling the solution. Together, these drive the photon occupation number towards a blackbody at the electron temperature, $\nbb(\xe)=1/[\expf{\xe}-1]$ with $\xe=h\nu/k\Te$. At low redshifts ($z\lesssim 10^4$), Compton scattering becomes extremely inefficient and it is thus most important to ask at which frequency the medium becomes optically thick to photon emission and absorption processes. We include both DC and BR into the estimate, but BR dominates over DC emission at $z\lesssim \pot{4}{5}$. Neglecting Compton scattering, at low frequencies we have the evolution equation\footnote{For the estimates presented in this section, we neglect any source terms and also the small changes to the blackbody induced by differences in the electron and photon temperature. All these effects are included in the main computation using {\tt CosmoTherm}.} for $\Delta n=n-\nbb(\xe)$
%----------------------------------------------------------
\begin{align}
\label{eq:em_ab_low}
\frac{\partial \Delta n}{\partial \tau}
\approx -\frac{\Lambda(\xe, \tau)}{\xe^3} (1-\expf{-\xe}) \Delta n,
%\approx -\frac{\Lambda(\xe, \tau)}{\xe^2}  \Delta n
\end{align}
%----------------------------------------------------------
where $\id \tau = \Ne \, \sigT\,c \id t$ determines the Thomson scattering optical depth and $\Lambda(\xe)$ describes the photon production (note the minus sign for absorption) rate by DC and BR, which depends on the number density of free electrons and ions, as well as the electron temperature \citep[see][for more details]{Chluba2011therm, Chluba2015GreensII}. For estimates, we use the standard solutions from {\tt CosmoRec/Recfast++} \citep{Chluba2010b}  for the ionization history, distinguishing scenarios with and without reionization at $z\simeq 10$. To compute the BR emission coefficient, we use {\tt BRpack} \citep{BRpack2020} for the free-free Gaunt factor. Reionization is modeled as in \cite{Short2019}, with a refined treatment of both singly- and doubly-ionized helium following the approach of \verb|CosmoSpec| \citep{Chluba2016CosmoSpec}.

The relevant characteristics of the solution to Eq.~\eqref{eq:em_ab_low} are then determined by the absorption optical depth \citep[see also][]{Chluba2015GreensII}
%----------------------------------------------------------
\begin{align}
\tau_{\rm abs}(\nuinj, \zi)&\approx \int_0^{\zi} \frac{\Lambda\left(\xe(z'), z'\right)}{\xe^2(z')} \frac{\Ne(z') \sigT c \id z'}{H(z')(1+z')}
\end{align}
%----------------------------------------------------------
where $\zi$ is the injection redshift and $\xinj=h\nuinj/k \Tg(\zi)$ is determined by the injection frequency, $\nuinj$. Since we are considering late times, $\Te$ can depart significantly from $\Tg=\TCMB(1+z)$. This effect has to be included when carrying out the optical depth integral, and amounts to writing $\xe(z)=\xinj(z) \,\phi(z)$ with $\phi=\Tg/\Te$ given by {\tt CosmoRec/Recfast++}. 

In Fig.~\ref{fig:tau_abs}, we illustrate the absorption optical depth for some examples. The cases without reionization are similar to those presented in \citet{Chluba2015GreensII}, but here we used {\tt BRpack} for the BR Gaunt factors\footnote{We assumed that the electron temperature never drops below $\Te=1\,\Kel$ to avoid unphysical contributions.} and also keep all non-linear terms in $\xe$. 
Reionization significantly increases the free-free opacity at $z \lesssim 10$, causing photons to be efficiently absorbed at $x\lesssim \pot{{\rm few}}{-8}$ for all redshifts. Quantitatively, this is shown in Fig.~\ref{fig:regimes}, where we highlight the domain with $\tau_{\rm abs}\geq 1$ as a function of $\xinj$ and $\zi$. Photons injected inside this domain are quickly converted into heat, sourcing $\mu$ and $y$-distortions but also hindering electrons from recombining. Both aspects can be used to place limits on these cases.

\subsection{Absorption of photons at high frequencies}
\label{sec:high_freq_abs}
%----------------------------------------------------------
The photo-ionization optical depth for the hydrogen atoms can be computed by \citep[e.g.,][]{Chluba2007b}
%----------------------------------------------------------
\begin{align}
\tau_{\rm Ly-c}(\nuinj, \zi)&= \int_{z_{\rm min}}^{\zi} \frac{N_{\rm 1s}(z') \sigma_{\rm 1s}\Big(\nuinj\frac{(1+z')}{(1+\zi)}\Big) \,c \id z'}{H(z')(1+z')},
\end{align}
%----------------------------------------------------------
where $z_{\rm min}=\max[0, \nu_{\rm 1sc}(1+\zi)/\nuinj]$. Similar expressions apply for \HeI and \HeII \citep[e.g.,][]{Chluba2009c}. 
To estimate when photo-ionization processes are important we again compute the domains with $\tau\geq 1$. The corresponding regions for \HI, \HeI and \HeII are shown in Fig.~\ref{fig:tau_abs}. 
In contrast to the free-free process, for each species a {\it low-} and {\it high-energy} boundary appears: the low-energy boundary is determined by photons never being above the \HI photon-ionization threshold of $\simeq 13.6\,\eV$ (curved region in the post-recombination Universe) or redshifting below this threshold before significant absorption can occurs (vertical part on the low energy side). The high-energy boundary is determined by the fact that on their journey through the Universe the photons never reach close enough to the corresponding continuum threshold to be absorbed.\footnote{When computing the opacity, we only included redshifting along the photon trajectory. In principle we should also add the effect of electron recoil which is relevant at early times and high energies. This would change the shape of the domain slightly, essentially tilting the vertical parts in Fig.~\ref{fig:regimes}.}
Broadly speaking, this means that photons injected at $\xinj\gtrsim 10^6$ and $z\gtrsim 1$ do not lead to significant ionizations.

As also visible from Fig.~\ref{fig:regimes}, without reionization, the overall optically-thick domain is a slightly bigger (dashed-purple line) since after recombination neutral \HI and $\HeI$ atoms are abundant at all times. Including reionization, allows the medium to become optically-thin to \HI photo-ionization at redshifts $z\lesssim 0.3$, although this transition depends on the specific reionization model.
The biggest region is determined by neutral helium, while \HeII photo-ionization is only important in the pre-recombination era. 

To be mainly absorbed in the \HI Lyman-continuum, photons have to be injected in a narrow region within a factor of $\nu^{\rm HeI}_{\rm 1sc}/\nu^{\rm HI}_{\rm 1sc}\simeq 1.8$ of the Lyman-continuum threshold energy. However, the full optically-thick \HI photo-ionization region overlaps significantly with the \HeI region (dotted-line visible within the \HeI domain) although \HeI photo-ionization usually occurs more rapidly.

\vspace{-0mm}
\subsection{Energy exchange at high frequencies}
\label{sec:energy_exchange}
%----------------------------------------------------------
The last aspect we are interested in is related to the cooling of particles at high energies. In particular, we want to ask the question for which initial photon energy a significant amount of non-thermal electrons is created, requiring another treatment. To estimate this, two steps are necessary: we first have to estimate for which energy of the electron cooling is dominated by interactions with photons rather than Coulomb interactions. This defines the kinetic energy threshold for electrons, $E_{\rm nt}$, to remain non-thermal for a significant time. The next question then is what minimal energy an injected photon produces a Compton electron above this threshold, $E_{\rm nt}$. 

To determine $E_{\rm nt}$, we simply have to compare the Coulomb scattering rates with the Compton cooling rates at each redshift. In both cases, the thermal particles provide the targets for the non-thermal electron to scatter with, as the scattering between non-thermal particles has a very low probability. The $e$-$p$ Coulomb scattering rates at a given temperature are several orders of magnitudes lower than the $e$-$e$ scattering rates and are thus neglected \citep{Stepney1983, Dermer1989}. For the $e$-$e$ Coulomb scattering rates, we use expressions from \citet{Dermer1989}, which are valid up to mildly relativistic temperatures of the thermal particles. We first reproduced Fig.~1 of \citet{Dermer1989} and then took the limit to low temperatures. At kinetic energies of a few $\keV$, the energy exchange rate obtained then becomes roughly independent of the plasma temperature and is well approximated by
%----------------------------------------------------------
\begin{align}
\frac{\id \epsilon_{ee}}{\id \tau}&\approx - 21.0  \,
\frac{\left( 1+ 0.570\,\epsilon_{\rm kin} -\pot{1.745}{-2} \epsilon_{\rm kin}^2\right)}{\sqrt{\epsilon_{\rm kin}}}
\,\left[\frac{\ln \Lambda}{20}\right],
\label{eq:dEdt_ee}
\end{align}
%----------------------------------------------------------
where we have used the Coulomb logarithm $\ln \Lambda = 20$ as a fiducial value and expressed the kinetic energy in units of the electron rest mass, $\epsilon_{\rm kin}=E_{\rm kin}/\me c^2$. 
Equation~\eqref{eq:dEdt_ee} works well when the kinetic energy of the projectile electrons is much larger than the typical thermal energy of the background electrons, and should be valid up to $E_{\rm kin}\simeq 10\,\MeV$. To lowest order, this agrees with the non-relativistic result of \citet[][see Eq.~19 therein]{Haug1988Coulomb} but it roughly a factor of 2 lower than what is given in \citet{Swartz1971}.

As for the Compton scattering of photons by bound electrons, we can again assume that above the threshold energies of $\simeq 2.6\,\keV$, $3.5\,\keV$ and $5.3\,\keV$ (see Sect.~\ref{ss:high_energy_photons}) for the three atomic species $e$-$e$ Coulomb scattering occurs in the same way whether the electron is bound or free.\footnote{For $e$-$p$ scattering, the corresponding thresholds are enhanced by a factor of $m_{\rm p}/\me \simeq 1836$, due to the reduction of the recoil effect on protons.} At energies below these thresholds, we also expect collisional ionization and excitation to contribute, further increasing the rate at which electrons loose their energy, however, these are neglected for our estimates.

The $e$-$e$ Coulomb cooling rate has to be compared to the energy loss rate of non-thermal electrons on the CMB blackbody. This can be approximated as \citep{Blumenthal1970}
%----------------------------------------------------------
\begin{align}
\frac{\id \epsilon_{e\gamma}}{\id \tau}&\approx -\frac{4}{3}\,\frac{(\gamma^2-1)}{\Ne}\, \frac{\rho_{\rm CMB}}{\me c^2},
\end{align}
%----------------------------------------------------------
where we neglected relativistic corrections \citep[for additional approximations see][]{CSpack2019} and set $\rho_{\rm CMB}\approx 2.7 k\TCMB N_{\rm CMB}$. The Lorentz factor $\gamma$ furthermore yields $\gamma^2-1=\epsilon_{\rm kin}(2+\epsilon_{\rm kin})$.

%----------------------------------------------------------
\begin{figure}
\includegraphics[width=\columnwidth]{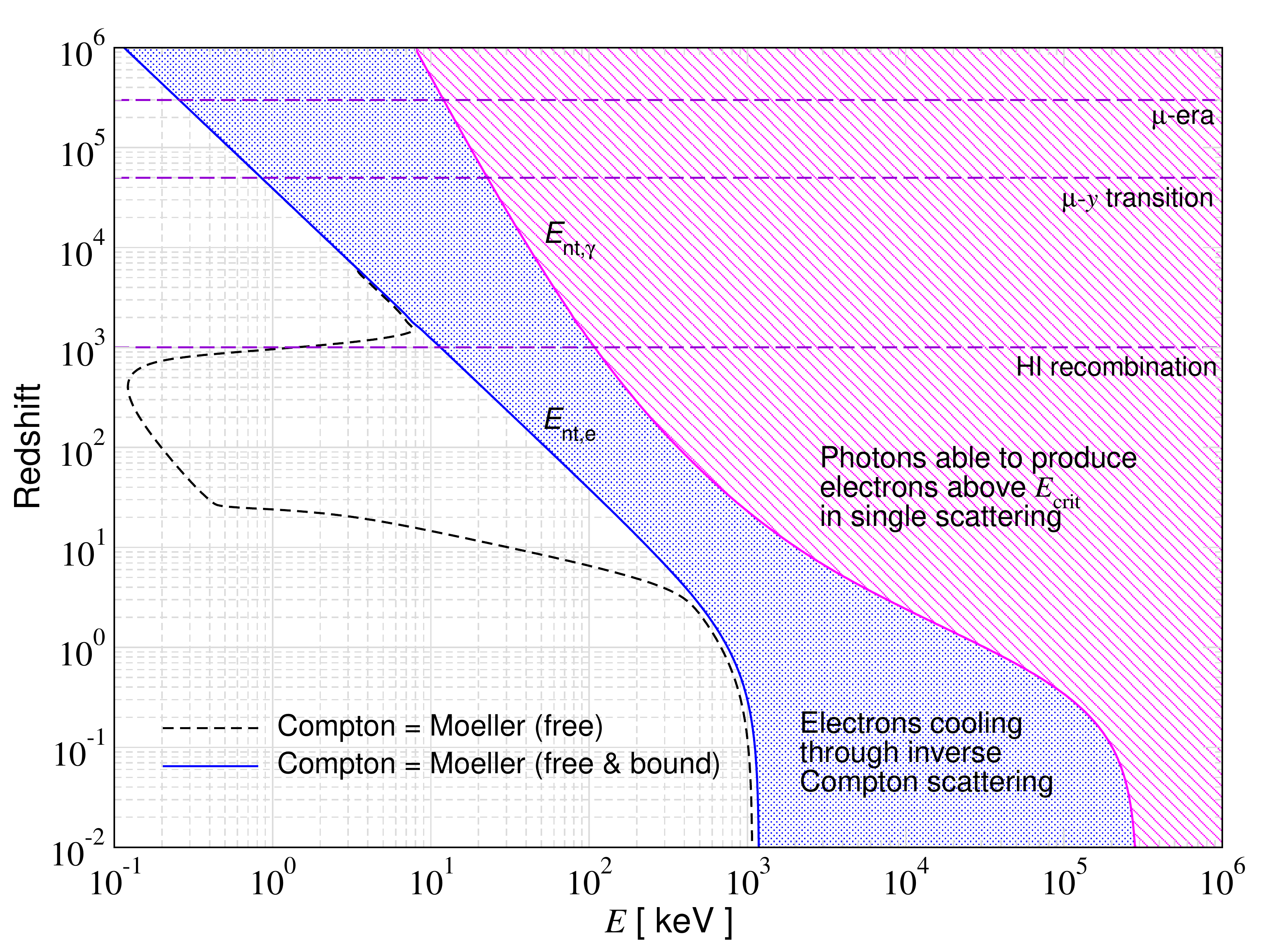}
\vspace{-3mm}
 \caption{Domains of importance for production of non-thermal electrons. In blue we show the domain in which an energetic electron cools by Compton scattering of CMB blackbody photons. The magenta region marks the domain in which an energetic photon can up-scatter an electron into the Compton cooling domain, i.e., $E_{\rm kin}>E_{\rm nt, e}$. The energy required for the photon is typically more than an order of magnitude larger than $E_{\rm nt, e}$.} 
\label{fig:regimes_electron}
\vspace{-2mm}
\end{figure}
%----------------------------------------------------------
In Fig.~\ref{fig:regimes_electron}, we show the critical energy, $E_{\rm nt, e}$, at which the $e$-$e$ Coulomb cooling rate (i.e., M{\o}ller scattering) equals the Compton cooling rate. For illustration, we show the difference when only considering Coulomb scattering off of free electrons, which greatly underestimates the total loss rate in the post-recombination era. 

We next ask the question what initial photon energy, $E_{\rm nt, \gamma}$, is required to produce a non-thermal electron above the critical energy $E_{\rm nt, e}$ in a single Compton scattering event. To compute the corresponding energy exchange we use {\tt CSpack} \citep[][Eq.~16b]{CSpack2019}. For $E_{\rm nt, \gamma}\ll \me c^2$, to leading order this implies the condition $E_{\rm nt, e}\approx E_{\rm nt, \gamma}^2/\me c^2$; however, Klein-Nishina corrections become important at low redshifts, where $E_{\rm nt, e}\gtrsim 1~\MeV$. 

The domain above the critical photon energy, $E_{\rm nt, \gamma}$, is illustrated in Fig.~\ref{fig:regimes_electron} and also Fig.~\ref{fig:regimes} (magenta). Photons injected inside the magenta regions would caused the production of non-thermal electrons, which then through subsequent Compton scattering cause non-thermal distortion corrections \citep{Ensslin2000, Colafrancesco2003, Slatyer2015, Acharya2018}.
In the calculations presented here, we avoid the production of non-thermal electrons by restricting ourselves to photon energies $h\nuinj\lesssim \mXmax$. This also avoids complications related to the expected soft photon production by DC from the high energy particle cascade \citep{DCpack2020}.

We furthermore point out that the Universe becomes transparent to high energy photons for Compton interactions even in the pre-recombination era (outside of red region, Fig.~\ref{fig:regimes}). This implies that for photon energies above $E_{\rm nt, \gamma}$, no Compton scattering event may occur within a Hubble time.
However, many other processes (e.g., pair-production and photon-photon scattering) become important in that regime \citep{Svensson1984, Zdziarski1989, Chen2004, Padmanabhan2005}, but these refinements are avoided for the cases considered here.

\vspace{-3mm}
\subsection{Anticipating the final spectrum}
\label{sec:anticipate}
%----------------------------------------------------------
Now that we determined all critical regions for photons injected at various redshifts, we can already anticipate the main features of the solutions. For this step, Fig.~\ref{fig:regimes} provides considerable insight. The crucial aspect is that for a given injection energy, the photon source moves along lines $\xinj(z)=\xinjc/(1+z)$ (brown lines in Fig.~\ref{fig:regimes}). Assuming a certain lifetime of the particle or excited state, one can determine the redshift at which most of the photons are injected (see Fig.~\ref{fig:zx_Ginj}). Moving along the corresponding trajectory $\xinj(z)$ then explains what general features the final spectrum will have.

For combinations of particle lifetimes and injection energies that mainly target the white areas in Fig.~\ref{fig:regimes}, a spectral distortion that closely tracks the time-dependence of the injection process is expected. In this case, the direct constraint from spectrometers will apply and the distortion is not well represented by a simple $y$- or $\mu$-type distortion, but rather has a form
%----------------------------------------------------------
\begin{align}
\label{eq:sol_direct}
\Delta I^{\rm inj}_\nu&\approx 
%-\frac{hc}{4\pi}\, \frac{f_\gamma  \dot N_X(\zi)}{H(\zi) (1+\zi)^3}=
%
\frac{hc}{4\pi}\, \frac{f_\gamma \Ginj N_{X,0}\exp\left[-\Ginj t(\zi)\right]}{H(\zi)},
\end{align}
%----------------------------------------------------------
with $1+\zi=\Einj/h\nu\geq 1$. This expression can be obtained as for recombination line emission, where extremely narrow line injection is assumed (see \citet{Jose2006} or  \cite{Masso:1999wj} for a different approach).

If the injection energy is $\xinjc\lesssim 10^{-8}$, we can assume that {\it all} the energy is converted into heat. For lifetimes $\tau_X\lesssim 10^{13}\,{\rm s}$, this means that the standard $\mu$ and $y$-distortion constraints from heating apply. For longer lifetimes, a $y$-distortion is created; however, in this case the ionization history is also directly affected and, hence, CMB anisotropy constraints apply, as we show below.
For injection at $z_{\rm inj}\gtrsim \pot{3}{5}$, the resultant $\mu$-distortion can be estimated analytically using the expressions from \citet{Chluba2015GreensII}. If photons are injected primarily at $\xinj < 3.6$, this can lead to $\mu<0$ (see Sect.~\ref{sec:mu_analytic}). 

If most of the photons are injected in the purple bands of Fig.~\ref{fig:regimes}, we again expect that the energy is quickly converted into heat and hence causes significant $\mu$ and $y$-type contributions. However, at this time also direct ionizations of atoms will play a role which again can be constrained using the CMB anisotropies.
Finally, for photons mainly injected in the red regions of Fig.~\ref{fig:regimes}, we expect a partially-Comptonized SD similar to $\Delta I^{\rm inj}_\nu$ but broadened and with small $y$-distortion contributions due to the associated heating.
Focusing at the region $\xinj \gtrsim 10^6$, we can also anticipate that only weak CMB anisotropy constraints can be derived if injection primarily occurs at $z\lesssim 50$. In this case, direct constraints from the X-ray background are expected to be more stringent.

%-------------------------------------------------------------------
\vspace{-5mm}
\section{Distortions in decaying particle scenarios}
\label{sec:num_sols}
%-------------------------------------------------------------------
In this section, we present the solutions for the photon injection problem from continuous decay at various energies and particle lifetimes. The main goal here is to illustrate the properties of the solutions with an eye on the various physical processes.
We numerically solve the photon injection problem using {\tt CosmoTherm} with the appropriate modifications to account for the effects discussed above. The main character of the solution can be deduced from Fig.~\ref{fig:regimes} as explained in Sect.~\ref{sec:anticipate}. 

%----------------------------------------------------------
\begin{figure}
\includegraphics[width=\columnwidth]{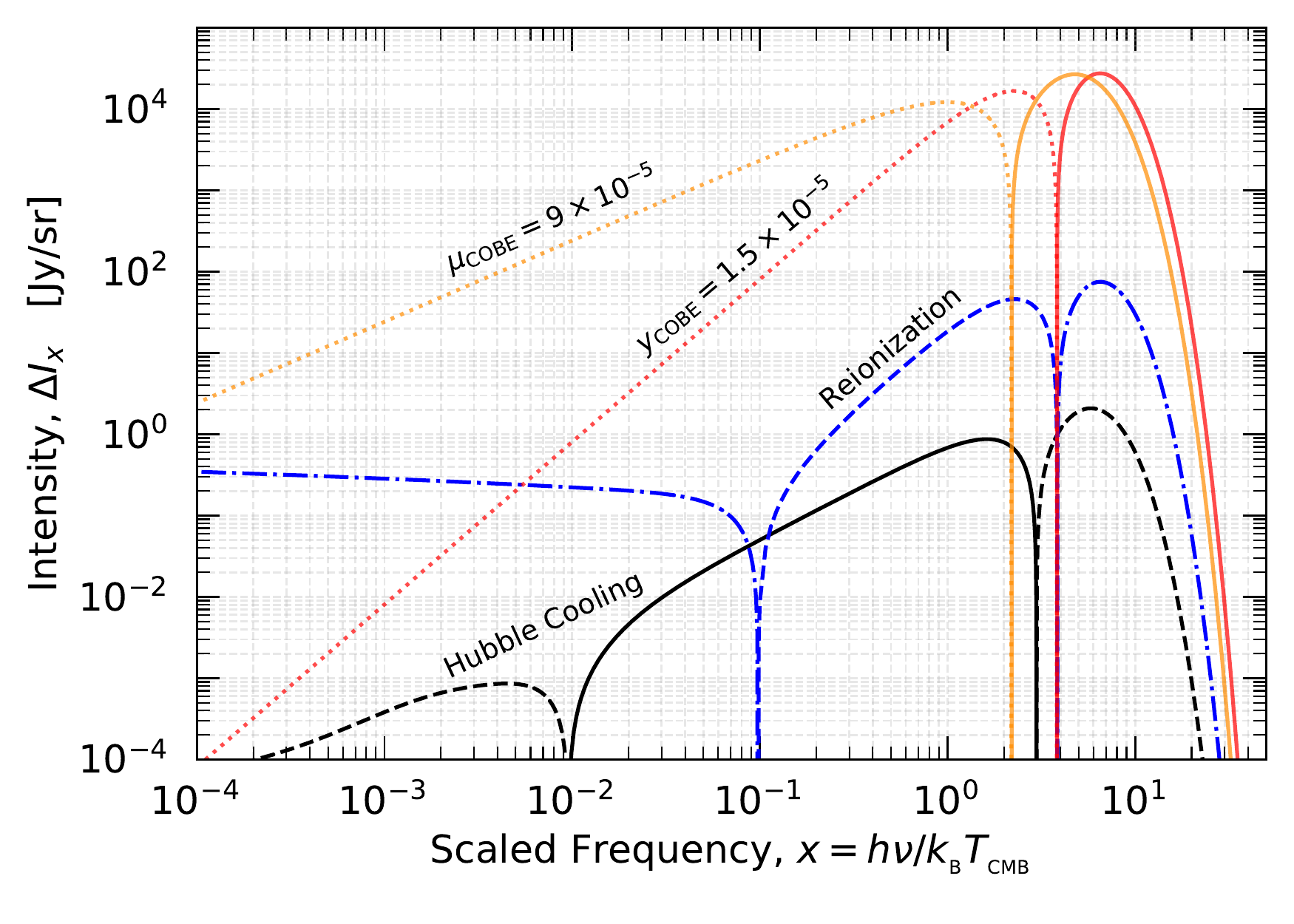}
\vspace{-5mm}
\caption{Standard distortion signals expected from reionization and the Hubble cooling. The standard $\mu$ and $y$ distortions for the \textit{COBE/FIRAS} limits are shown for comparison. Negative branches are shown as dashed lines.} 
\vspace{-3mm}
\label{fig:standard_cooling}
\end{figure}
%----------------------------------------------------------

\vspace{-4mm}
\subsection{Cooling and reionization distortion}
\label{sec:distortion_results_cool_reion}
%-------------------------------------------------------------------
To set the stage for the photon injection cases, we start with the standard distortions created in $\Lambda$CDM by adiabatic cooling \citep{Chluba2011therm} and reionization \citep{Hu1994pert}. In the calculations presented below, these signals have to be subtracted in order to reliably estimate the photon-injection parameters. We do not include any extra heating from the dissipation of acoustic modes or the cumulative contributions associated with Sunyaev-Zeldovich effect from galaxy clusters \citep[see][for overview]{Chluba2016} as these do not affect the overall picture in terms of ionization history or thermal history. 
The adiabatic cooling and reionization distortions obtained with {\tt CosmoTherm} are shown in Fig.~\ref{fig:standard_cooling} together with the standard $y$- and $\mu$-type distortion. 

The adiabatic cooling process causes a small negative $\mu$- and $y$-type distortion with $\mu\simeq -\pot{3}{-9}$ and $y\simeq-\pot{5}{-10}$, because the electrons, which are cooling faster than radiation, are continuously extracting energy from the photons. For our computations of photon injection distortions the adiabatic cooling process is included to leave the ionization history at late stages comparable to the standard {\tt CosmoRec} computation. 
In Fig.~\ref{fig:standard_cooling}, we can also observe that at very low frequencies ($x\lesssim 0.1$ or $\nu\lesssim 6\,\GHz$), the standard $\Lambda$CDM distortions departs notably from the analytic $\mu$ and $y$ formulae (see Sect.~\ref{sec:FIRAS_setup} for more details). This is related to free-free absorption at late times, which has significant time-dependence \citep[e.g.,][]{Illarionov1974, Chluba2011therm}.
\\
For the reionization distortion, a $y$-distortion with $y\simeq 10^{-7}$ is created by the late-time heating at $z\lesssim 10$. The same heating leads to a low-frequency free-free distortion which is visible at $x\lesssim 0.1$ \citep[see also][]{Cooray2004, Trombetti2014MNRAS}, indicating that $\Te>\Tg$. In particular the low-frequency CMB spectrum can be thought of as an {\it electron thermostat}.  

%----------------------------------------------------------
\begin{figure*}
\begin{centering}
\includegraphics[trim={0 0 100mm 0}, clip, width=1.92\columnwidth]{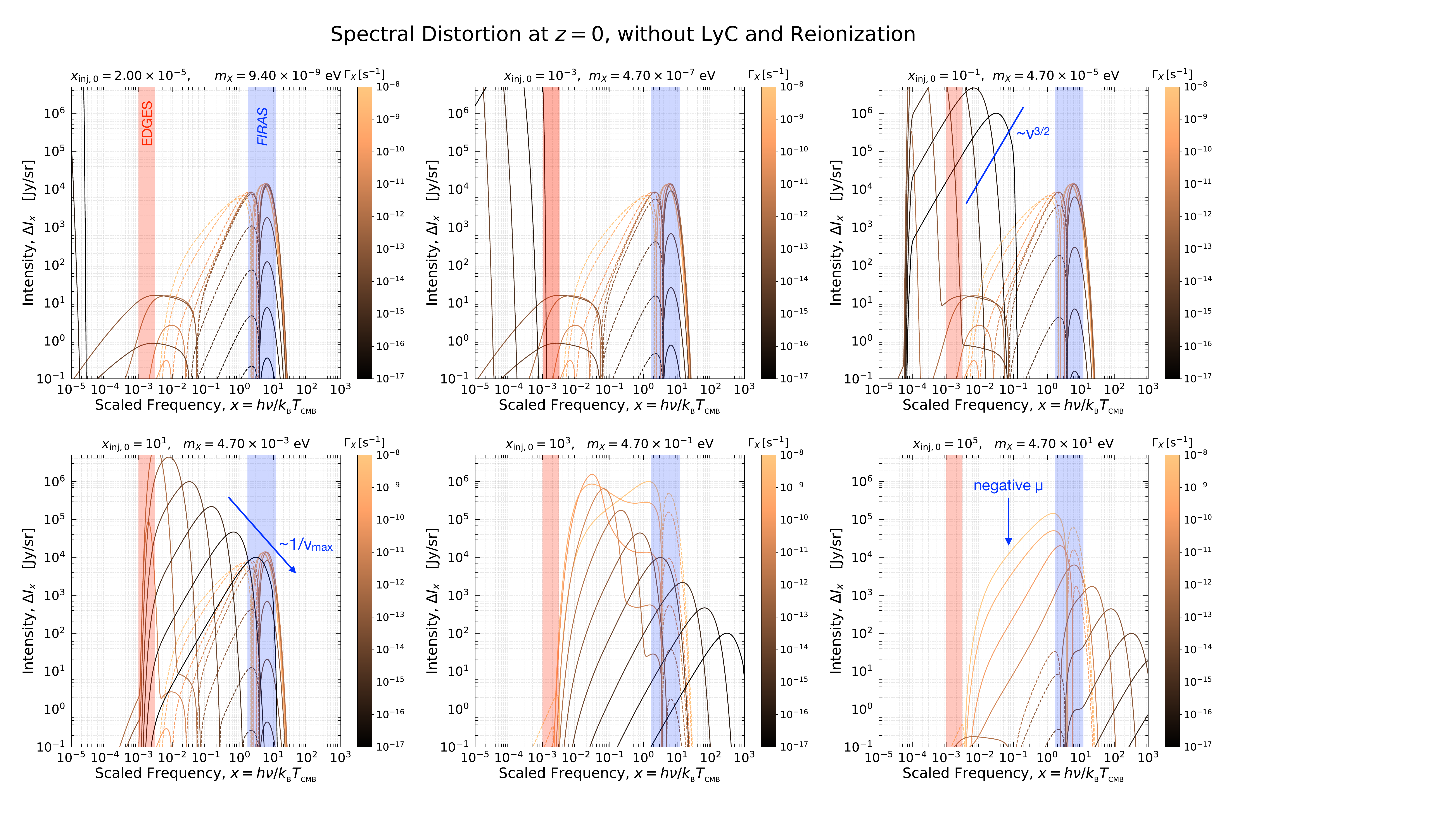}
\par\end{centering}
\vspace{-5mm}
\caption{Solutions for the spectral distortion created by photon injection from decaying particles for several masses, $m_X$, and lifetimes, $\Gamma_X$. The main features of the solutions can be understood in combination with Fig.~\ref{fig:regimes} (see Sect~\ref{sec:anticipate} and \ref{sec:num_sols} for details). In all shown cases, photo-ionization corrections were neglected and reionization is also not included. All spectra are normalized such that $\Delta \rho/\rho|_\mathrm{inj}=3\times 10^{-5}$. Negative parts of the signal are shown as dashed lines.
}
\vspace{-3mm}
\label{fig:distortion_sols}
\end{figure*}
%----------------------------------------------------------

\vspace{-4mm}
\subsection{Numerical spectral distortion results}
\label{sec:distortion_results}
%-------------------------------------------------------------------
In Fig.~\ref{fig:distortion_sols} we present several solutions for the spectral distortion created by photon injection from decaying particles with varying masses and lifetimes. For now, we omit the effect of atomic photo-ionization and reionization and reintroduce them in the subsequent sections. All solutions are normalized such that 
%----------------------------------------------------------
\begin{align}
\label{eq:sol_norma}
\frac{\Delta\rho_\gamma}{\rho_\gamma}\Bigg|_{\rm inj}&= \int \frac{\id \ln \rho_\gamma}{\id z}\,\mathcal{J}_{\rm bb}(z)\id z.
\end{align}
%----------------------------------------------------------
is fixed to $\Delta \rho/\rho|_\mathrm{inj}=3\times 10^{-5}$. Here, the distortion visibility function $\mathcal{J}_{\rm bb}(z)\approx \expf{-(z/\zmu)^{5/2}}$ with $\zmu=\pot{1.98}{6}$ accounts for the reduction of the distortion amplitude by thermalization processes, efficient at $z\gtrsim \zmu$ (see Sect.~\ref{sec:mu_analytic} for more details). 

Starting with injections at low frequencies ($\xinjc=\pot{2}{-5}$), we can see that the overall distortion is dominated by $\mu$ and $y$-type contributions in the usual CMB bands. This is naturally expected from the fact that for $z_X \gtrsim 50$ photons are mostly injected in the optically thick BR absorption band (Fig.~\ref{fig:regimes} without reionization), thus always creating heating. At low frequencies, we can also notice the effect of free-free emission, which rise the CMB spectrum to the temperature of the electrons, which for long lifetimes can become noticeable in the \EDGES band. However, none of the direct decay photons are visible in the domains of observational interest.

Moving to $\xinjc=10^{-3}$, we find solutions that are overall similar to those for $\xinjc=\pot{2}{-5}$. However, for late decay (i.e., rates $\Ginj<\pot{{\rm few}}{-14}\,{\rm s}^{-1}$), we now notice the appearance of a direct injection distortion, with a shape resembling  Eq.~\eqref{eq:sol_direct} at $x\lesssim 10^{-3}$. This is even more visible for $\xinjc=10^{-1}$: redward of the emission peak, the slope is given by $\Delta I_\nu\propto 1/H(\zi)\propto \nu^{3/2}$, while for $\Ginj t_0<1$ the normalization scales as $\Delta \rho/\rho \propto 1/\nu_{\rm max}$ (see Fig.~\ref{fig:distortion_sols}).
Further increasing $\xinjc$, enhances the visibility of this direct injection distortion. In particular, cases with $\xinjc\simeq 10^{-3}-10$ (or masses $m_X\simeq 0.5\,\mu\eV$ to $5\,\meV$) and long particle lifetimes, $\Ginj \lesssim \pot{{\rm few}}{-14}\,{\rm s}^{-1}$ can be directly constrained with \EDGES.

For $\xinjc\gtrsim 10-10^3$, the $y$- and $\mu$-type contributions and direct photon injection emission both play significant roles, and the constraint on these cases are expected to be dominated by CMB spectrometer measurements. For $\xinjc=10^5$, we observe that the distortion response in the CMB bands becomes extremely small for the longest lifetimes, since here we did not include any photo-ionization heating and the Universe essentially is transparent for these energies in the post-recombination era (Fig.~\ref{fig:regimes} without purple region). Thus very weak distortion constraints are expected in these cases.
For the shortest lifetimes and $\xinjc=10^5$, we can furthermore see that the distortion signal is given by a {\it negative} $\mu$-distortion with significantly enhanced amplitude, reaching $\mu\simeq 10^{-3}$ with our normalization condition. This interesting aspect will be explained in Sect.~\ref{sec:mu_analytic} and is related to differences between distortions sourced by pure energy release and photon injection \citep{Chluba2015GreensII}.

\vspace{-4mm}
\subsubsection{Effect of photo-ionization}
\begin{figure*}
\begin{centering}
\includegraphics[width=1.92\columnwidth]{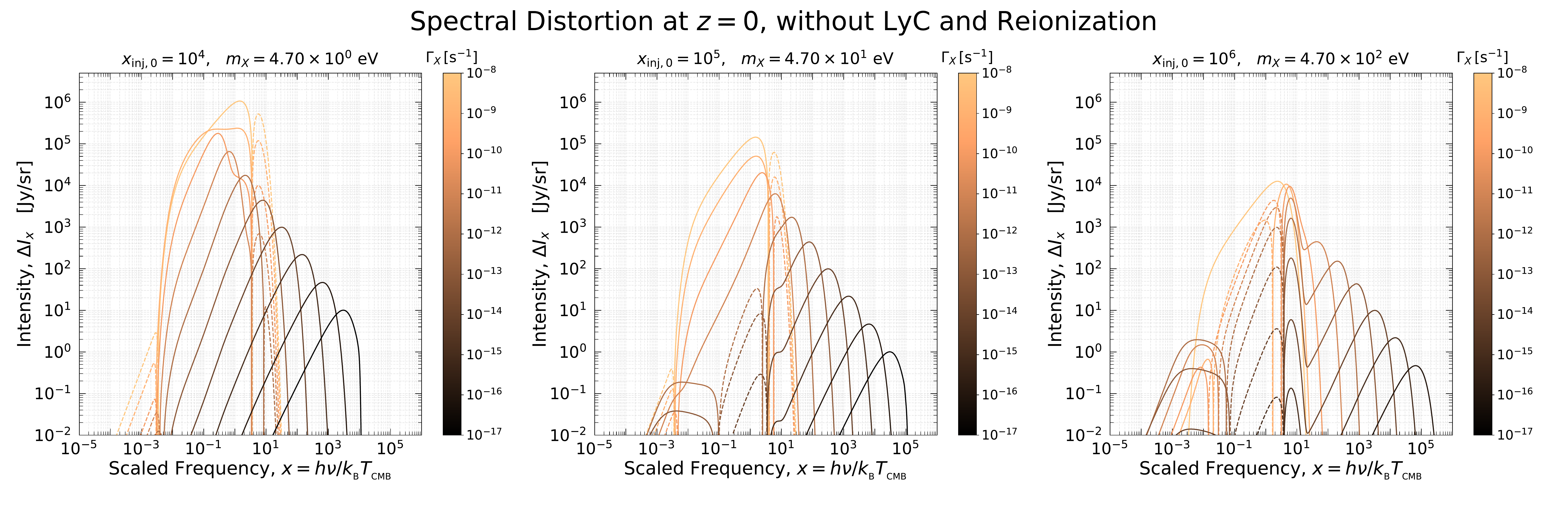}
\\[-2mm]
\includegraphics[trim={0 140mm -10mm 0}, clip, width=1.92\columnwidth]{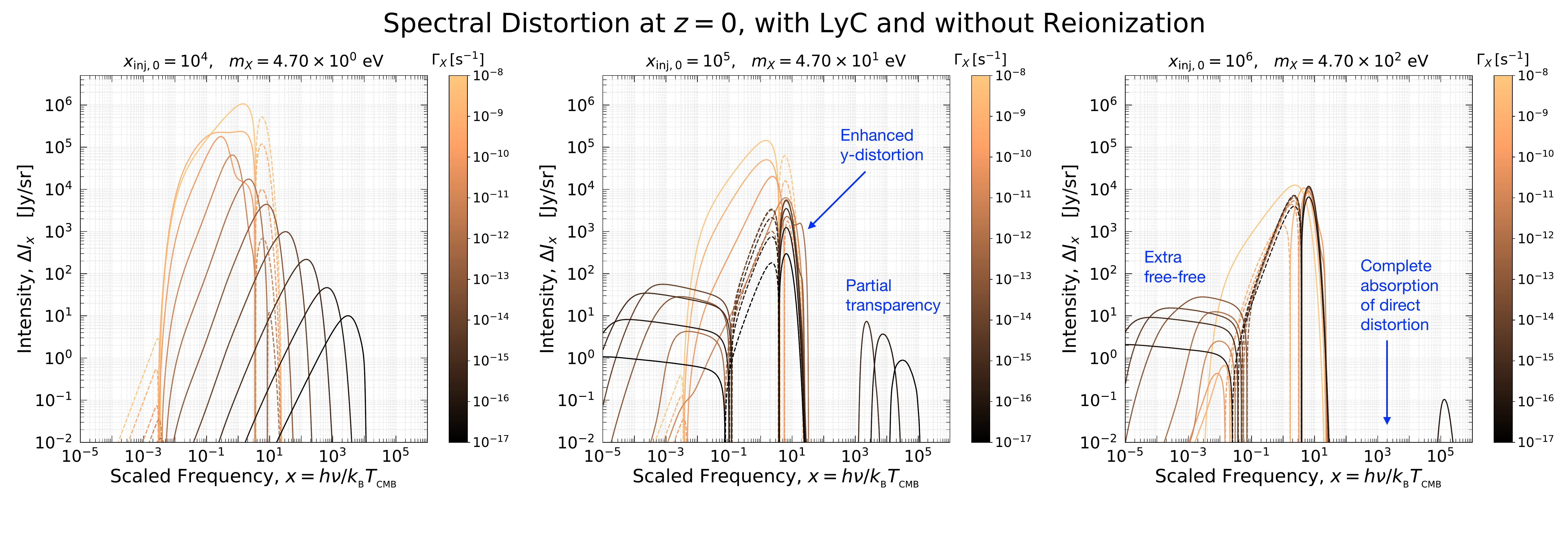}
\par\end{centering}
\vspace{-5mm}
\caption{Similar to Fig.~\ref{fig:distortion_sols} but for $\xinjc=\{10^{4}, 10^{5}, 10^{6}\}$ and with \HI, \HeI and \HeII photo-ionization taken into account. The differences are mainly visible at high frequencies for cases with $\xinjc\gtrsim \pot{5.8}{4}$ and lifetimes $\Ginj \lesssim 10^{-13}\,{\rm s}^{-1}$. Including continuum absorption leads to strong photon absorption at high frequencies and a spectral distortion at low frequencies as discussed in the text. Negative parts of the signal are shown as dashed lines.
} 
\vspace{-3mm}
\label{fig:distortion_sols_Lyc}
\end{figure*}

%-------------------------------------------------------------------
As the next step, we include the effect of \HI, \HeI and \HeII photo-ionization on the distortion evolution. While it is clear that this will change the distortions at high frequencies in {\it all} cases, this indirect effect remains small until\footnote{In our computation, this transition is more gradual, since we inject photons in a line with finite width. Furthermore, we slightly smooth the continuum cross section close to the ionization threshold to ease the numerical treatment. This  does not have a  major effect on the main conclusions.} $\xinjc\gtrsim \pot{5.8}{4}$, corresponding to the ionisation energy for hydrogen.
Thus, as is clear from Fig.~\ref{fig:regimes} in cases with injection at $\xinj\gtrsim \pot{5.8}{4}/(1+\zi)$ in the post-recombination era, we expect significant extra heating from the absorbed photons. Indeed, this is visible in Fig.~\ref{fig:distortion_sols_Lyc}, where the heating-related $y$-type distortion is significantly enhanced over the case without photo-ionization for $\xinjc=10^5$ (compare bottom and top rows of plots). Moreover, an additional enhancement of the low-frequency distortion due to extra free-free emission appears.
Hence, it is clear that the distortion bounds for decaying particles with rest mass $m_X c^2\gtrsim 2\times 13.6\,\eV$ will be noticeably tightened, once Lyman continuum absorption is taking into account.
However, partial transparency of the plasma to photons is restored at $\xinj \gtrsim \pot{{\rm few}}{6}$ and $z\lesssim 10^2$, when the injection happens so far above the ionization thresholds that no absorption can occur until today (see Fig.~\ref{fig:regimes}).

We highlight that we also included the effect of the ionizations on the recombination history, since the related modifications affect the thermal contact between photons and baryons, as we explain below. This was achieved by adding the corresponding rate equations to the {\tt CosmoTherm} setup using {\tt CosmoRec/Recfast++}.
When deriving the constraint in Sect.~\ref{sec:constraints}, we assumed that the effects remain linear to leading order. We will discuss the validity of this assumption below.
We also mention again that here we do not include the re-emission of photons by recombination processes. These introduce additional spectral features in the Wien-tail of the CMB \citep[e.g.,][]{Chluba2008c}, outside of the regime that is currently directly constrained by \FIRAS, but otherwise should not affect the results significantly.

%----------------------------------------------------------
%----------------------------------------------------------

%----------------------------------------------------------
\begin{figure*}
\begin{centering}
\includegraphics[width=1.92  \columnwidth]{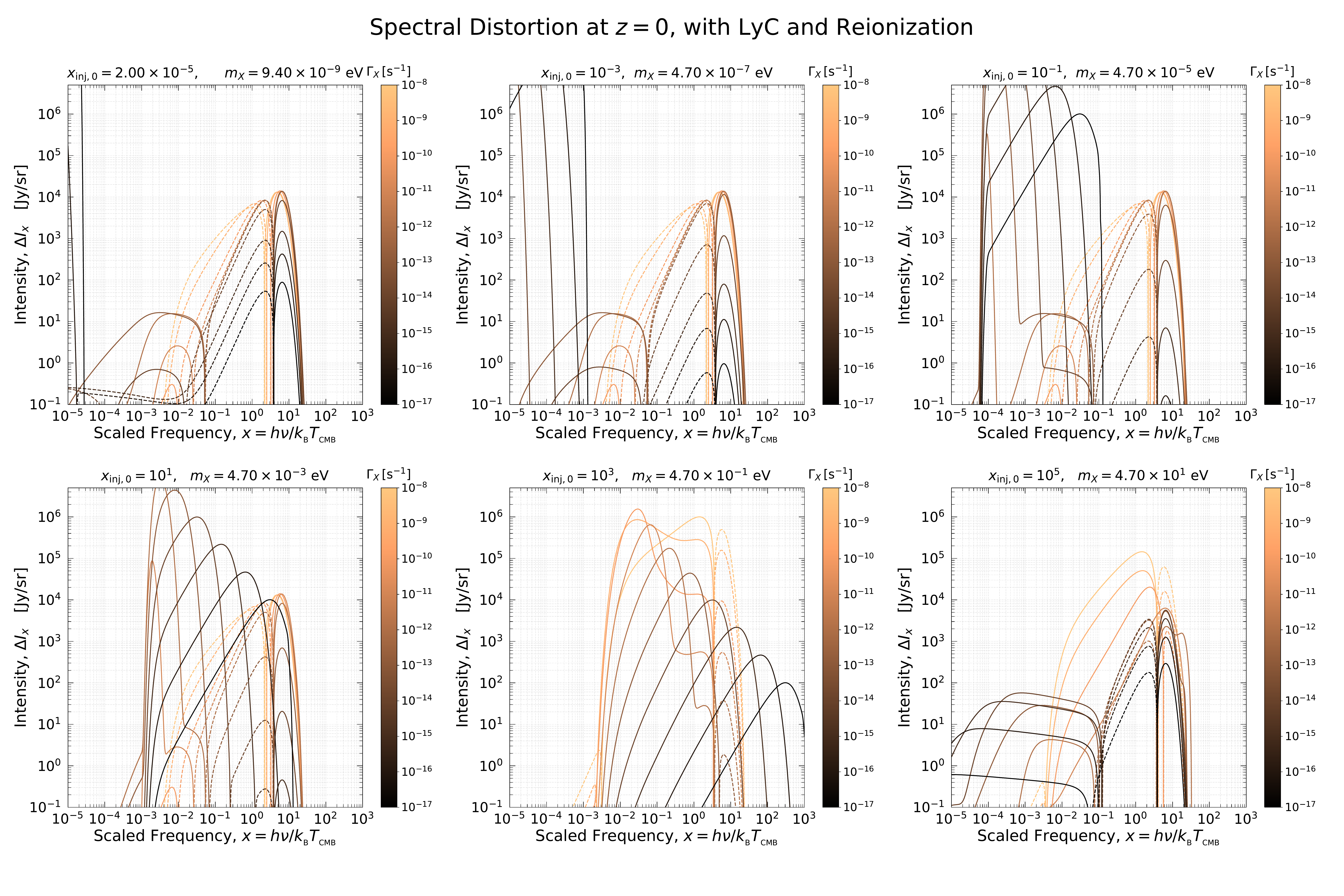}
\par\end{centering}
\vspace{-5mm}
\caption{Same as Fig.~\ref{fig:distortion_sols}, but including simple modelling for reionization in addition to \HI, \HeI and \HeII photo-ionization. In contrast to the case with only Lyman continuum absorption included (Fig.~\ref{fig:distortion_sols_Lyc}), we can now observe differences for both very low ($\xinjc\lesssim 10^{-3}$) and high frequency injection ($\xinjc \gtrsim 10^{5}$) at long lifetimes, $\Ginj \lesssim 10^{-13}\,{\rm s}^{-1}$. Negative parts of the signal are shown as dashed lines.} 
\vspace{-3mm}
\label{fig:distortion_sols_Lyc_reion}
\end{figure*}
%----------------------------------------------------------

\vspace{-3mm}
\subsubsection{Effect of reionization}
%-------------------------------------------------------------------
As a final illustration, we now also include the effect of reionization on the distortion signal (see Fig~\ref{fig:distortion_sols_Lyc_reion}).  The most important difference with respect to the previous cases is expected at low frequencies, since the free-free emissivity of the plasma ($\propto \Ne N_{\rm p}$) is greatly enhanced once reionization occurs. Indeed, by comparing Fig.~\ref{fig:distortion_sols} with Fig.~\ref{fig:distortion_sols_Lyc_reion}, we notice that for $\xinjc=\pot{2}{-5}$ and $\xinjc=10^{-3}$ the $y$-distortion contribution for post-recombination injection is noticeably enhanced. This just signifies the fact that the optically-thick domain, due to free-free absorption, is increased (see Fig.~\ref{fig:regimes}, blue region, dashed versus solid boundary) and conversion into heat is very efficient inside this region. One important consequence of this is that the heating of electrons by soft photon injection stays significant for a wider range of masses. Hence, distortion and ionization history limits should become more stringent.

We note that in our computation the changes of the ionization history are consistently propagated. Even if the general picture does not change, the domains estimated in Fig.~\ref{fig:regimes} for the standard ionization history are indeed modified by these effects. We return to this point below (Sect.~\ref{sec:collisions_effect}) when discussing various effects related to collisional ionization.

\vspace{-3mm}
\subsection{Analytic description of the distortion in the $\mu$-era}
\label{sec:mu_analytic}
%-------------------------------------------------------------------
Using the Green's function method of \citet{Chluba2015GreensII}, we can in principle describe the spectral distortions created by photon injection in the pre-recombination era ($z\gtrsim 10^3$) analytically. 
For scenarios with short lifetimes $\tau_X\lesssim \pot{2}{8}\,{\rm s}$ ($z_X\gtrsim \pot{3}{5}$), the signal is approximately given by a classical $\mu$-distortion and an analytic treatment is straightforward. 
For energy release distortions, it is well known that $\mu \approx 1.401\,\Delta\rho_\gamma/\rho_\gamma\big|_\mu$ \citep{Sunyaev1970mu}, where $\Delta\rho_\gamma/\rho_\gamma\big|_\mu$ is the effective energy release during the $\mu$-era
%----------------------------------------------------------
\begin{align}
\label{eq:Drho_rho_mu}
\frac{\Delta\rho_\gamma}{\rho_\gamma}\Bigg|_\mu&= \int \frac{\id \ln \rho_\gamma}{\id z}\,\mathcal{J}_\mu(z)\id z.
\end{align}
%----------------------------------------------------------
with $\mathcal{J}_\mu(z)$ describing the $\mu$-distortion visibility \citep[see][for various approximations]{Chluba2016}. Assuming that energy is only released at $z\gtrsim \pot{3}{5}$, one has $\mathcal{J}_\mu(z)\approx \mathcal{J}^*_{\rm bb}(z)$, where the distortion visibility function, $\mathcal{J}^*_{\rm bb}(z)$, can be approximated using \citep{Chluba2015GreensII}
%----------------------------------------------------------
\begin{align}
\label{eq:J_bb}
\mathcal{J}^*_{\rm bb}(z)&\approx 0.983\left[1-0.0381(z/\zmu)^{2.29}\right] \expf{-(z/\zmu)^{5/2}}
\end{align}
%----------------------------------------------------------
with $\zmu=\pot{1.98}{6}$, or $\mathcal{J}^*_{\rm bb}(z)\approx \expf{-(z/\zmu)^{5/2}}$ for even simpler estimates. We note that for large initial distortions ($\mu\gtrsim 10^{-2}$), this approximation is no longer valid and the visibility is significantly increased \citep{Chluba2020large}, however, we do not consider these cases here as they are not consistent with current observational bounds. 

In contrast to energy release distortions, photon injection distortions need to take the extra photons added by the injection process into account. In this case, the $\mu$ parameter can be estimated using \citep[see Eq.~15 of][]{Chluba2015GreensII}:
%----------------------------------------------------------
\begin{align}
\label{eq:sol_mu}
\mu_{\rm inj}&\approx 1.401\int \left[\xinj-x_{\rm null}\,\mathcal{P}_s(\xinj, z)\right]\,\alpha_\rho \frac{\id \ln N_\gamma}{\id z}\,\mathcal{J}^*_{\rm bb}(z)\id z.
\end{align}
%----------------------------------------------------------
Here, $\alpha_\rho=\mathcal{G}_2/\mathcal{G}_3\approx 0.370$ and $x_{\rm null}=(4/3)\mathcal{G}_3/\mathcal{G}_2\approx 3.602$.
We assume that the injection of photons occurs at $\xinj=\xinjc/(1+z)$, with the injection rate given by Eq.~\eqref{eq:dNg_dt_inj} in our case. In addition, the probability of low-frequency photons being converted into heat is given by $\mathcal{P}_s(x, z)\approx\expf{-\xc/x}$, with the critical frequency
%----------------------------------------------------------
\begin{align}
\label{eq:xc}
\xc\approx \pot{8.6}{-3}\,\sqrt{\frac{1+z}{\pot{2}{6}}}\,\sqrt{1+\left[\frac{1+z}{\pot{3.8}{5}}\right]^{-2.344}},
\end{align}
%----------------------------------------------------------
which accounts for contributions from DC and BR.

%----------------------------------------------------------
\begin{figure}
\includegraphics[width=\columnwidth]{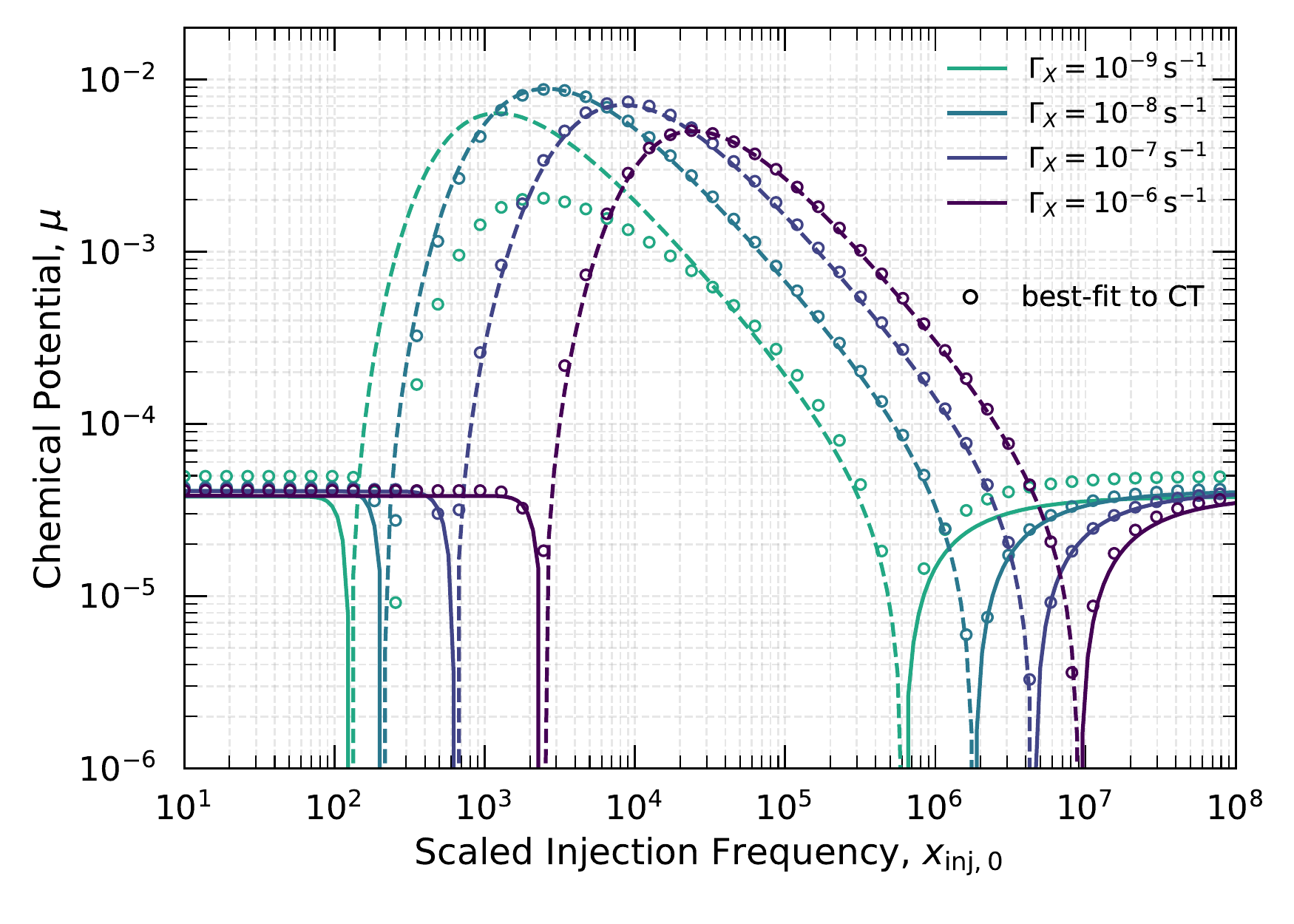}
\vspace{-5mm}
\caption{
Analytic modeling in the $\mu$-era using Eq.~\eqref{eq:sol_mu} to compute the chemical potential for various lifetimes and injection frequencies. Negative branches are represented by dashed lines. For comparison we show the $\mu$ values obtained by fitting to the SDs generated using {\tt CosmoTherm}. All distortions are normalized such that $\Delta \rho/\rho|_\mathrm{inj}=3\times 10^{-5}$. The analytic approximations appear to work well for short lifetimes, $\Ginj\gtrsim 10^{-8}\,{\rm s}^{-1}$.
} 
\vspace{-4mm}
\label{fig:mu_analytic_demo}
\end{figure}
%----------------------------------------------------------

%----------------------------------------------------------
\begin{figure*}
\includegraphics[width=1.6\columnwidth]{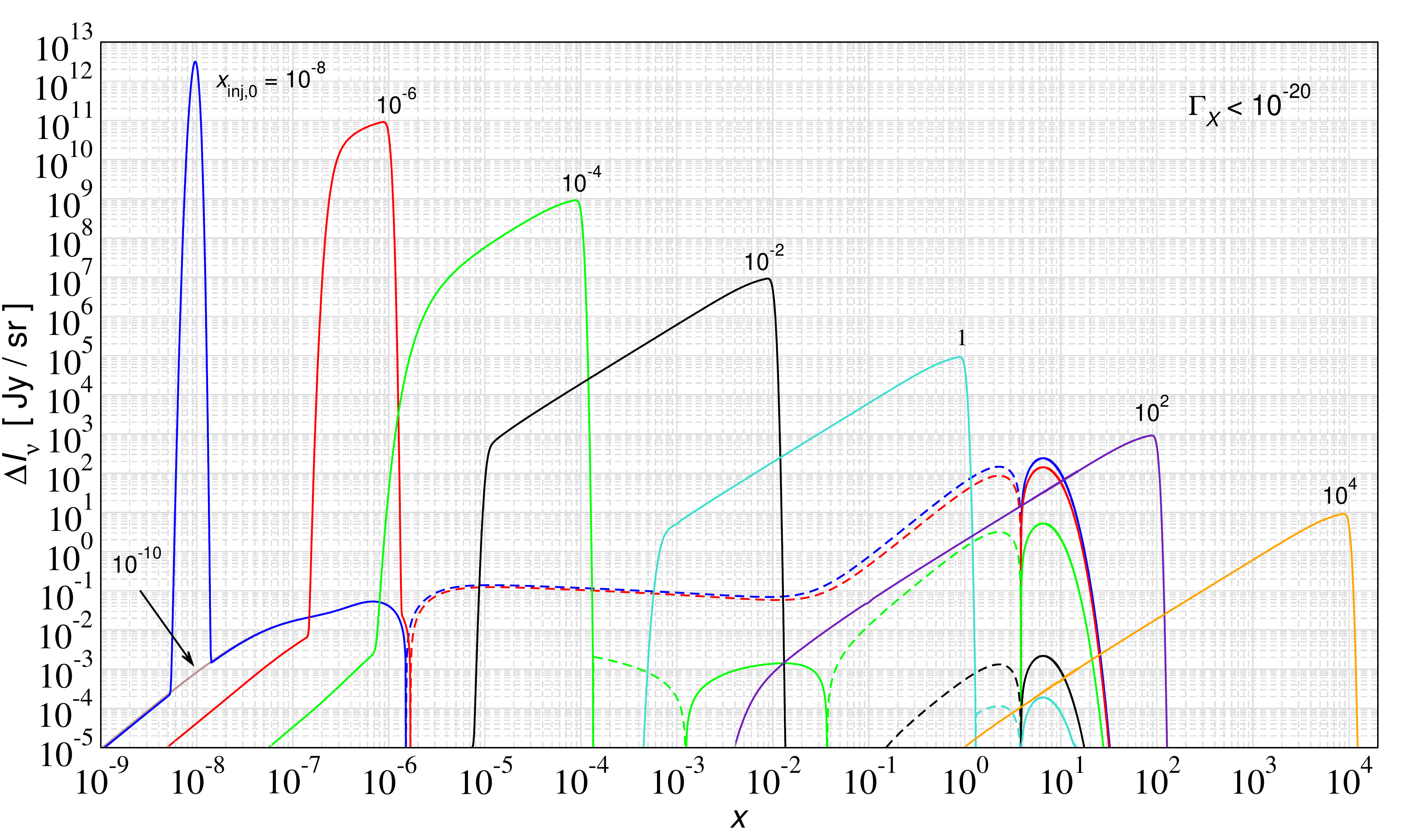}
\vspace{-0mm}
\caption{Universal distortion template for quasi-stable decay scenarios (i.e., $\Ginj\leq 10^{-20}\,{\rm s^{-1}}$). The results shown here are computed for $\Delta \rho_{\gamma}/\rho_{\gamma}=\pot{3}{-5}$ and $\Ginj=10^{-20}\,{\rm s^{-1}}$. Negative parts of the signal are shown as dashed lines. 
For $\xinjc\leq 10^{-4}$, we can observe a noticeable high-frequency $y$-distortion due to heating via free-free absorption at low frequency. The case $\xinjc=10^{-10}$ essentially coincides with that for $\xinjc=10^{-8}$ at all frequencies, but close to $x\simeq 10^{-8}$, where for $\xinjc=10^{-8}$ a direct injection distortion is visible. For all cases with $\xinjc\leq 1$, the abrupt low-frequency drop in the distortion redward of their respective peaks stems from free-free absorption, while the upper boundary simply marks the end of photon injection within the age of the Universe.} 
\vspace{-3mm}
\label{fig:universal_profile}
\end{figure*}
%----------------------------------------------------------
The most important difference to distortions created purely by heating is that injections at $x\lesssim x_{\rm null}$ can cause a negative chemical potential \citep{Chluba2015GreensII}. This is because the redistribution of photons over the full CMB spectrum on average requires more energy than was added. In addition for injection at $x\lesssim \xc/8$ \citep{Chluba2015GreensII}, the photon absorption process becomes so rapid that the added photons have no time to contribute to the shaping of the spectrum at high frequencies, thus essentially generating heat and again a positive chemical potential. 

These aspects are recovered in Fig.~\ref{fig:mu_analytic_demo}, were we compare the analytic results for $\mu$ directly with the solutions from {\tt CosmoTherm}. For lifetimes $\Ginj\gtrsim 10^{-8}$, the estimates reproduce the numerical result very well. Both at very low and very high energies we find $\mu\simeq 1.4\, \Delta \rho/\rho \simeq \pot{4.2}{-5}$ as expected from pure energy release. At intermediate injection frequencies, $\mu$ becomes negative and is significantly enhanced. This is because for fixed $\Delta \rho/\rho$, the corresponding $\Delta N/N$ is enhanced by a factor of $f\simeq 2.7(1+\zi)/\xinjc$, which can become large. For example, for $\xinjc=10^4$ and $\Ginj=10^{-8}\,{\rm s}$, we have $\zi\simeq \pot{4.1}{5}$ and hence $f\simeq 110$. Assuming pure photon injection, we then have $\mu_{\rm inj}\approx - 1.9 \Delta N/N\approx -1.9\,f \,\Delta \rho/\rho\approx -\pot{6.3}{-3}$, which is in very good agreement with Fig.~\ref{fig:mu_analytic_demo}. This also explains the enhancements seen in Fig.~\ref{fig:distortion_sols} through \ref{fig:distortion_sols_Lyc_reion} for cases with $\xinjc>10^3$. Needless to say, these cases are already ruled out by \FIRAS.

\vspace{-3mm}
\subsection{Universal distortion shapes for ultra-long lived particles}
\label{LLp}
%-------------------------------------------------------------------
For extremely long lifetimes, significantly exceeding the age of the Universe, i.e., $\Ginj\lesssim \pot{2}{-18}\,{\rm s}^{-1}$, the decay process never enters the exponential phase of the evolution (i.e., $\Ginj t_0\ll1$), such that $\id \ln \rho/\id t \propto \efdm\Ginj/(1+z)$ [cf., Eq.~\eqref{eq:drho_dt_inj}]. In this case, a {\it universal distortion shape} is obtained, which becomes only a function of $\xinjc$, such that we may write
%----------------------------------------------------------
\begin{align}
\label{eq:Universal}
\Delta I_\nu(\xinjc, \efdm, \Ginj)\approx \efdm \Ginj\, \Delta \hat{I}_\nu(\xinjc).
\end{align}
%----------------------------------------------------------
Here, the distortion template, $\Delta \hat{I}_\nu(\xinjc)$, is obtained numerically for a sufficiently long lifetime (in practice we use $\Ginj^{\rm ref}=10^{-20}\,\mathrm{s}^{-1}$). To obtain constraints on quasi-stable particle decays we can use the universal distortion template to determine $\fdm$ from the data by simply rescaling the solution for $\Ginj^{\rm ref}$. 

The distortion created by extremely long-lived particle is essentially formed by the combined action of photon injection, free-free absorption at low frequencies and photo-ionization of neutral atoms at high frequencies. Electron scattering can be neglected in terms of redistributing photon in energy, but has to be included for the thermal balance with electrons.
In Fig.~\ref{fig:universal_profile}, we show the obtained universal distortion template for various values of $\xinjc$. Injection at low frequencies leads to significant heating while for $\xinjc>10^{-4}$ mostly the direct distortion becomes visible. The free-free absorption edges are visible as abrupt rise of the signal redward of the direct injection maxima for $\xinjc=10^{-8}, 10^{-6}, 10^{-4}, 10^{-2}$ and $1$. For cases with $\xinjc\simeq 10^5-10^6$ (not shown in the figure), photo-ionizations lead to heating and, hence, increased $y$-distortions.

%----------------------------------------------------------
\begin{figure}
\includegraphics[width=\columnwidth]{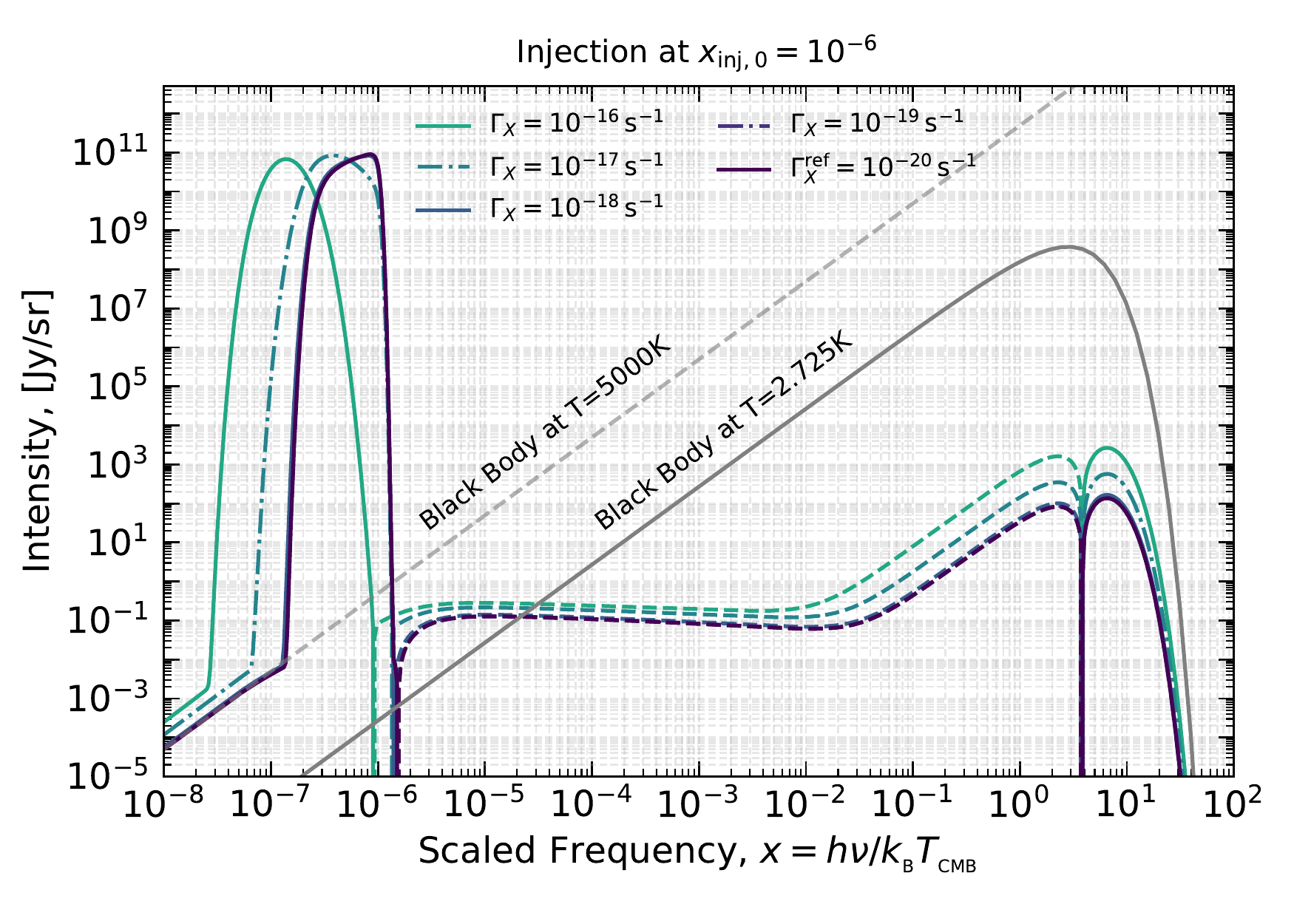}
\\[-3mm]
\includegraphics[width=\columnwidth]{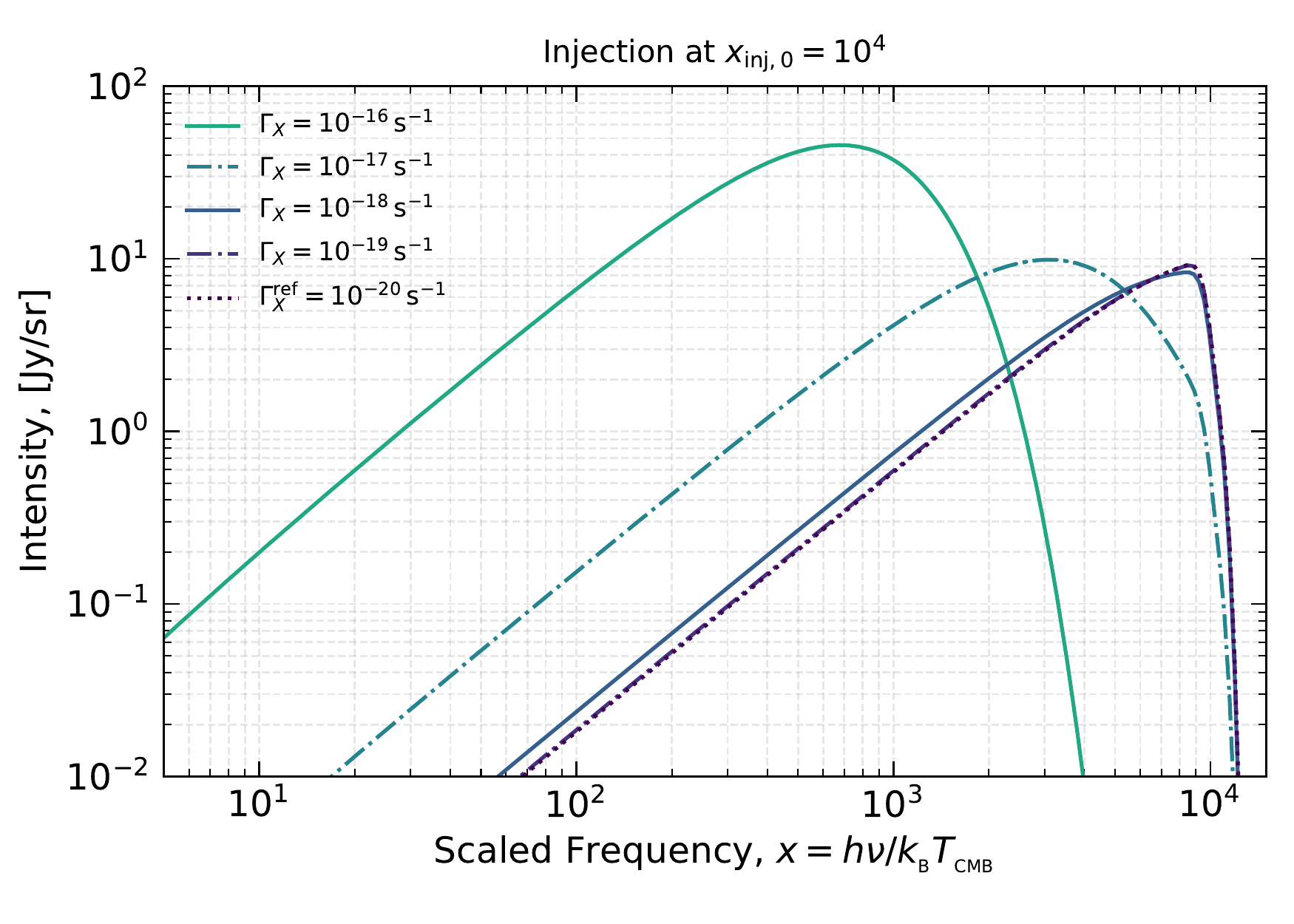}
\vspace{-5mm}
\caption{Ultra long-lived decaying particles for $\xinjc=10^{-6}$ and $10^4$. The solution in each case is scaled to match the overall amplitude of the reference case with $\Ginj^{\rm ref}=10^{-20}\,{\rm s^{-1}}$. For each $\xinjc$, the universal shape is obtained independent of the lifetime once $\Ginj\lesssim 10^{-19}\,{\rm s}^{-1}$.} 
\vspace{-3mm}
\label{fig:universal_profile_convergence}
\end{figure}
%----------------------------------------------------------
Figure~\ref{fig:universal_profile_convergence} illustrates how the distortion approaches the universal distortion shape for $\xinjc=10^{-6}$ and $10^4$ while increasing the lifetime. The distortion shape freezes once $\Ginj\lesssim 10^{-19}\,{\rm s^{-1}}$. 
One giveaway signature of the universal case is the abrupt drop of the signal blueward of the direct injection peak, which is simply due to the fact that the injection process is computed for a finite time. For cases with lifetimes shorter than the age of the Universe, the onset of the exponential decay phase is visible in the shape of the signal around the injection maximum (blue and red lines). 

In the upper panel of Fig.~\ref{fig:universal_profile_convergence}, we also show the spectra of the CMB at $\TCMB=2.725\,\Kel$ and for a $5000\,\Kel$ blackbody. For low-frequency injection, the medium is indeed significantly heated and the direct injection distortion can exceed the blackbody spectrum by many orders of magnitude. While current direct distortion constraints do not exclude these cases, the effect on the ionization history places more stringent limits in this regime.

We note here in passing that the computation for low-frequency injection ($x\lesssim 1$) becomes highly non-linear in the distortion itself. {\tt CosmoTherm} includes blackbody-induced stimulated scattering effects \citep[e.g.,][for analytic discussion]{Chluba2008d} but these would need to be augmented by terms $\propto \Delta n^2$ to obtain fully consistent results. This could lead to Bose-Einstein condensation of photons \citep{Levich1969}, which we cannot include accurately in the current version of {\tt CosmoTherm}. While a detailed treatment is beyond the scope of this paper, the general results are not expected to be changed by this omission. A recent related discussion can be found in \citet{Brahma2020}.

%----------------------------------------------------------
\begin{figure}
\includegraphics[width=\columnwidth]{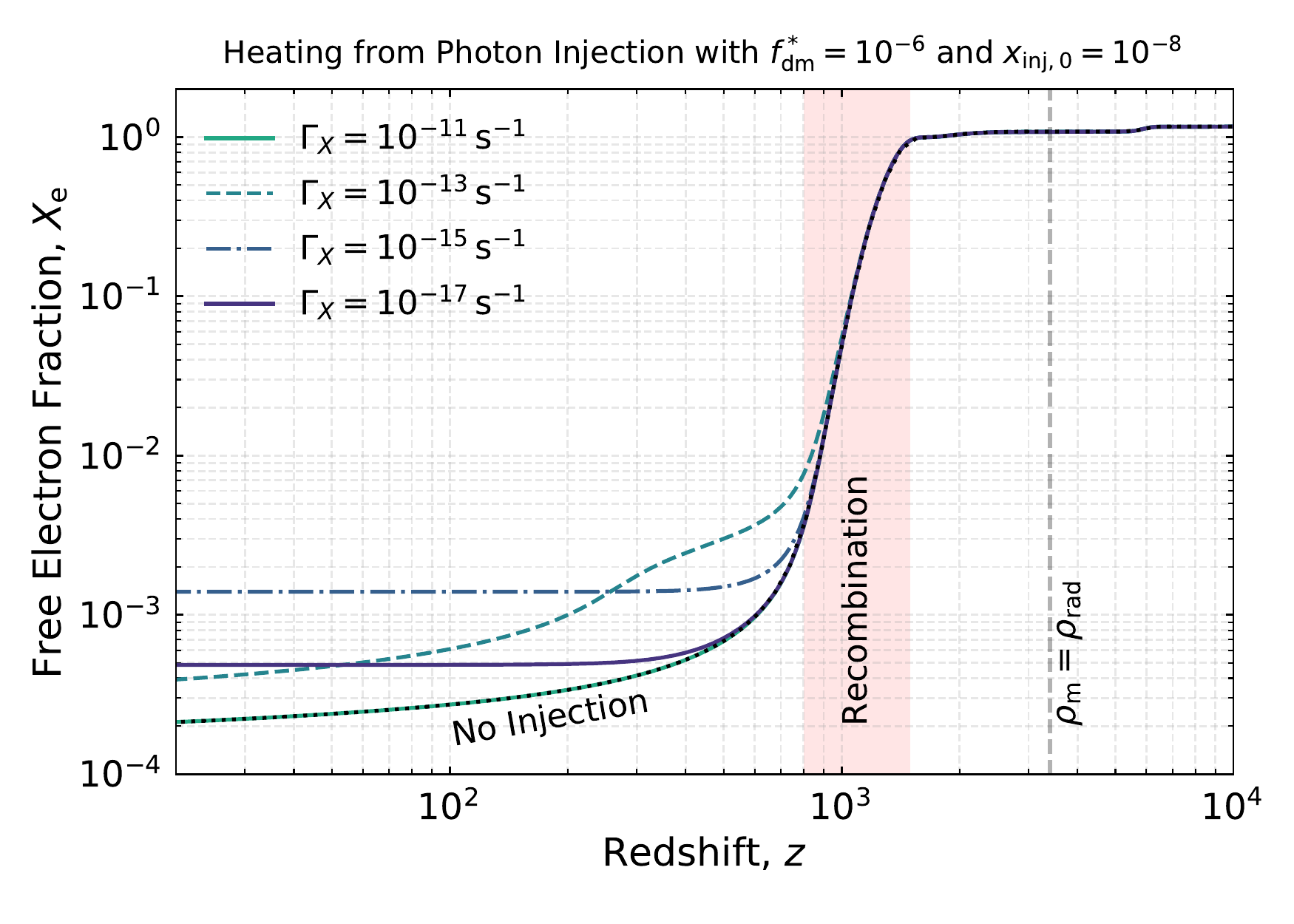}
\vspace{-6mm}
\caption{The pre-reionization history with heating from ultra soft photon injection. Computation were done with {\tt{CosmoRec/Recfast++}}. The dotted line shows the free electron fraction in the case of no injection.} 
\vspace{-5mm}
\label{fig:soft_heating}
\end{figure}
%----------------------------------------------------------

\subsection{Effects on the ionization history}\label{sec:xe}
%-------------------------------------------------------------------
So far we focused on the final distortion as it would be observed today. Another important effect of photon injection is the change associated to the ionization history. For high energy decays, well above the ionization thresholds of hydrogen and helium atoms ($E>1\,\MeV$), a high-energy particle cascade is induced, leading to many secondary particles causing atomic ionizations, excitations and heating \citep[e.g.,][]{Shull1985, Valdes2010, Slatyer2015}. This problem has been studied several times \citep[e.g.,][]{Chen2004, Padmanabhan2005, Galli2009, Slatyer2009}. 
Here we investigate injections close to the ionization threshold and at low energies. Even in the latter case, a significant effect on the ionization history can be observed and, hence, can be constrained using the CMB temperature and polarization anisotropy, as we explain now.

To have a significant effect on the ionization history and CMB anisotropy, energy needs to be released at $z\lesssim 1400$, implying relevant lifetimes $\Ginj\lesssim \pot{{\rm few}}{-13}\,{\rm s}^{-1}$. Earlier, the plasma quickly adjusts to the extra energy input, but the ionization history is hardly affected, such that the only witness of the injection process is the spectral distortion \citep[e.g.][]{Chluba2008c, Chluba2010a}. 
If we consider injections at $\xinjc\lesssim 10^{-8}$ we can furthermore be sure that the main effect is through heating of the medium. 
In Fig.~\ref{fig:soft_heating}, we illustrate the results of this calculation for $\fdm=10^{-6}$. The curves were obtained with {\tt CosmoRec/Recfast++} \citep{Chluba2010b} including the extra heating source, Eq.~\eqref{eq:drho_dt_inj}, in the electron temperature equation. Since we assume very soft photon injection, no direct ionizations or excitations of atoms occurs. Thus, the main effect is a reduction of the recombination rate due to hotter electrons. It is important to note that the photo-ionization rates are not affected significantly, as was also discussed in the context of heating from primordial magnetic fields \citep{Chluba2015PMF}.
As expected (see Fig.~\ref{fig:soft_heating}), the heating by soft photon injection causes a delay of recombination. This induces direct changes to the CMB anisotropies, which we will constrain using an ionization history principal component projection method \citep{Hart2020PCA}.

\begin{figure}
\includegraphics[width=\columnwidth]{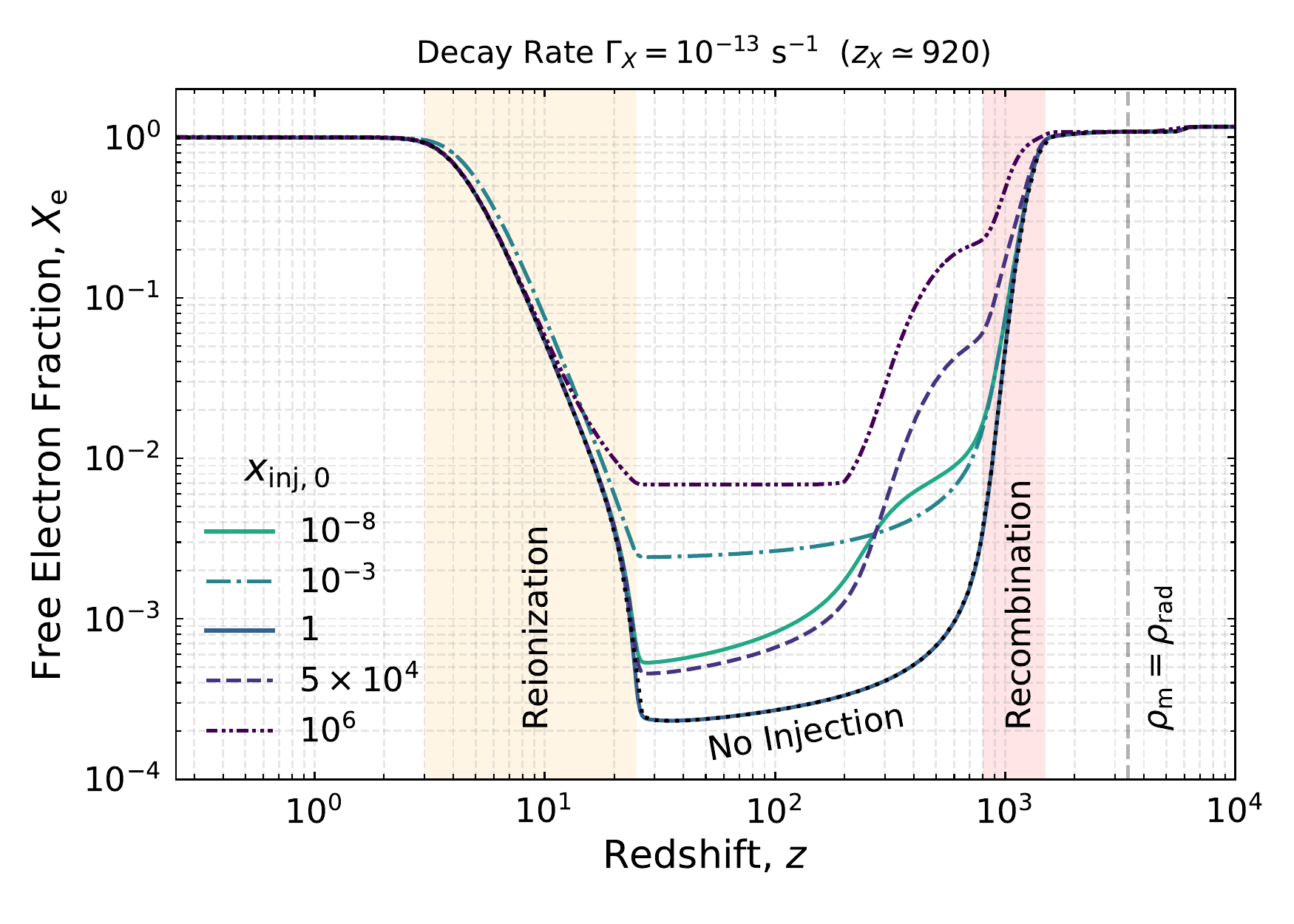}
\includegraphics[width=\columnwidth]{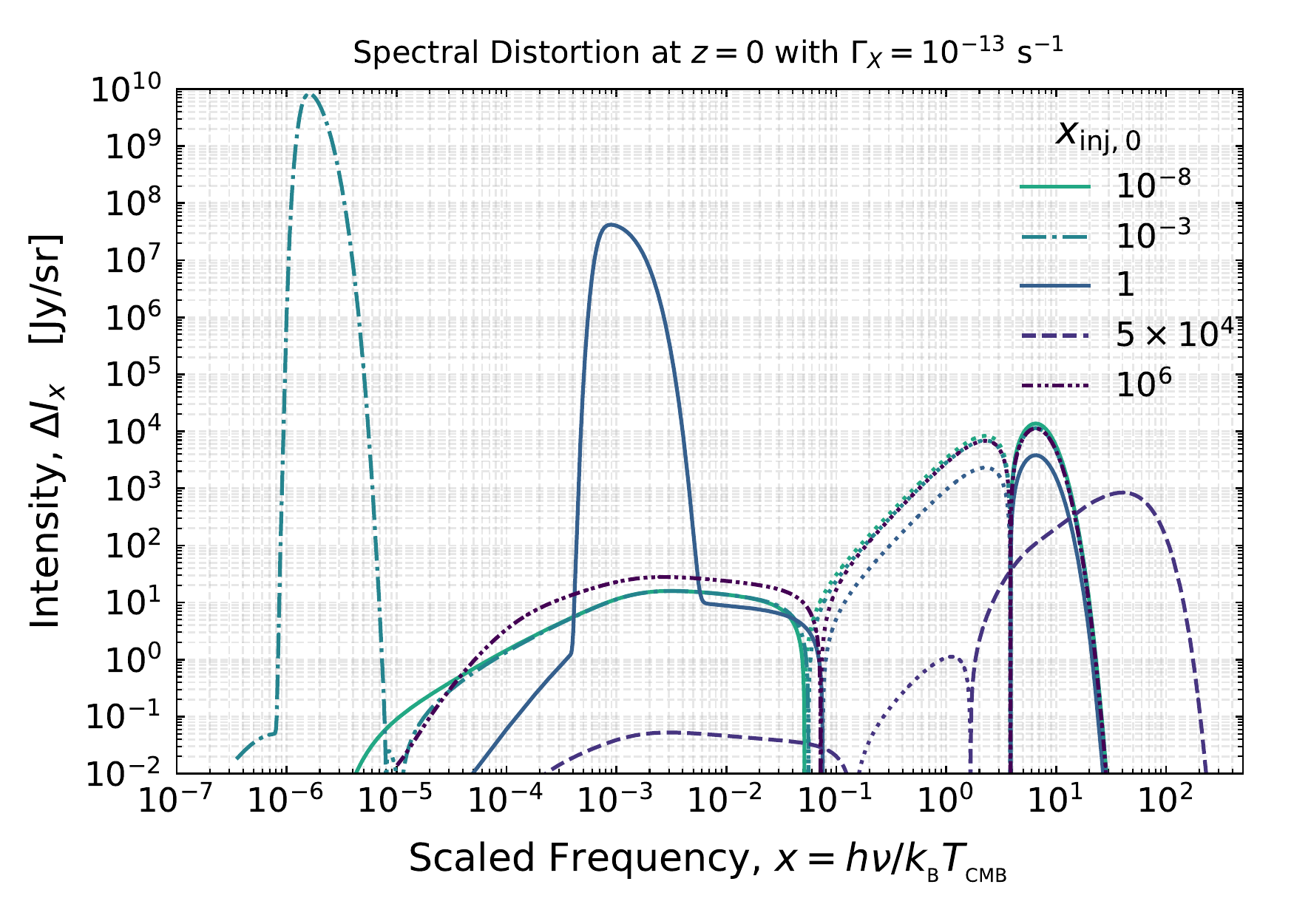}
\vspace{-5mm}
\caption{Changes in ionization history with photon injection from a decaying particle with $\Ginj = 10^{-13}\,{\rm s}^{-1}$ and $(\Delta \rho/\rho)_\mathrm{inj}=3\times10^{-5}$ as a function of the injection energy (upper panel). The corresponding distortion is shown in the lower panel. The dotted line in the upper panel is case without photon injection. The computations were performed with \tt{CosmoTherm}.}
\label{fig:high-energy-Xe}
\end{figure}

Looking at Fig.~\ref{fig:regimes}, even in cases with $\Ginj\lesssim \pot{{\rm few}}{-13}\,{\rm s}^{-1}$ we expect a gradual reduction of the effect on the ionization history as we increase $\xinjc$ to about $\simeq 10^{-2}$. This is because a diminishing fraction of the injected  energy causes a direct effect on the electrons, leaving the ionization history mostly unaffected. This effect can be seen in Fig.~\ref{fig:high-energy-Xe}, where $\xinjc=10^{-8}$ shows a large effect that gradually reduces as $\xinjc=1$ is reached. 
This behavior continues until $\xinjc\simeq \pot{5.8}{4}$, which corresponds to the \HI Lyman continuum threshold, allowing for direct ionizations from the ground state.
In Fig.~\ref{fig:high-energy-Xe}, we see a significant response in ionization history caused by both heating and direction ionizations. Aside from the scenarios for $\xinjc=1$, all other cases are already in tension with current CMB data from \Planck, showing how a combination of anisotropy and distortion measurements provides complementary information about particle physics.

%----------------------------------------------------------
%----------------------------------------------------------

%----------------------------------------------------------
\begin{figure}
\includegraphics[width=\columnwidth]{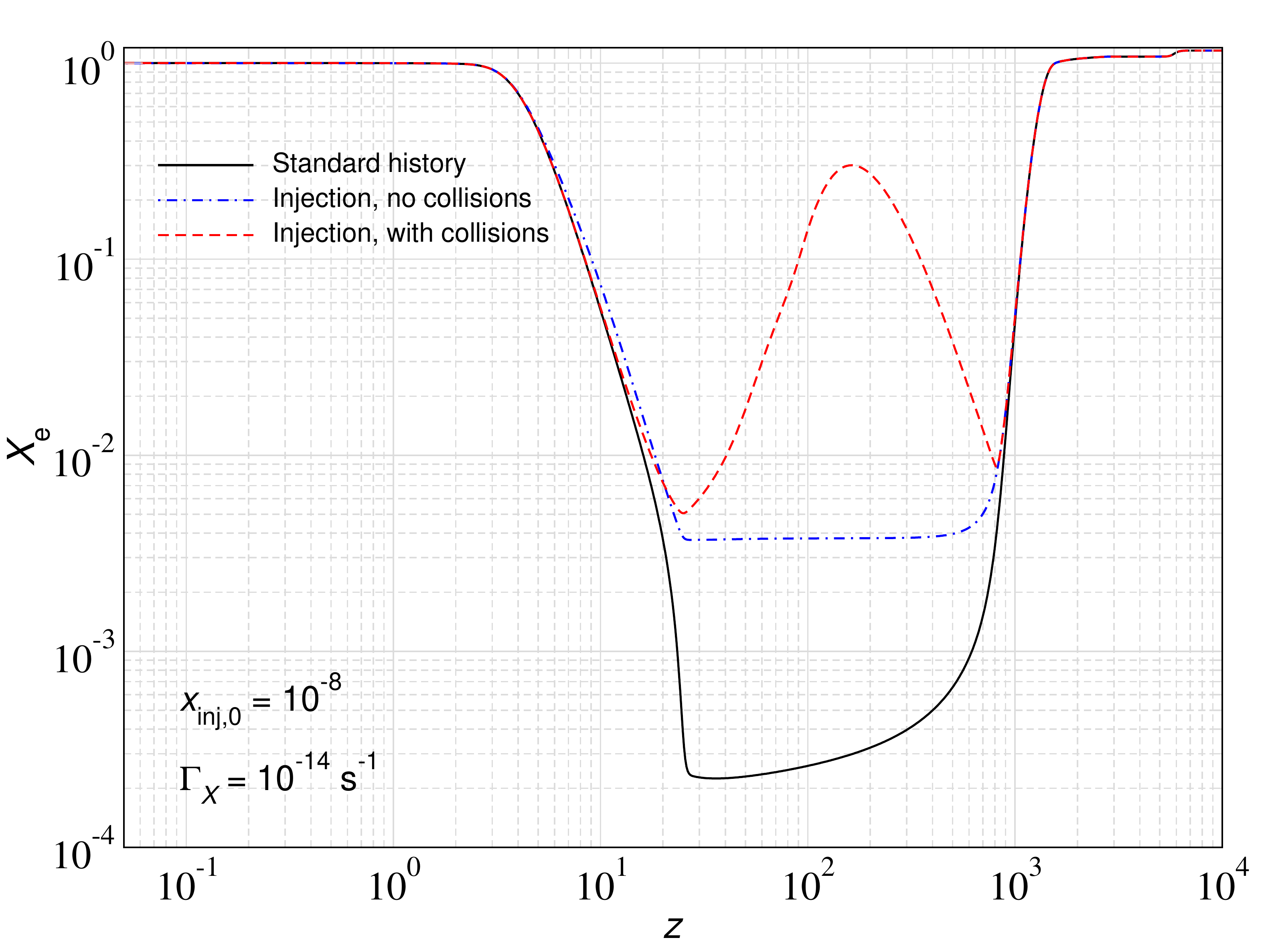}
\vspace{-4mm}
\caption{An illustration of the modifications to the recombination history due to collisions. We injected photons at $\xinjc=10^{-8}$ for $\Ginj=10^{-14}\,{\rm s}^{-1}$ and $\Delta \rho_\gamma/\rho_\gamma=\pot{3}{-5}$ switching the effect of collisions on and off. For reference, we also show the standard ionization history with reionization.} 
\vspace{-4mm}
\label{fig:collisions}
\end{figure}
%----------------------------------------------------------
\subsubsection{Importance of collisional processes}
\label{sec:collisions_effect}
%----------------------------------------------------------
It turns out that collisional processes play an important role for the evolution of the distortion and ionization history in particular when significant injection occurs at very low frequencies. 
To illustrate the role of collisions, in Fig.~\ref{fig:collisions} we computed the ionization history for soft photon injection switching the effect of collisions on and off.
Due to the heating by free-free absorption, collisional ionizations become important and lead to a strong increase in the free electron fraction, which without collisional ionizations shows a flat response in the freeze-out tail of recombination. 

%----------------------------------------------------------
\begin{figure}
\includegraphics[width=\columnwidth]{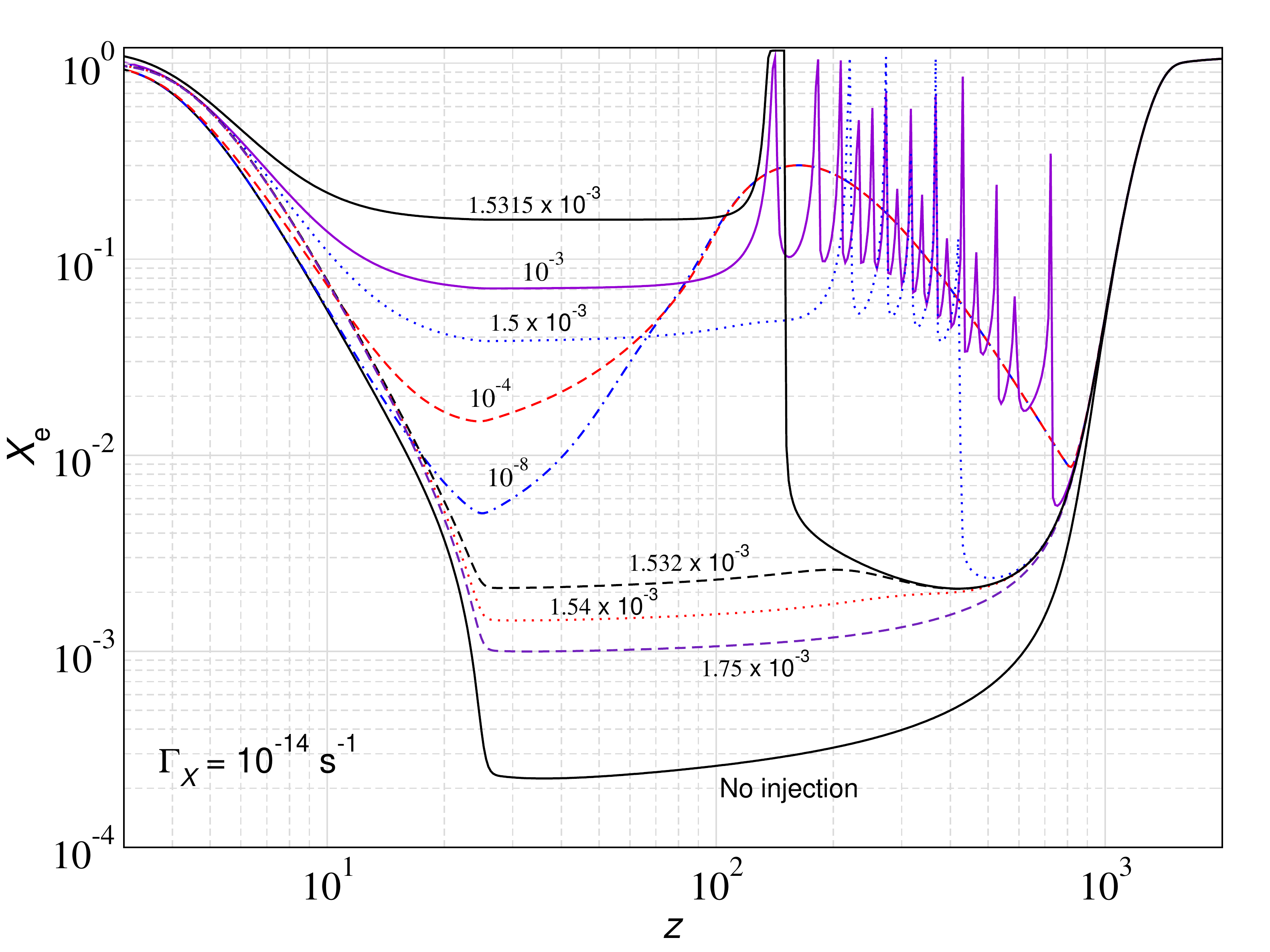}
\vspace{-4mm}
\caption{Non-linear responses in the ionization history around critical injection frequencies for $\Ginj=10^{-14}\,{\rm s}^{-1}$ and $\Delta \rho_\gamma/\rho_\gamma=\pot{3}{-5}$. For $10^{-3}\lesssim \xinjc \lesssim \pot{1.53}{-3}$ the free electron fraction shows intermittent bursts due to photon injection, free-free absorption and collisional processes.} 
\vspace{-3mm}
\label{fig:collisions-critical}
\end{figure}
%----------------------------------------------------------
In several of our computations, we included the effects of collisions to study the changes in signals. In fact, when transitioning from very soft photon injection to higher energies, we find a critical, highly non-linear behaviour of the ionization history, mimicking a phase-transition in the recombination mode. This is illustrated in Fig.~\ref{fig:collisions-critical}, where we compute several histories for $\Ginj=10^{-14}\,{\rm s}^{-1}$. The computations for these cases indeed push the treatment within {\tt CosmoTherm} to its limit. For injections at $\xinjc\leq 10^{-4}$, as before, the heating leads to extra collisional ionizations and a significant increase in the free electron fraction at $z\lesssim 800$. Raising the injection energy to $\xinjc \simeq 10^{-3}$, we observe strong intermittent changes in $X_{\rm e}$, with episodes of very high, followed by more moderate, levels of ionization. The general intermittent behaviour with varying levels in the number of bursts continues until $\xinjc\simeq \pot{1.532}{-3}$ is reached, when the recombination response becomes more moderate and smooth again.

{\it What causes this erratic behavior?} While the numerical treatment of this transition is certainly challenging, we identified the interplay between photon injection heating and increased collisional ionization as cause. Photons are efficiently absorbed in the optically-thick regime of the free-free process. Once crossing over to the optically-thin regime, the response in the ionization history becomes very sensitive to the injection process. Absorbed photons are converted into heat, which increases the ionization fraction through collisions. This in turn increases the amount of free-free absorption leading to a positive feedback loop. During the critical behavior, we thus see alternating phases of strong free-free absorption and recombination. 

The onset of this new recombination mode 
depends on the selected lifetime and also the amount of photon injection.
Changing the lifetime modifies the critical injection energies.
Reducing the photon injection process causes the non-linear response to stop, as no runaway collisional ionization phase followed by increased free-free absorption is produced. We were furthermore unable to produce the critical behaviour without explicitly following the photon injection process and buildup of low-frequency distortions. Thus, the full treatment implemented here in {\tt CosmoTherm} is required.

Since the corresponding ionization histories are already in strong tension with existing limits from \Planck, we avoid the complications due to the interesting non-linear physics by i) reducing the photon injection rate and ii) omitting collisions in the main computation. Unless pushed to more extreme cases with a significant rise in the baryon temperature ($\Te\gtrsim 10^4\,\Kel$), the non-linear recombination mode is not usually excited, such that this should not be a significant limitation for our main conclusions.

%-------------------------------------------------------------------
\vspace{-3mm}
\subsection{Blackbody-stimulated decay}
\label{sec:stim}
%-------------------------------------------------------------------
Until now, we have neglected the effect of stimulated decay; however, as we shall see next, it can play an important role in modifying the phenomenology of photon injection processes. By `stimulated decay' we mean the enhancement of the decay rate with respect to vacuum case due to the presence of background photons. This process is only relevant if the coupling to the photon field is direct, without an intermediate unstable mediator.
The net decay rate then depends on the Boltzmann terms 
%---------------------------------------------
\begin{equation}
    \label{eq:stim-boltz}
    \mathcal{F}=
    f_X(1+n_\gamma)(1+n_\gamma^\prime)-n_\gamma n_\gamma^\prime(1\pm f_X),
\end{equation}
%---------------------------------------------
where $f_X$ is the distribution function of the decaying particle. The last term reflects the inverse process, in which the `$+$' applies for bosonic particles and `$-$' for Fermions. Again assuming that the particles are cold, any broadening of the emission line due to thermal motion of the particle can be neglected and $x\approx x'\approx \xinj(z)$. Collecting terms, we then have 
%---------------------------------------------
\begin{equation}
    \label{eq:stim-boltz_simp}
    \mathcal{F}\approx
    \begin{cases}
    f_X(1+2n_\gamma) &\text{(Boson)}
    \\
    f_X(1+2n_\gamma+n^2_\gamma) &\text{(Fermion)}
    \end{cases}
\end{equation}
%---------------------------------------------
with the implicit assumption $n_\gamma^2\ll f_X$. Below we will consider the bosonic case as an example \citep[see][for related discussion in the context of ALPs]{Jaeckel2020}. Therefore, the decay process in the stimulated case has a rate $\Gamma_X
^\mathrm{stim}\approx[1+2n_\gamma(x_\mathrm{inj})]\Gamma_X$, where $\Gamma_X$ is the vacuum decay rate. We note, however, that for the fermionic case the effects could be significantly stronger, with the enhancement essentially scaling like $\propto n_\gamma^2$.

For the ambient photon occupation number we use the Planck law, $n_\gamma (x)=[\expf{x}-1]^{-1}$. For small distortions, this will be extremely accurate down to redshift $z\simeq 10$, when one expects a significant radio background to arise due to structure formation. In addition, for large low-energy injection, one does expect the distortion itself to induce further emission and potentially even violate the condition $n_\gamma^2\ll f_X$. Hence, our calculations below are mostly for illustration.
For $x_\mathrm{inj}\gg 1$, there are no differences between the stimulated and vacuum decay cases, since $n_\gamma (x_\mathrm{inj})\ll 1$. However, for $x_\mathrm{inj}\lesssim 1$, the photon occupation number, $n_\gamma (x_\mathrm{inj})\approx (1+z)/x_\mathrm{inj,0}$, can be large and lead to very different injection histories. 

Taking stimulated decay into account, the time evolution of the particle number density has same form as in vacuum, provided we change the time coordinate using $\mathrm{d}t_\mathrm{stim}=[1+2n_\mathrm{Pl}(x_\mathrm{inj})]\mathrm{d}t$, which we compute explicitly in {\tt CosmoTherm}.
In this case, we have $N_X^\mathrm{stim}(t)=N^{\rm vac}_X(t_\mathrm{stim})$, where $N^{\rm vac}_X(t)=N_{X,0}\,(1+z)^3\exp(-\Ginj t)$ denotes the vacuum decay solution. For instance, in the {low-frequency} regime, $\xinj\ll 1$, during radiation domination, we have
%---------------------------------------------
\begin{align}
\label{eq:tstim_approx_rad}
t_\mathrm{stim}&\approx\frac{2\sqrt{2 t}}{\xinjc\Omega_\mathrm{rad}^{1/4}\sqrt{H_0}}
\approx \pot{2.0}{10}\,\frac{\sqrt{t}}{\xinjc}\,\secs^{1/2}.
\end{align}
%---------------------------------------------
Since $\dot{N}_\gamma = - f_\gamma \dot{N}_X=f_\gamma \Gamma_X[1+2n_\gamma]N_X$, it follows that the photon injection source term for the spectral distortion can also be written in the same way as the vacuum case, provided one includes the enhancement factor for the decay rate and uses $N_X^\mathrm{stim}(t)=N^{\rm vac}_X(t_\mathrm{stim})$. 

\begin{figure}
\includegraphics[width=\columnwidth]{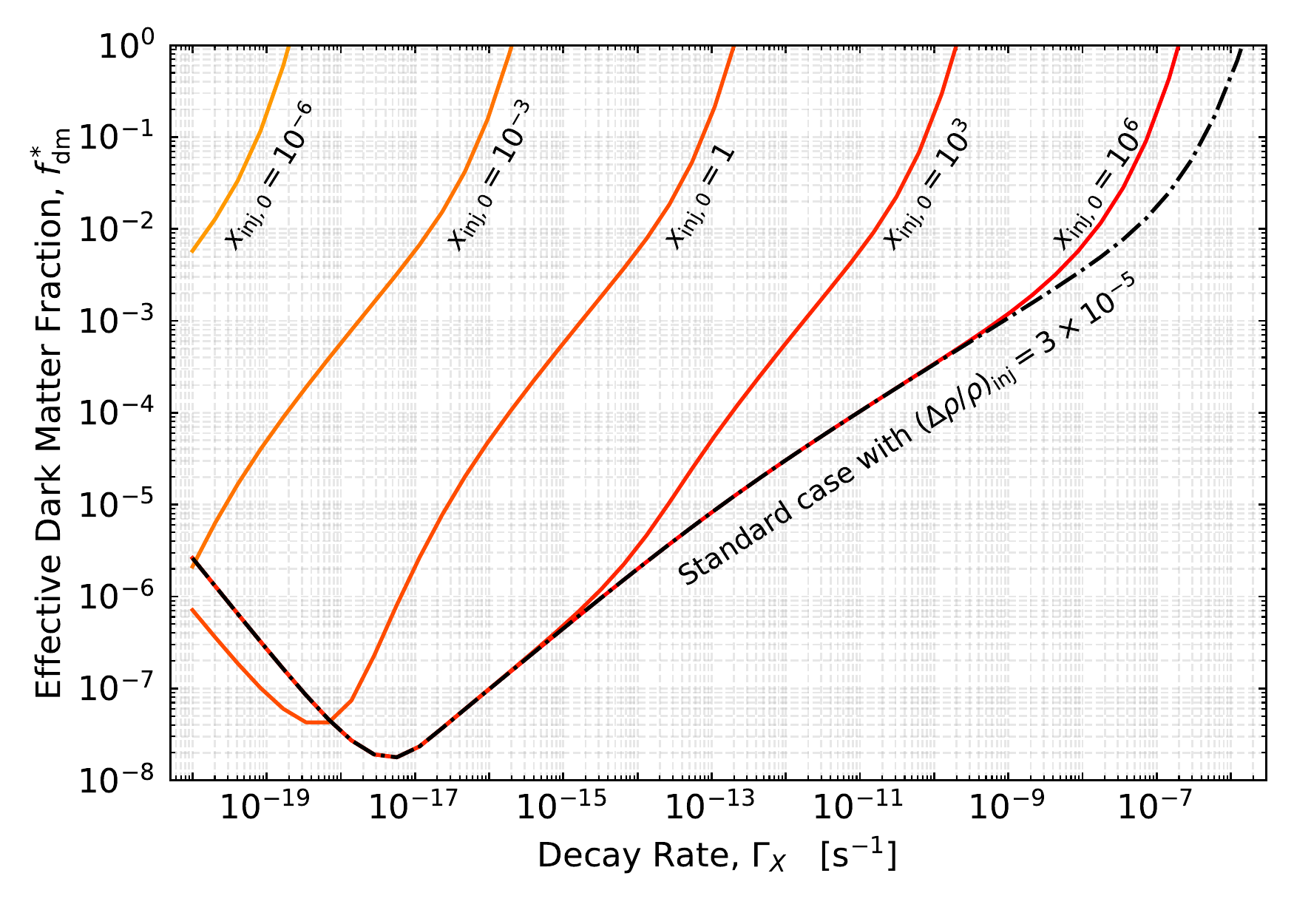}
\\[-2mm]
\includegraphics[width=\columnwidth]{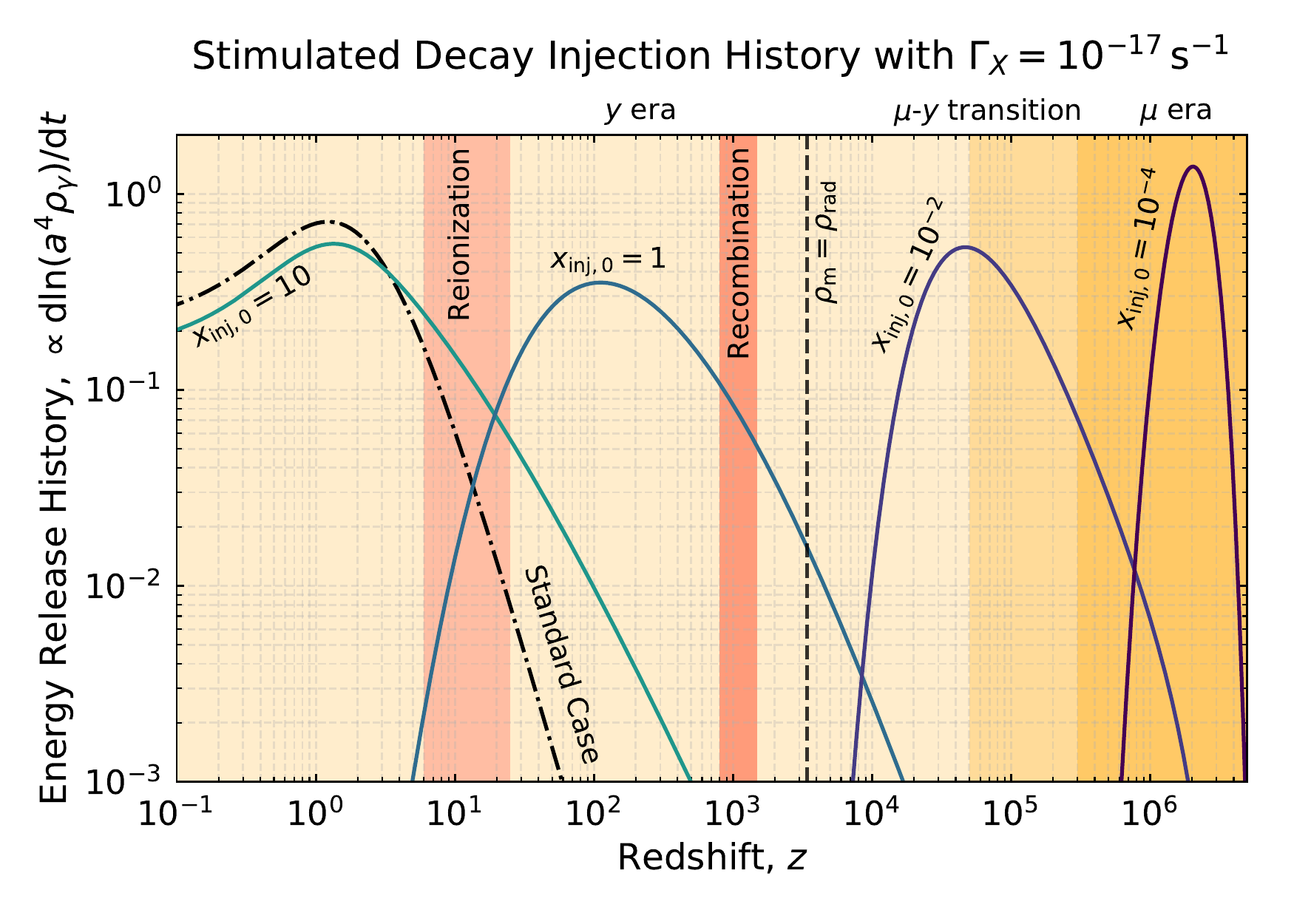}
\vspace{-3mm}
\caption{Effect of stimulated decay on the injection process. {\it Upper panel}: Allowed values of $\fdm^*=\efdm$ satisfying the condition $\Delta\rho/\rho|_{\rm inj}=\pot{3}{-5}$ as function of $\Ginj$ for different values of $\xinjc$. {\it Lower panel}: Energy release history for a lifetime $\Ginj=10^{-17}\,\mathrm{s}^{-1}$ and several values of $x_\mathrm{inj,0}$. The maximum of the injection history moves towards higher redshifts when decreasing $\xinjc$ because blackbody-induced effects accelerate the decay. }
\vspace{-3mm}
\label{fig:Gamma_stim}
\end{figure}

However, one must be cautious with the normalization condition. Indeed, for a given $\Delta \rho/\rho|_\mathrm{inj}$ the value of $f_\mathrm{inj}$ in the stimulated decay case differs from the vacuum case, affecting $\fdm^*=\efdm$. The modified values can be obtained using the integral\footnote{The standard case is recovered for $n_\mathrm{Pl}(x_\mathrm{inj})=0$, also implying $t_\mathrm{stim}=t$.}
%---------------------------------------------
\begin{align}
\frac{\Delta\rho_\gamma}{\rho_\gamma}\Bigg|^{\rm stim}_{\rm inj}
&= \int [1+2n_\mathrm{Pl}(x_\mathrm{inj})] \frac{\id \ln \rho^{\rm vac}_\gamma}{\id t}\Bigg|_{t=t_\mathrm{stim}} \mathcal{J}_{\rm bb}(z) \id t
\nonumber
\\
&= \frac{\mathcal{G}_3}{\mathcal{G}_2}x_\mathrm{inj,0}\,\Gamma_X
\int \frac{\mathcal{J}_{\rm bb}(z)}{(1+z)}\,\expf{-\Gamma_X t_\mathrm{stim}} \id t_\mathrm{stim}.
\end{align}
%---------------------------------------------
The upper panel of Fig.~\ref{fig:Gamma_stim} illustrates the initial conditions in terms of $\fdm^*=\efdm$ for $\Delta \rho/\rho|_\mathrm{inj}=3\times10^{-5}$. The dashed-dotted curve is for vacuum decay, while the other curves are with stimulated decay for various injection frequencies $x_\mathrm{inj,0}$ as labeled. As anticipated, there is hardly any difference for large $x_\mathrm{inj,0}$. As $x_\mathrm{inj,0}$ decreases, and the particles decay faster due to stimulated effects, it is necessary to increase the particle abundance in order to keep the same value of $\Delta \rho/\rho|_\mathrm{inj}$. This can be understood when realizing that stimulated decays essentially reduce the effective lifetime of the particle. Thus, for stronger stimulated effects, injection occurs in the regime where thermalization is extremely effective ($z\gtrsim \pot{2}{6}$) and the effective $\Delta \rho/\rho|_\mathrm{inj}$ drops rapidly for a fixed value of $\fdm^*$.

%----------------------------------------------------------
%----------------------------------------------------------

The bottom panel of Fig.~\ref{fig:Gamma_stim} shows the energy release history for a long-lived particles with $\Gamma_X=10^{-17}\,\,\mathrm{s}^{-1}$ with stimulated decays (solid lines) compared to the vacuum decay case (dashed dotted line). These curves are normalized with respect to the total integral, and they include the effect of the distortion visibility function, which exponentially suppresses the energy injection in the \textit{temperature era}, i.e., at $z\gtrsim 2\times 10^{6}$. In the vacuum decay case, the energy release history is independent of the injection frequency and for the chosen long lifetime mostly occurs at low redshifts, with a maximum at a redshift $z_X$ that can be estimated using the approximation in Eq.~\eqref{eq:zx_max}. With stimulated decay, the energy release history directly depends on $\xinjc$ because the effective decay rate, $\Gamma_\mathrm{stim}$, becomes a function of the injection frequency. The ratio $\Gamma_\mathrm{stim}/\Gamma_X$ is always larger than unity, so the particles decay faster than in vacuum, which means that the energy injection happens earlier, with lower $\xinjc$ corresponding to earlier injection. 
Thus, while an injection was happening mostly in the post-reionization phase in vacuum, the injection with stimulated decays may now occur at an earlier time, depending on the injection energy.

\begin{figure}
\includegraphics[width=\columnwidth]{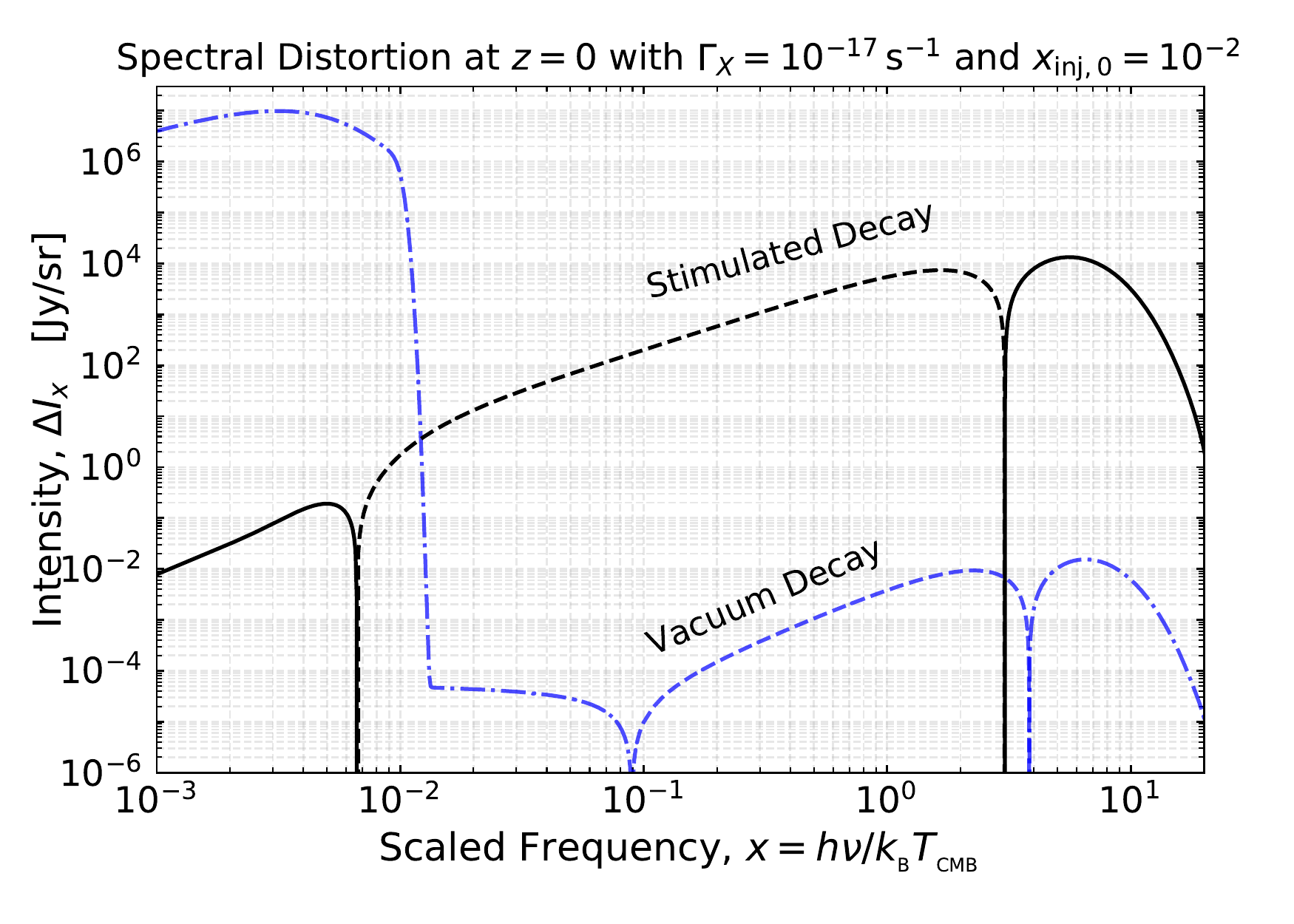}
\\[-2mm]
\includegraphics[trim={0 0 -3mm 0}, clip, width=\columnwidth]{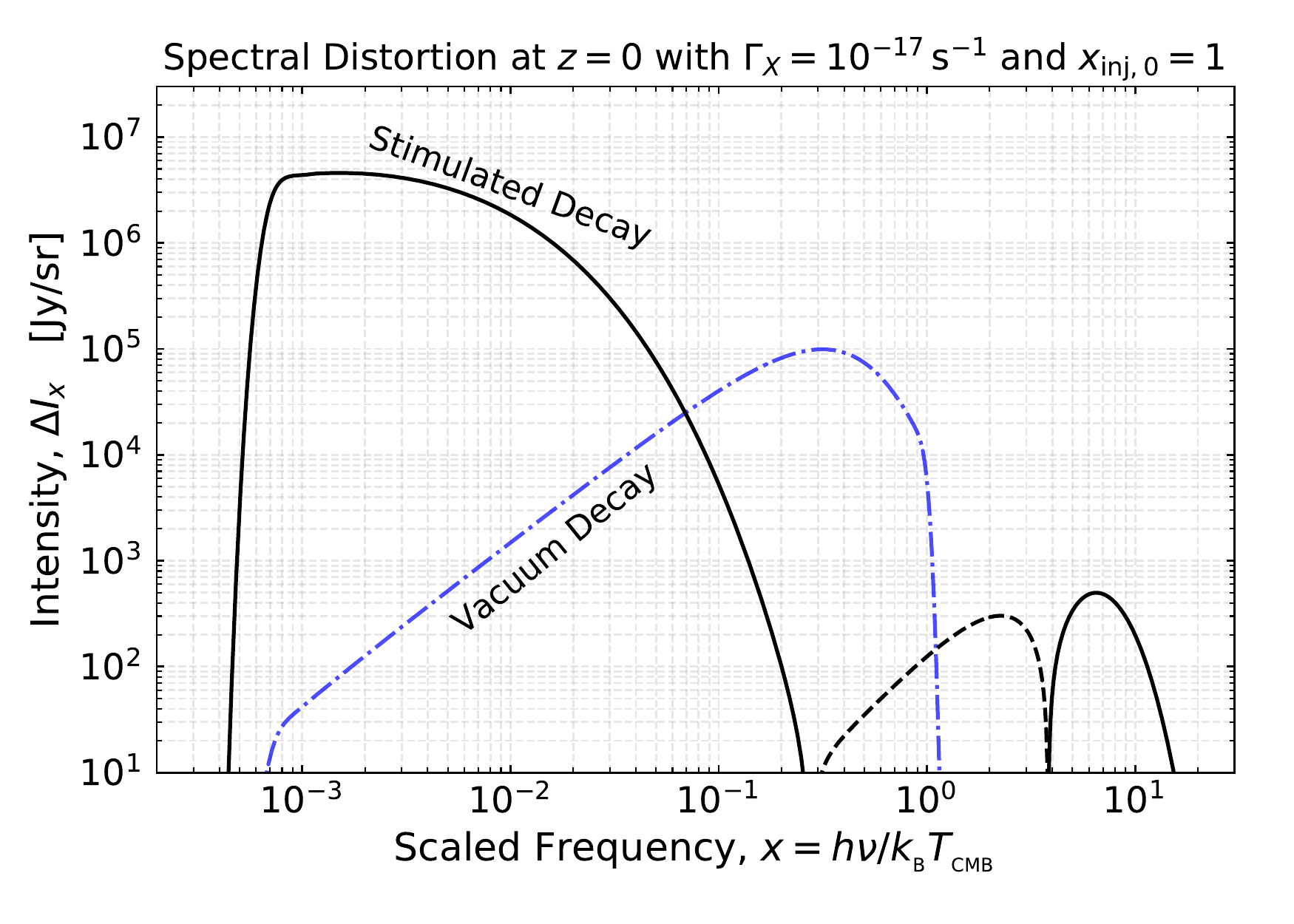}
\\[-2mm]
\includegraphics[width=\columnwidth]{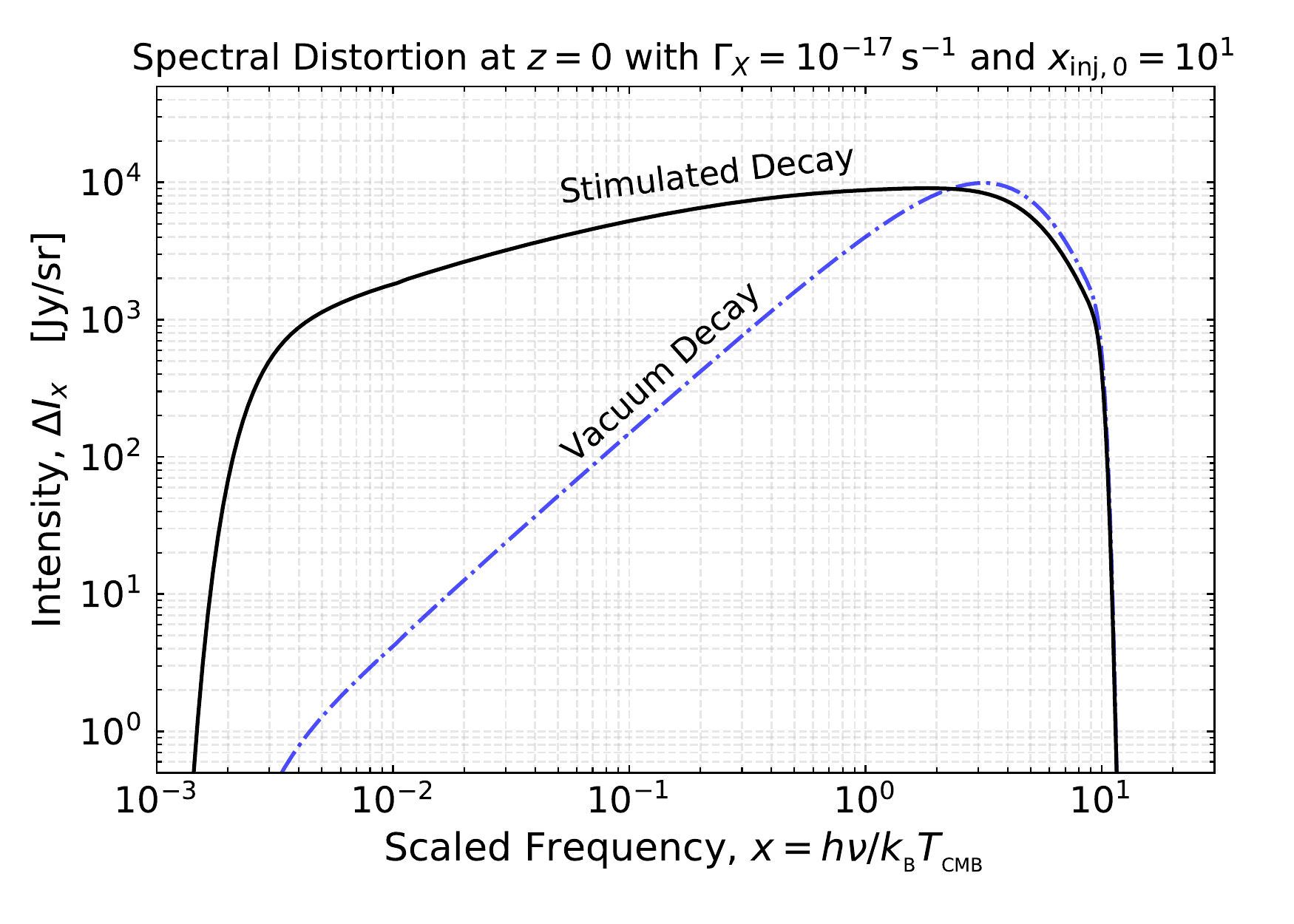}
\vspace{-5mm}
\caption{SDs for stimulated and vacuum decay cases. The curves are for $\Gamma_X=10^{-17}\,\mathrm{s}^{-1}$ and $\Delta \rho/\rho|_\mathrm{inj}=3\times10^{-5}$, with Lyman continuum absorption and reionizaton included, but collisions switched off.}
\vspace{-3mm}
\label{fig:DI_Gamma_stim}
\end{figure}

In Fig.~\ref{fig:DI_Gamma_stim}, we illustrate the distortions obtained when including stimulated decay. These correspond to the cases shown in Fig.~\ref{fig:Gamma_stim} for $x_\mathrm{inj,0}=\{10^{-2}, 1, 10\}$. 
For $x_\mathrm{inj,0}=10^{-2}$, most of the injection occurs at the early stage of the $y$-era, while for $x_\mathrm{inj,0}=1$, the recombination/post-recombination era is targeted. 
%----------------------------------------------------------
%----------------------------------------------------------
For relatively high injection energy (see bottom panel in Fig.~\ref{fig:DI_Gamma_stim}), the differences between stimulated and vacuum decay are less pronounced, as the injection occurs roughly at the same time (i.e., post reionization, for $\Gamma_X=10^{-17}\,\,\mathrm{s}^{-1}$) and with a similar time-dependence. Nevertheless, we see a different slope for the distortions at frequencies below the peak. As we explained in Sect.~\ref{sec:distortion_results}, for vacuum decay the scaling is $\Delta I \propto x^{3/2}$. With stimulated decay at relatively high redshift ($z \gtrsim 10 \,\xinjc$ such that $x_\mathrm{inj}(z)\ll 1$) the amplitude of the injection is  enhanced by a factor $1+2n_\gamma\simeq 2/x$ and hence the scaling of the distortion changes to $\Delta I\propto x^{1/2}$.

%------------------------------------------
\vspace{-3mm}
\subsubsection{Critical lifetime for early injection}
%------------------------------------------
When the injection occurs at early time, in the $\mu$-era, for redshifts $z>z_\mu=\pot{3}{5}$ we can again describe the distortion analytically, following the procedure of Sect.~\ref{sec:mu_analytic}. In the cases without stimulated decay, this was for photon injection from decaying particles with $\Gamma_X\gtrsim 10^{-8}\mathrm{s}^{-1}$, irrespective of the injection frequency, $x_\mathrm{inj,0}$. But in the stimulated decay case, even particles with relatively low $\Gamma_X$ can decay at high redshift, provided $x_\mathrm{inj,0}$ is small enough around the time where most of the decay occurs. 

To estimate the critical vacuum decay rate, we can simply solve the condition $\Gamma^\mu_X t_\mathrm{stim}(x_\mathrm{inj,0}, z_\mu)\approx 10^{-8}\mathrm{s}^{-1} \, t(z_\mu)$ at $z_\mu=\pot{3}{5}$. Using $t(z_\mu)\simeq \pot{2.6}{8}\,\secs$ together with Eq.~\eqref{eq:tstim_approx_rad} valid during radiation domination, we then find 
$\Gamma^{\mu}_X\simeq \pot{8.1}{-15}\,\mathrm{s}^{-1}\,\xinjc$. For $\Ginj\gtrsim\Gamma^{\mu}_X(\xinjc)$, we can compute the SD analytically using the expressions of Sect.~\ref{sec:mu_analytic}, while for $\Ginj<\Gamma^{\mu}_X(\xinjc)$, we use {\tt CosmoTherm}.

%------------------------------------------
\subsubsection{Critical lifetime for late injection}
\begin{figure}
\includegraphics[width=\columnwidth]{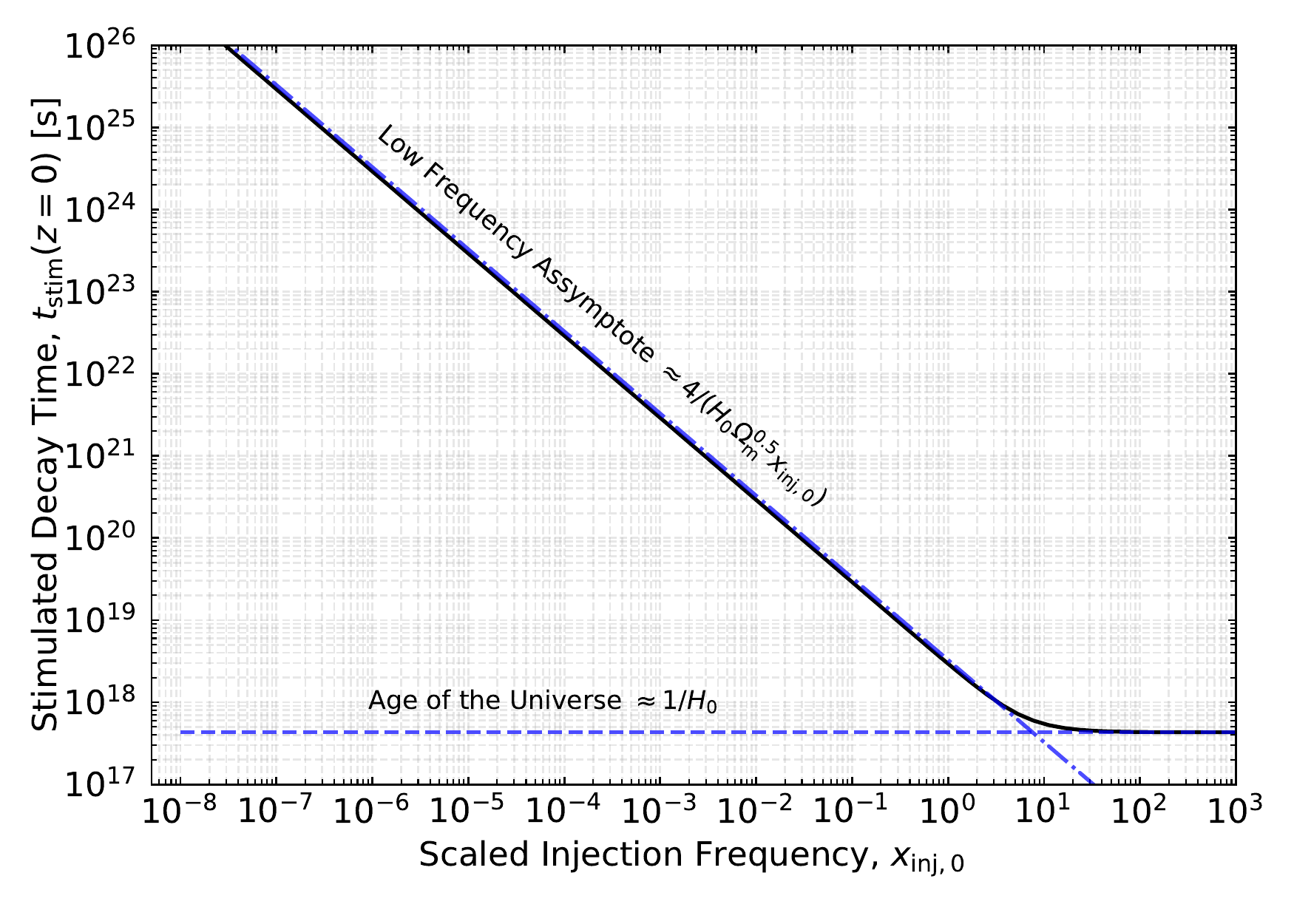}
\vspace{-4mm}
\caption{Stimulated decay time at $z=0$ as a function of $x_\mathrm{inj,0}$.}
\vspace{-3mm}
\label{fig:t_stim_age}
\end{figure}

%------------------------------------------
To further limit the computational requirements, it is useful to estimate the critical lifetime at which one can expect the distortion shape to become universal. In the vacuum decay case, this was shown to be a good approximation once $\Gamma_X\lesssim \Gamma_X^{\rm uni}=10^{-20}\,\mathrm{s}
^{-1}$ (see Sect.~\ref{LLp}).
The same reasoning applies when including stimulated decays simply using $t_\mathrm{stim}(z=0)$ instead of $t_0$. 
As illustrated on Fig.~\ref{fig:t_stim_age}, the stimulated decay time  $t_\mathrm{stim}(z=0)$ is the same as age of the universe for $x_\mathrm{inj,0} \gtrsim 10$, i.e., roughly $1/H_0$. For lower $x_\mathrm{inj,0}$, a simple derivation using the matter domination relation between time and redshift allows one to find an  approximation for the asymptote, yielding $t_\mathrm{stim}(z=0)\approx 4/[H_0\Omega_{\rm m}\xinjc]$ (see Fig.~\ref{fig:t_stim_age}). Therefore, when including stimulated effects, for decaying particles with $x_\mathrm{inj,0} \gtrsim 10$ we shall still use $\Gamma_X^{\rm uni}=10^{-20}\,\mathrm{s}
^{-1}$ as the lower lifetime, while for particles with $x_\mathrm{inj,0} \lesssim 10$ we need to solve for a wider range of lifetimes. 
Using fiducial values for $H_0$ and $\Omega_\mathrm{m}$, we find $\Gamma^{\rm uni}_X(\xinj)\approx 10^{-21}\,\secs^{-1}\,x_\mathrm{inj,0}$. 

%----------------------------------------------------------
%----------------------------------------------------------

%-------------------------------------------------------------------
\vspace{-3mm}
\section{Constraints from \FIRAS, \EDGES and CMB anisotropies}
\label{sec:constraints}
%-------------------------------------------------------------------
In this section, we explain how we use existing data from \FIRAS (Sect.~\ref{sec:FIRAS_setup}) and \EDGES (Sect.~\ref{ss:edges}) to derive constraints on the decaying particle scenarios from the previous section. In addition, we consider CMB anisotropy limits from \Planck using a principal component analysis method for changes to the ionization history (Sect.~\ref{xeconstraints}). We present model-independent constraints in Sect.~\ref{ss:mic} and then illustrate how to apply the limits to axion and ALP scenarios in Sect.~\ref{ss:mdc}.

\vspace{-3mm}
\subsection{General setup for the distortion database}
\label{ss:database}
\begin{figure}
\includegraphics[width=\columnwidth]{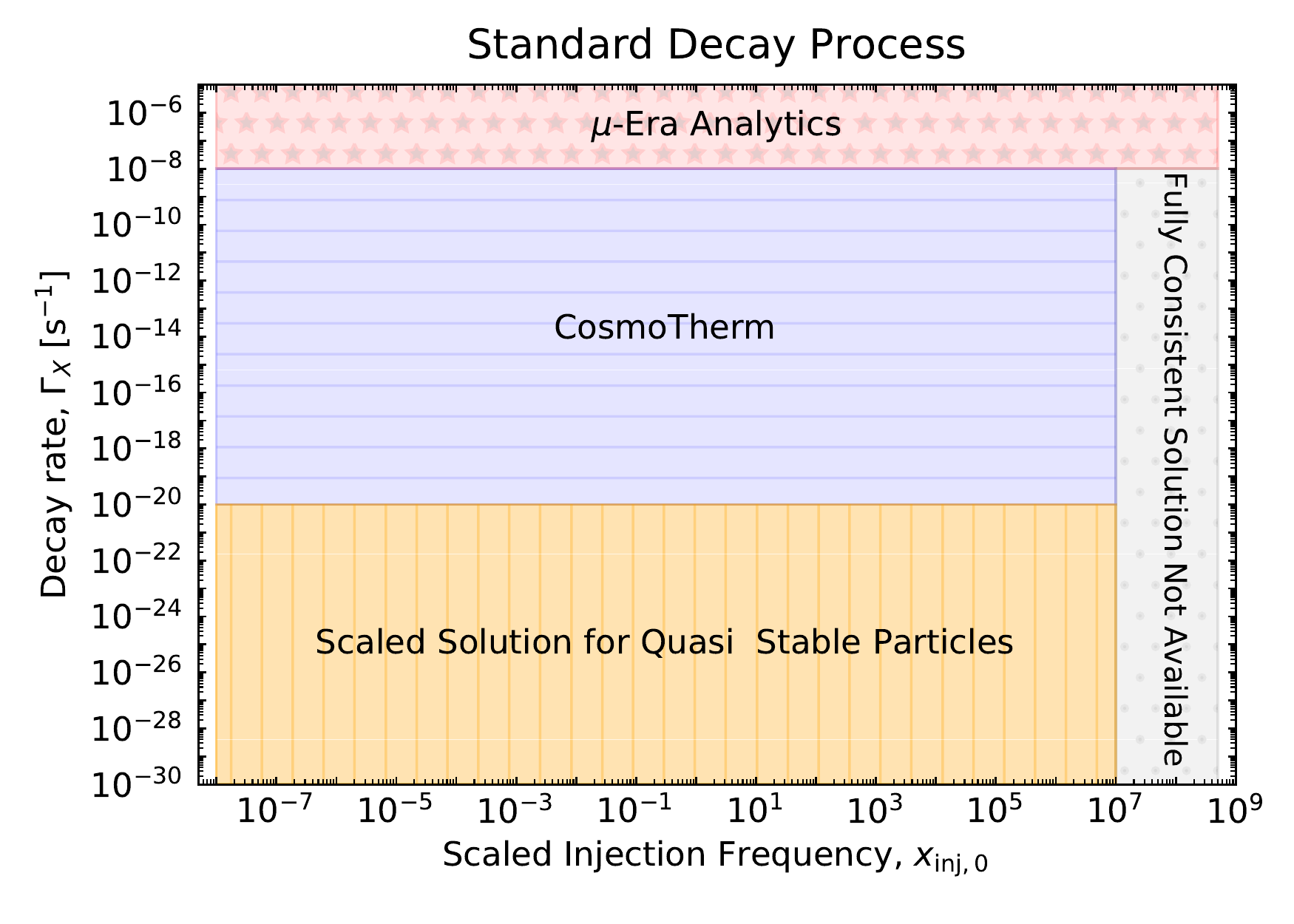}
\\[-2mm]
\includegraphics[width=\columnwidth]{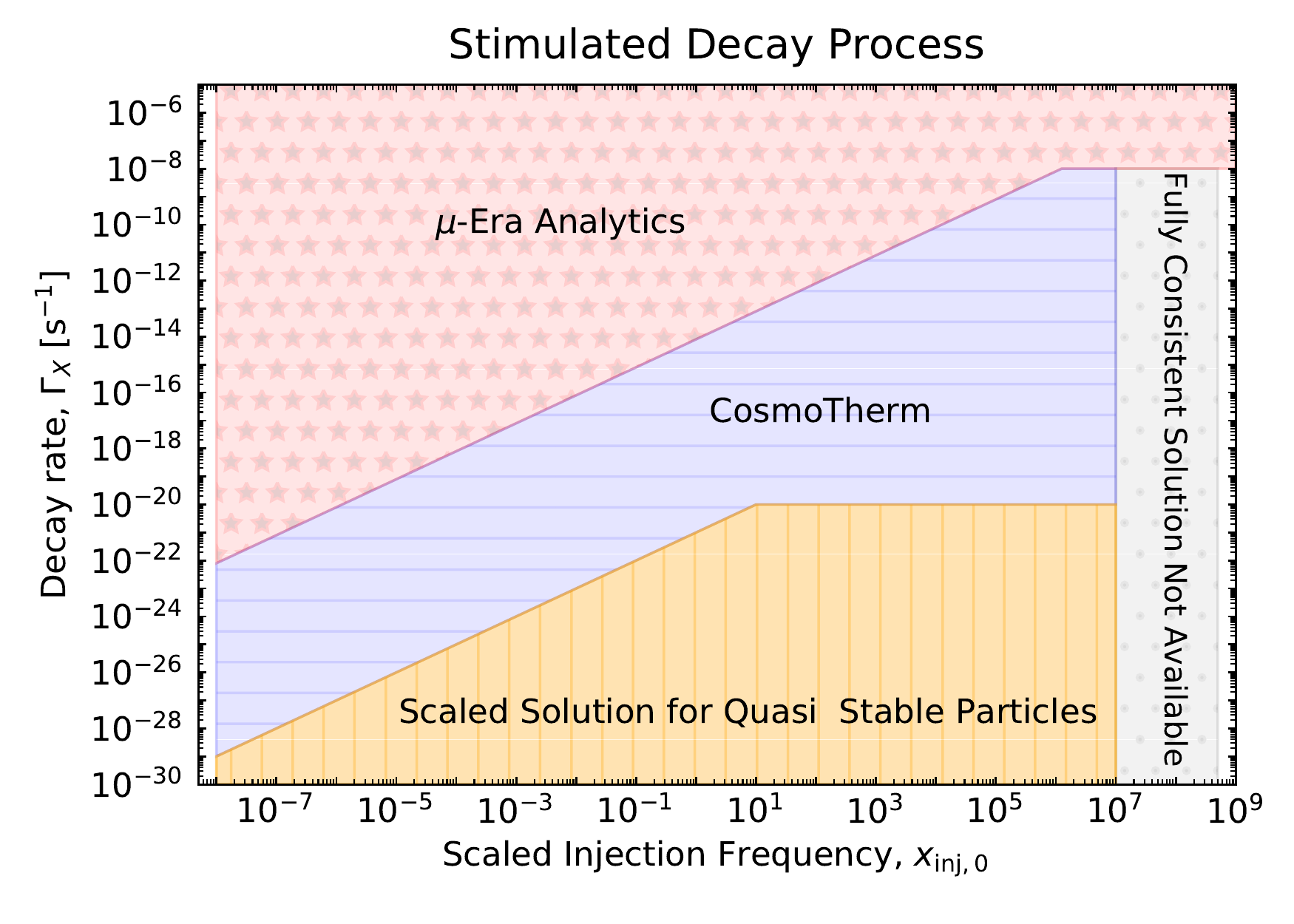}
\vspace{-6mm}
\caption{Strategies to solve the distortion without (upper panel) and with (lower panel) stimulated decay effects.}
\vspace{-4mm}
\label{fig:region_age_stim}
\end{figure}

%-------------------------------------------------------------------
Since a single run of {\tt CosmoTherm} can take several minutes, we generated a database of distortion spectra for decaying particle scenarios. We can then load and interpolate the spectra from our pre-computed database and compare them with measurement of the CMB spectrum in order to extract constraints in a relatively short time. The overall computational strategy of the distortion spectra is summarized in Fig.~\ref{fig:region_age_stim} for cases with and without stimulated decay effects. In the grey domains, a more detailed treatment of non-thermal particle production is required. We avoid this regime by limiting the injection energies when extracting constraints.
%----------------------------------------------------------
%----------------------------------------------------------

As in Sect.~\ref{sec:distortion_results}, we have focused on three cases: i) `bare', for which neither the effects of photo-ionization, nor the effects of reionization were included; ii) `lyc', for which we included photo-ionizations; and iii) `lyc+reio' where both reionization and photo-ionization were taken into account. For each setup, we computed $O(10^4)$ spectra using {\tt CosmoTherm} for $\approx300$ values of the scaled injection frequency $x_\mathrm{inj,0}$, logarithmically spaced between $10^{-8}$ and $10^7$, and $\approx 80$ values of particle lifetime $\Gamma_X$, logarithmically-spaced between $10^{-8}\,\rm{s^{-1}}$ and $10^{-20}\,\rm{s^{-1}}$. By interpolating the database we can then obtain accurate spectra for any set of values ($x_\mathrm{inj,0}, \Gamma_X$).
The computational domain is modified when considering stimulated decay effects, as illustrated for comparison in Fig.~\ref{fig:region_age_stim}.

For $\Gamma_X>10^{-8}\,\rm{s^{-1}}$, i.e., short-lived particles, the injection occurs at high redshift, in the $\mu$-era, and we can use the analytical formulae of subsection \ref{sec:mu_analytic} to predict the value of the chemical potential corresponding to the requested para. 
For $\Gamma_X\lesssim 10^{-20}\,\rm{s^{-1}}$, i.e., long-lived particles, the lifetime is larger than the age of the universe, and the universal SD shape described in \ref{LLp} can be used.
The range of injection energy, i.e., $x_\mathrm{inj,0}$ is chosen such that it covers the wide phenomenology associated with the regimes in highlighted in Fig.~\ref{fig:regimes}.
The modifications to the short and long-lifetime regimes when including stimulated decay are also illustrated in Fig.~\ref{fig:region_age_stim}.

We mention a few details relevant to the creation of the distortion database. For baseline calculations with photons injected mostly at $\xinj=10^{-5}-150$, we require $\simeq 3000$ frequency points. Depending on the injection frequency, the grid was extended at low or high frequencies. For this we used a log-density with $300$ points per decade. Since we did not want to perturb the background evolution too much, we included the standard heating and cooling terms in all calculations. As mentioned above, the corresponding distortion signal had to be subtracted from our result (see Fig.~\ref{fig:standard_cooling}). Similar statements apply to the contributions from reionization. Finally, the calculated spectra for a given lifetime are then assumed to depend linearly on the normalization parameter, $\finj$ or $\fdm$. In detail, this is {\it not} expected to be perfect, because changes to the ionization history can lead to non-linear responses. However, the results are not affected dramatically by this assumption, as we discuss below.

When computing cases with long lifetimes, we determine the redshift at which a fraction (we chose one  percent) of the total energy has been injected due to the decay. We identify this redshift numerically by solving the energy release history prior to the thermalization calculation and then use the result as our starting redshift, if it is found to be higher than the decoupling redshift. Otherwise, we chose the decoupling redshift in order to not miss some important effects associated with photo-ionizations.

\vspace{-2mm}
\subsection{Data sets from various experiments}
%-------------------------------------------------------------------
In this section, we briefly explain how we use data from \FIRAS, \EDGES and \Planck to constrain photon injection scenarios. The constraints will be presented in Sect.~\ref{ss:mic} and \ref{ss:mdc}.

\vspace{-2mm}
\subsubsection{\FIRAS constraints on $\mu$ and $y$}
\label{sec:FIRAS_setup}
%-------------------------------------------------------------------
The \FIRAS monopole measurement spans frequencies between 68.05 GHz and 639.5 GHz ($1.20\leq x\leq 11.26$) in 43 bands\footnote{We used the 2005 release of the \FIRAS monopole spectrum measurement: \href{https://lambda.gsfc.nasa.gov/data/cobe/firas/monopole_spec/firas_monopole_spec_v1.txt}{firas\_monopole\_spec\_v1.txt}\label{firas-data}}. This dataset is absolutely-calibrated and, as of today, still provides the most stringent constraints on the CMB spectrum \citep{Mather1994, Fixsen1996, Fixsen2003}. Below, we use this data to derive constraints on photon injection problems, but as a first step we repeat the calculation for the constraints on $\mu$ and $y$.

\begin{figure}
\includegraphics[width=\columnwidth]{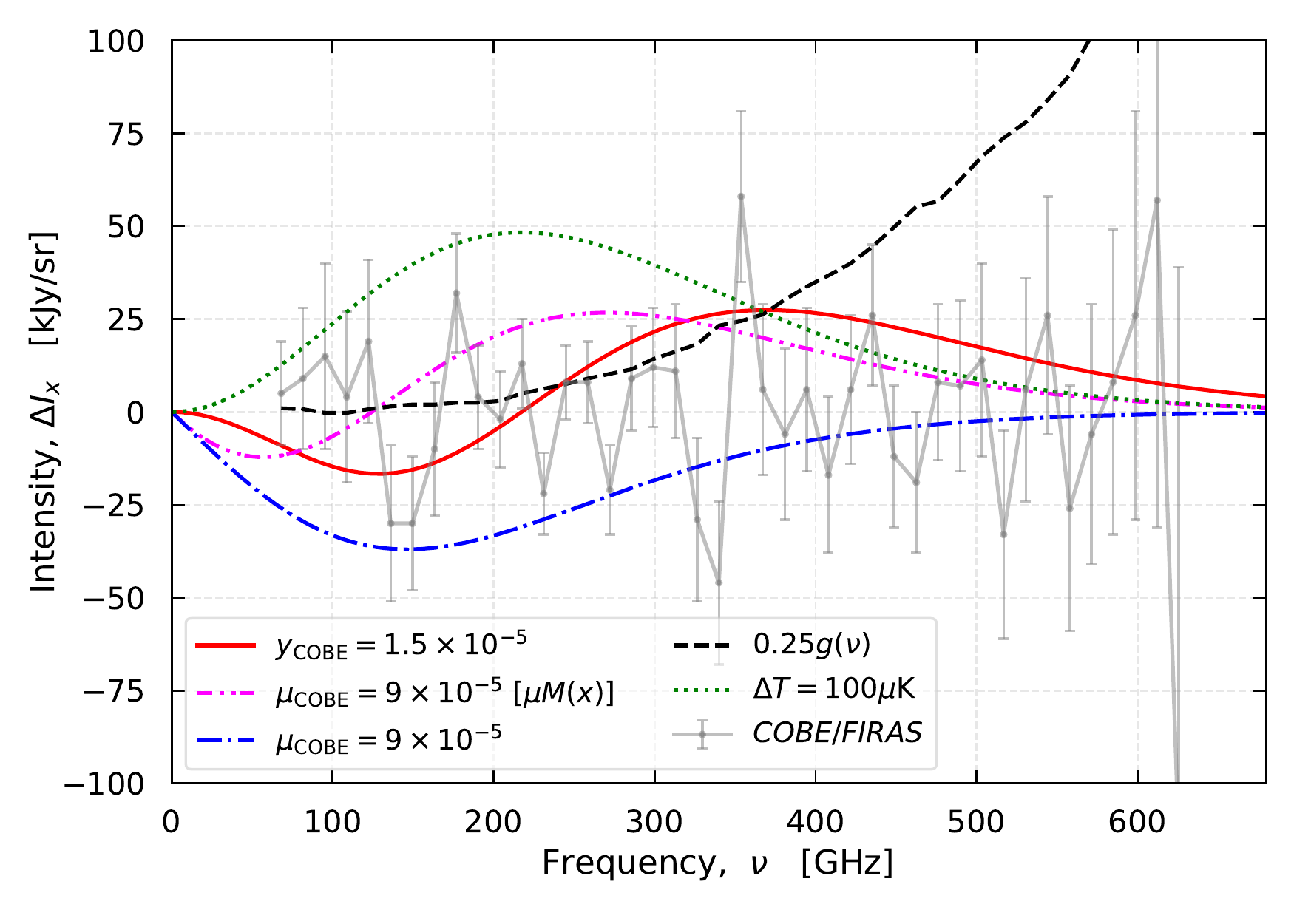}
\vspace{-2mm}
\caption{Residuals, distortions and galactic contamination term for the \FIRAS monopole measurement. The green dotted line is the distortion associated with a temperature shift of $\Delta T=100\mu \mathrm{K}$. For this figure the pivot temperature is $T_0 = 2.725\,\Kel$.}
\vspace{-2mm}
\label{fig:FIRAS}
\end{figure}

Following \cite{Fixsen1996}, the \FIRAS constraints on $\mu$ and $y$ are obtained by fitting the monopole measurement with a blackbody law at a pivot temperature $T_0$, the $\mu$ and $y$ distortions and a galactic contamination term:
%-------------------------------------------------------------------
\begin{equation}
I(\nu) = B(T_0) +\Delta T \left.\frac{\partial B}{\partial T}\right|_{T_0} + \mu \left.\frac{\partial S_{\mu}}{\partial \mu}\right|_{T_0} + y \left.\frac{\partial S_{y}}{\partial y}\right|_{T_0} + G_0 g(\nu).\label{eq:linear-fit}
\end{equation}
%-------------------------------------------------------------------
Here, $G_0 g(\nu)$ is the galactic contamination term with free parameter $G_0$ and frequency dependence characterized by $g(\nu)$. The $\mu$ and $y$ distortions are proportional to the frequency dependent functions
%-------------------------------------------------------------------
\begin{equation}
\frac{\partial S_{\mu}}{\partial \mu}\Bigg|_{T_0} = -\frac{T_0}{x} \frac{\partial B}{\partial T}\Bigg|_{T_0}, 
\;\;\;\; 
\frac{\partial S_{y}}{\partial y}\Bigg|_{T_0} =  \left[x\coth\left(\frac{x}{2}\right)-4\right]T_0\frac{\partial B}{\partial T}\Bigg|_{T_0}
\label{eq:mu+y}
\end{equation}
%-------------------------------------------------------------------
where $x=h\nu/kT_0$ and 
%-------------------------------------------------------------------
\begin{equation}
\frac{\partial B}{\partial T}\Bigg|_{T_0} = \frac{B_0}{T_0}\frac{x\, \expf{x}}{\expf{x}-1}, \quad\mathrm{with} \quad B(T_0) = \frac{2h}{c^2}\frac{\nu^3}{\expf{x}-1}
\nonumber
\end{equation}
%-------------------------------------------------------------------
denoting the blackbody spectrum at temperature\footnote{When $h$, $c$ and $\nu$ are expressed in standard units, one needs to multiply $B(\nu, T)$ by a factor $10^{26}$ in order to obtain the intensity $I(\nu)$ in units of Jy/sr.} $T_0$. An alternative description of the $\mu$ distortion is 
%-------------------------------------------------------------------
\begin{align}
\label{eq:mu-correct}
\mu \,M(x) &= \mu \frac{\partial S_{\mu}}{\partial \mu}\Bigg|_{T_0} \left(1-\frac{x}{\beta_\mu}\right),
\end{align}
%-------------------------------------------------------------------
with $\beta_\mu = 3{\cal G}_2/(2{\cal G}_1)\simeq 2.1923$ and where photon number conservation is enforced by $\int x^2 M(x)=0$. However, the main modification in this case is a shift in the monopole temperature and the final results are not significantly affected by this choice. With this formula and $\mu>0$, the $\mu$ distortion is negative at low frequency and changes sign at $x=\beta_\mu$ or $\nu\simeq 124.5\mathrm{GHz}$ (magenta line on Fig.~\ref{fig:FIRAS}), while with Eq.~\eqref{eq:mu+y} it remains negative at all $\nu$ (blue line on Fig.~\ref{fig:FIRAS}). We note that foregrounds such as synchrotron, anomalous microwave emission (AME) and free-free are neglected as sub-dominant in the \FIRAS analysis. These can have important effects on the $\mu$-distortion constraints \citep{Abitbol2017}, but a more careful reanalysis of the \FIRAS data that also includes valuable information of galactic foregrounds from \Planck and uses modern foreground separation methods \citep[e.g.,][]{Rotti2020} is beyond the scope of this work.

%----------------------------------------------------------------
%----------------------------------------------------------------

%----------------------------------------------------------------
\begin{figure}
\includegraphics[width=1.\columnwidth]{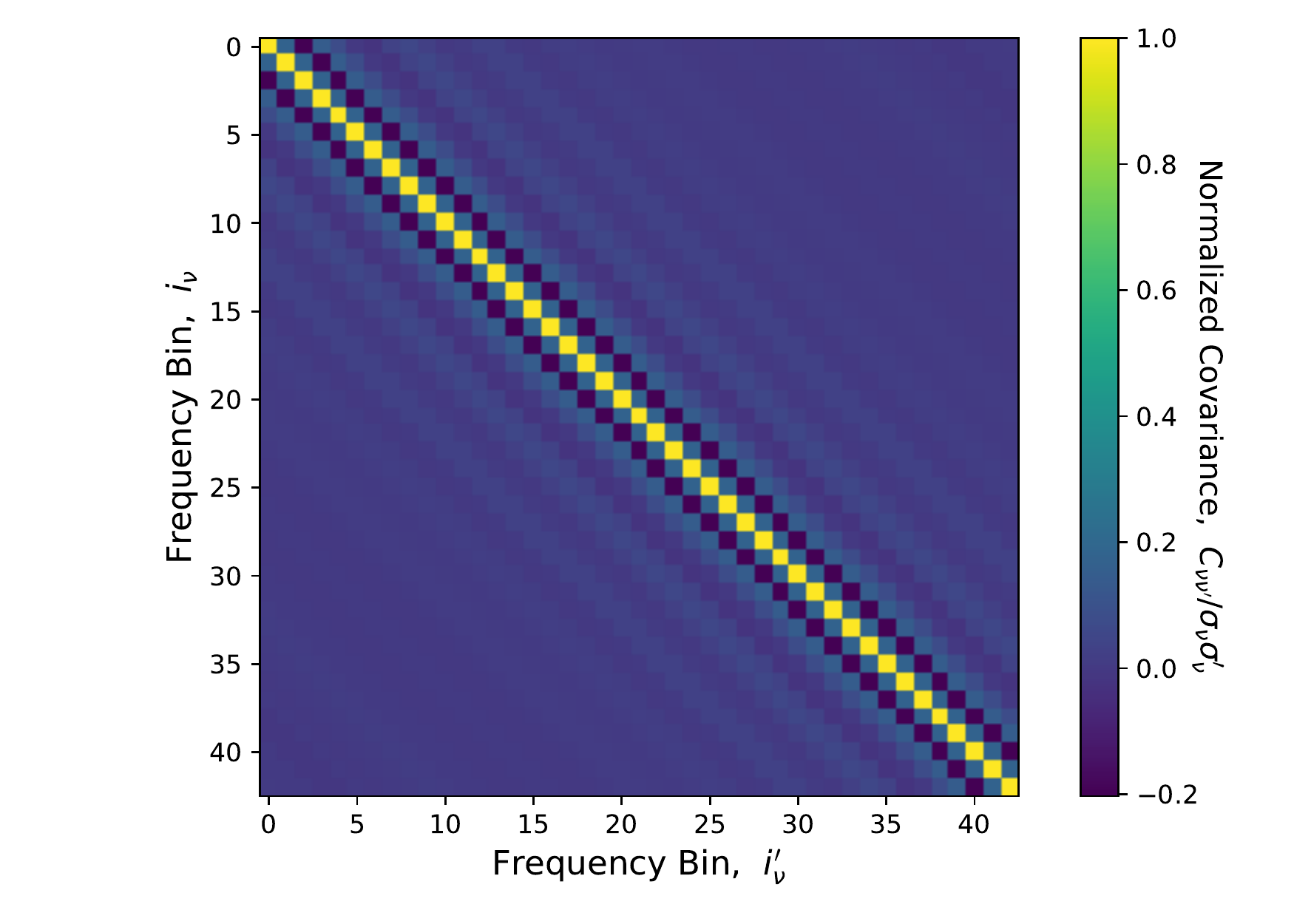}
\vspace{-2mm}
\caption{ Normalized data covariance matrix $Q(\left|\nu - \nu^\prime\right|)=C_{\nu\nu^\prime}/ \sigma_\nu \sigma_{\nu^\prime}$ as used in the \FIRAS analysis.}
\vspace{-0mm}
\label{fig:FIRAS_cov}
\end{figure}
%----------------------------------------------------------------

All the frequency-dependent functions in Eq.~\eqref{eq:linear-fit} are either fixed or depend on the pivot temperature $T_0$, such that the monopole model is linear with respect to the free parameters $\Delta T$, $G_0$, $\mu$ and $y$. Therefore, one can use a simple $\chi^2$ fit to find the best-fitting parameters and associated uncertainties, as was done in \cite{Fixsen1996}.  
To ensure that we use the \FIRAS data in a consistent way to set constraints on photon injection processes, we first attempted to reproduce the results for the original analysis for $\mu$ and $y$. The covariance matrix between the 43 frequency bins is given in \cite{Fixsen1996} as $C_{\nu\nu^\prime} = \sigma_\nu \sigma_{\nu^\prime} Q(\left|\nu - \nu^\prime\right|)$, where the function $Q$ is tabulated in Section 3.3 of the original paper (also see our  Fig.~\ref{fig:FIRAS_cov}). 
We write $\chi^2 = \vek{D}^TC^{-1}\vek{D}$ where $\vek{D}$ is the vector of difference between measurement and model monopole and $C$ is the covariance matrix. For the minimization we used either the Levenberg-Marquardt algorithm via the {\tt curve\_fit} method of {\tt scipy} or the Monte Carlo Markov Chain method (MCMC). Both methods agree, nevertheless we note that the MCMC method (which typically takes a few seconds to converge) is always reliable while the {\tt curve\_fit} method fails for some cases when we study photon injection constraints (see Table \ref{tab:chi2-firas} for a compilation of all the results).

Although overall our results on $\mu$ and $y$ are consistent with \cite{Fixsen1996}, we note minor differences which are probably associated with a slightly different treatment of the galactic contamination term. For the galactic contamination term $g(\nu)$ we used the data provided in the fourth column of Table 4 of \cite{Fixsen1996} and fit for the amplitude $G_0$ (see dashed black line in Fig.~\ref{fig:FIRAS}).  We find $\TCMB=2.725\mathrm{K}\pm 10\mu\mathrm{K}$ (68\%CL) when we fit for $\TCMB$ and $G_0$, or for  $\TCMB,G_0$ and $y$. The quoted error here is purely determined by the statistical noise.
When we fit for $\TCMB,G_0$ and $\mu$ we find a statistical uncertainty of  $28\mu\mathrm{K}$ (68\%CL) when we use Eq. \eqref{eq:mu+y} for the $\mu$ distortion (blue line in Fig.~\ref{fig:FIRAS}) and $22\mu\mathrm{K}$ (68\%CL) when we use Eq. \eqref{eq:mu-correct} (magenta line in Fig.~\ref{fig:FIRAS}). Fitting for  $\TCMB,G_0$ and $\mu$ or $y$ yields $\mu=(-1\pm 3.7)\times10^{-5}$  (68\%CL)  and $y=(0.2\pm 3.9)\times10^{-6}$  (68\%CL), which is in good agreement with the values quoted in  \cite{Fixsen1996}, namely $\mu=(-1\pm 4)\times10^{-5}$  (68\%CL)  and $y=(-1\pm 6)\times10^{-6}$  (68\%CL). Our statistical uncertainties on $\mu$ and $y$ are not sensitive to the choice for the pivot temperature ($T_0=2.728\mathrm{K}$ as suggested in \cite{Fixsen1996} or $T_0=2.725\mathrm{K}$, the best-fitting temperature), or the expression used for the $\mu$ distortion\footnote{We understand that \cite{Fixsen1996} have used Eq. \eqref{eq:mu+y} for the fit with  a $\mu$ distortion, although it seems that Eq. \eqref{eq:mu-correct} was used in their Fig.~5.}.

To obtain the final \FIRAS bounds on $\mu$ and $y$, the statistical uncertainties are supplemented by systematic uncertainties, provided in \cite{Fixsen1996}. The systematic uncertainty for $\mu$ is $1\times 10^{-5}$ (68\%CL) and for $y$ it is $4\times 10^{-6}$ (68\%CL). Propagating the errors, we find $\mu < 7.7\times 10^{-5}$ (95\%CL) and $y < 1.1\times 10^{-5}$ (95\%CL). These translates into the energy bounds $(\Delta \rho/\rho)_\mu < 5.5\times 10^{-5}$ and $(\Delta \rho/\rho)_y < 4.5\times 10^{-5}$ each at 95\%CL. Here, we applied the relations $(\Delta \rho/\rho)_\mu = \mu/1.401$ and $(\Delta \rho/\rho)_y = 4y$. Had we considered only the statistical uncertainty on $y$, the limit for the energy bound from $y$ would be $(\Delta \rho/\rho)^{\mathrm{stat}}_y < 3.1\times 10^{-5}$ (95\%CL), which motivates the normalization of the photon injection spectra that we computed.

Our results on $\mu$ and $y$ have to be compared with the well-known values quoted in  \cite{Fixsen1996}, namely $\mu < 9\times 10^{-5}$ (95\%CL) and $y<1.5\times 10^{-5}$ (95\%CL). Our bound on $\mu$ is 15\% tighter and our 2-$\sigma$ bound on $y$ is 30\% tighter than the original ones. As mentioned above, we believe the source of this differences is the treatment of the galactic contamination term. Nevertheless, the differences are relatively small and our treatment of the \FIRAS seems satisfactory enough for it to be used to set new constraints on photon injection processes. 
However, when presenting results below, we only include the statistical uncertainties in the analysis. It is not easy to estimate the systematic uncertainties for the wide range of spectra obtained in the considered scenarios. This likely means that our main constraints are uncertain by a factor of $\simeq 1.5-2$ due to \FIRAS systematic uncertainties.

Finally, we also mention the limits that are obtained when simultaneously varying $\mu$ and $y$ (see last row of Table~\ref{tab:chi2-firas}), which yields $\mu=(-3.6\pm6.5)\times10^{-5}$ and $y=(3.3\pm6.9)\times10^{-6}$ (68\%~CL). This implies $(\Delta \rho/\rho)^{\mathrm{stat}}_\mu < 9.3\times 10^{-5}$ and $(\Delta \rho/\rho)^{\mathrm{stat}}_y < 5.5\times 10^{-5}$ (95\% CL) for the energy injection in the $\mu$ and $y$ eras, respectively. Assuming continuous energy release, e.g., from decaying particles, this can be interpreted at a limit of $(\Delta \rho/\rho)^{\mathrm{stat}}_{\rm tot} < 4.9\times 10^{-4}$ (95\%CL) on the total energy release in the pre-recombination era, or about one order of magnitude weaker than the individual $\mu$ or $y$ distortion limits. Although here we only included statistical errors, adding the \FIRAS estimates for the systematic uncertainties does not modify the result significantly.

\vspace{-3mm}
\subsubsection{\EDGES measurement}
\label{ss:edges}
%-------------------------------------------------------------------
The low-band antenna of EDGES covers frequencies in the range $50-100$~MHz \citep{Bowman:2018vm}. This frequency range probes the 21cm hyperfine transition of neutral hydrogen from $z\simeq 27-13$. For a given frequency $\nu$, or equivalently redshift  $(1+z)=\nu_{21}/\nu$, within this range, \EDGES measures the brightness temperature associated with the global 21cm line, $T_{21}(z)$. Theoretically, the signal can be modeled as \citep{Zaldarriaga:2003du, Pritchard2012}
%-------------------------------------------------------------------
\begin{align}
    T_{21}(z)&=\frac{T_\mathrm{spin}-\Tg}{1+z}(1-\mathrm{e}^{-\tau_{21}}),
    \nonumber\\
    &\approx  0.023 X_{\rm HI}(z)\left[\frac{0.15}{\Omega_\mathrm{m}h^2}\frac{1+z}{10}\right]^{1/2}\left(\frac{\Omega_\mathrm{b}h^2}{0.02}\right)\left[1-\frac{\Tg}{T_\mathrm{spin}}\right],
    \nonumber
\end{align}
%-------------------------------------------------------------------
where $\Tg$ is the photon brightness temperature around the 21cm transition, $T_\mathrm{spin}$ is the spin temperature of neutral hydrogen \citep{1959ApJ...129..536F} and $\tau_{21}$ is the optical depth for the 21cm line, replaced explicitly in the second line and assumed to remain small for the linear expression to be a good approximation. Both  $T_\mathrm{spin}$ and $\tau_{21}$ depend on $\Tg$, since $\tau_{21}\propto \Tg/T_\mathrm{spin}$ and 
%-------------------------------------------------------------------
\begin{equation}
    T_\mathrm{spin} = \frac{x_\mathrm{rad}+x_\mathrm{c}+x_\alpha}{x_\mathrm{rad}\Tg^{-1}+x_\mathrm{c}T_\mathrm{k}^{-1}+x_\mathrm{\alpha}T_\mathrm{\alpha}^{-1}},\,\,\mathrm{with} \quad x_\mathrm{rad}=\frac{1-\mathrm{e}^{-\tau_{21}}}{\tau_{21}}.
\end{equation}
%-------------------------------------------------------------------
Here, $T_\mathrm{k}$ is the kinetic temperature, $T_\mathrm{\alpha}$ the Lyman-$\alpha$ brightness temperature, $x_\mathrm{c}$ and $x_\alpha$ are the spin-flip rates due to atomic collisions and resonant scattering of Lyman-$\alpha$ photons respectively \citep[Wouthuysen-Field effect;][]{1952AJ.....57R..31W,1959ApJ...129..536F}. We refer to \cite{Venumadhav:2018uwn} and \cite{Fialkov_2019} for more thorough discussions on the \EDGES results and 21cm physics, and \cite{Panci:2019zuu} for additional brief overview. 

\cite{Bowman:2018vm} reported a symmetric U-shaped absorption profile centred at $78$ MHz, corresponding to 21cm absorption at $z\simeq 17$, with an amplitude $\simeq -500$ mK and a full-width at half maximum of $20$ MHz. 
At redshift $z\simeq 150-200$, Compton scattering keeps the spin and CMB temperatures coupled. At lower redshifts, baryons and CMB are no longer thermally coupled so that $T_\mathrm{gas}\propto (1+z)^2$ (adiabatic cooling), while $\Tg\propto (1+z)$. Although Compton scattering may not be efficient, collisional coupling ensures $T_\mathrm{spin}\simeq T_\mathrm{gas}$ until $z\simeq 40$. Hence, at high redshift, $T_\mathrm{spin}\simeq \Tg\simeq \TCMB$ due to Compton scattering, so no net effect from the 21cm line is expected until $z\simeq 150$. 
At lower redshift ($40<z<150$), the spin temperature is coupled to the gas temperature and lower than $\Tg$, hence leading to an absorption in the 21cm line. However, this corresponds to frequencies $35-10$ MHz, which are below the EDGES band. When collisional coupling is inefficient at $z<40$, the spin temperature is re-coupled to radiation and no absorption of the 21cm line is expected. At cosmic dawn, $z<20$,  the redshifted UV photons emitted during star formations couple the neutral hydrogen to the gas via resonant scattering of Lyman $\alpha$ photons and X-ray heating, $T_\mathrm{spin}\simeq T_\mathrm{gas}<\TCMB$ creating an absorption profile. At lower redshift $z<13$, the gas becomes hotter than the background radiation due to large heating and a 21cm emission signal is expected. Finally, no global signal is expected when the Universe becomes fully reionised, as $x_\mathrm{HI}\simeq 0$ at $z\lesssim 7$. 

The absorption trough measured by \EDGES is roughly a factor two deeper than what is expected from standard astrophysics and cosmology, which gives $T_\mathrm{21}\gtrsim -0.21\mathrm{K}$.  It means that at $z\simeq 17$ either the spin temperature is much lower than the standard expectation, $T_\mathrm{spin}\simeq 7\mathrm{K}$, which is a possibility if, for instance, the gas is cooled down non-adiabatically due to interacting dark matter particles \citep[e.g.,][]{Barkana2018, Munoz:2018uu}, or that the CMB brightness temperature in the Rayleigh Jeans tail is much higher than $\Tg=\TCMB(1+17)\approx 50\,\mathrm{K}$ \citep[e.g.,][]{Feng2018}. Keeping the CMB temperature to the fiducial value, the spin temperature would need to be $T_\mathrm{spin}\approx T_\mathrm{gas}\approx 3.5\,\Kel$, to explain $T_{21}\lesssim-0.45\mathrm{K}$. Alternatively, keeping the spin temperature to the fiducial value, the CMB temperature in the RJ tail would need to be $\Tg \simeq 100\,\mathrm{K}$.  

Here, we are interested in the second solution, i.e., an excess of photons in the RJ tail of the CMB radiation with respect to the blackbody law. First indications of a possible low-frequency excess stems from the measurements of \ARCADE at $\simeq {\rm few}\,\GHz$ \citep{Fixsen2011excess, arcade2}, and is also supported by a recent analysis of the LAW1 data at 40-80 MHz \citep{Dowell_2018}. Even before the \EDGES measurement, it was already suggested that the \ARCADE signal could be the trace of some partially Comptonized soft photon emission or injection from decaying or annihilating particles \citep{Chluba2015GreensII}. A solution for the \EDGES result along these lines has been considered by several authors. For instance, \cite{Pospelov_2018} studied the decay of particles with milli-eV masses and lifetime longer than the age of the Universe.
Similarly, \citet{Brahma2020} considered models with decaying particles to explain the \EDGES observation.

Our main goal here is not to explain the \EDGES result but rather to use the measurement to derive constraints on various models. For our analysis of photon injection constraints from CMB spectral distortions, we use the \EDGES data in a very simple way: we supplement the \FIRAS measurements with one data point at frequency $\nu_\mathrm{EDGES}=78\,\mathrm{MHz}$ (i.e., $\xinjc\simeq \pot{1.4}{-3}$), where the CMB brightness temperature is simply $T\simeq (c^2/2k\nu^2)\,I_\nu\lesssim 2\TCMB$ at 68\% CL, consistent with the above discussion. 
We do not attempt to model low-frequency galactic and extra-galactic foregrounds carefully for this part of the spectrum, but simply use this bound {\it prima facie}. This procedure is adopted in Sect.~\ref{ss:mic} and \ref{ss:mdc} and provides a {\it conservative constraint} to the considered scenarios, but for clarity, we also discuss \FIRAS only limits.

\vspace{-0mm}
\subsubsection{Estimating limits from CMB anisotropy data}
\label{xeconstraints}
%-------------------------------------------------------------------
In addition to the constraints directly derived from \FIRAS and \EDGES, we can consider changes to the CMB anisotropies caused by modifications in the ionization history. In general this requires running a full MCMC analysis for each of the models. In real-time this is not feasible as some of the computations are quite time-consuming. However, possible departures away from the standard recombination history are already tightly constrained using a principal component (PC) analysis
\citep{Farhang2011, Farhang2013, Calabrese2013, Planck2015params}. Therefore, we can expect tight limits on photon injection scenarios in the post-recombination and recombination eras.

In \citet{Hart2020PCA}, improved PC constraints from \Planck 2015 were presented and a novel PC projection method that allows the derivations of simple constraints was introduced, without the need to run a full MCMC analysis for each ionization history. 
This method relies on the ionization history mode functions, $E_i(z)$, which are computed in the variable $\zeta(z)=\Delta \Ne/\Ne\ll 1$ around the standard ionization history. Computing $\zeta_{\rm inj}(z)$ from our $\Ne(z)$ outputs and then projecting this onto the eigenmodes $E_i(z)$ yields
%----------------------------------------------------------
\begin{align}
\label{eq:PCA_rhoi}
\rho^{\rm inj}_i= \int \zeta_{\rm inj}(z)\,E_i(z)\id z,
\end{align}
%----------------------------------------------------------
for the mode amplitudes. Comparing this to the \Planck 2015 limits $\rho_1=-0.08 \pm 0.12$, $\rho_2=-0.14 \pm 0.19$ and $\rho_3=-0.30 \pm 0.35$, we can then estimate $\fdms$ as follows. Since the covariance of the PCs is by construction very close to diagonal \citep{Hart2020PCA}, we find the limiting  $\fdms$ (95\% CL) by solving\footnote{In the first version of this manuscript we had assumed a linear scaling of $\rho_i^\mathrm{inj}$ with $\fdms$ to simplify the determination of the 95\% CL limit. Here we propose an exact solution, using the root finding method, which is slightly more computationally expensive but does not rely on a linear scaling assumption between $\rho^{\rm inj}_i$ and $\fdms$.}
%----------------------------------------------------------
\begin{align}
\label{eq:PCA}
2\left[\sum_i \frac{\left[\rho^{\rm inj}_i(\fdms)\right]^2}{\sigma(\mu_i)^2}\right]^{-1/2} = 1,
\end{align}
%----------------------------------------------------------
where $\sigma(\mu_i)=\{0.12, 0.19, 0.35\}$ denotes the measurement errors of $\mu_i$.% and $\hat{\rho}^{\rm inj}_i=\rho^{\rm inj}_i/\efdm^{\rm fid}$ with the fiducial value $\efdm^{\rm fid}$ used in the computation of the $\Ne$ response \textcolor{red}{$\zeta_{\rm inj}$}. 

%----------------------------------------------------------
\begin{figure}
\begin{centering}
\includegraphics[width=\columnwidth]{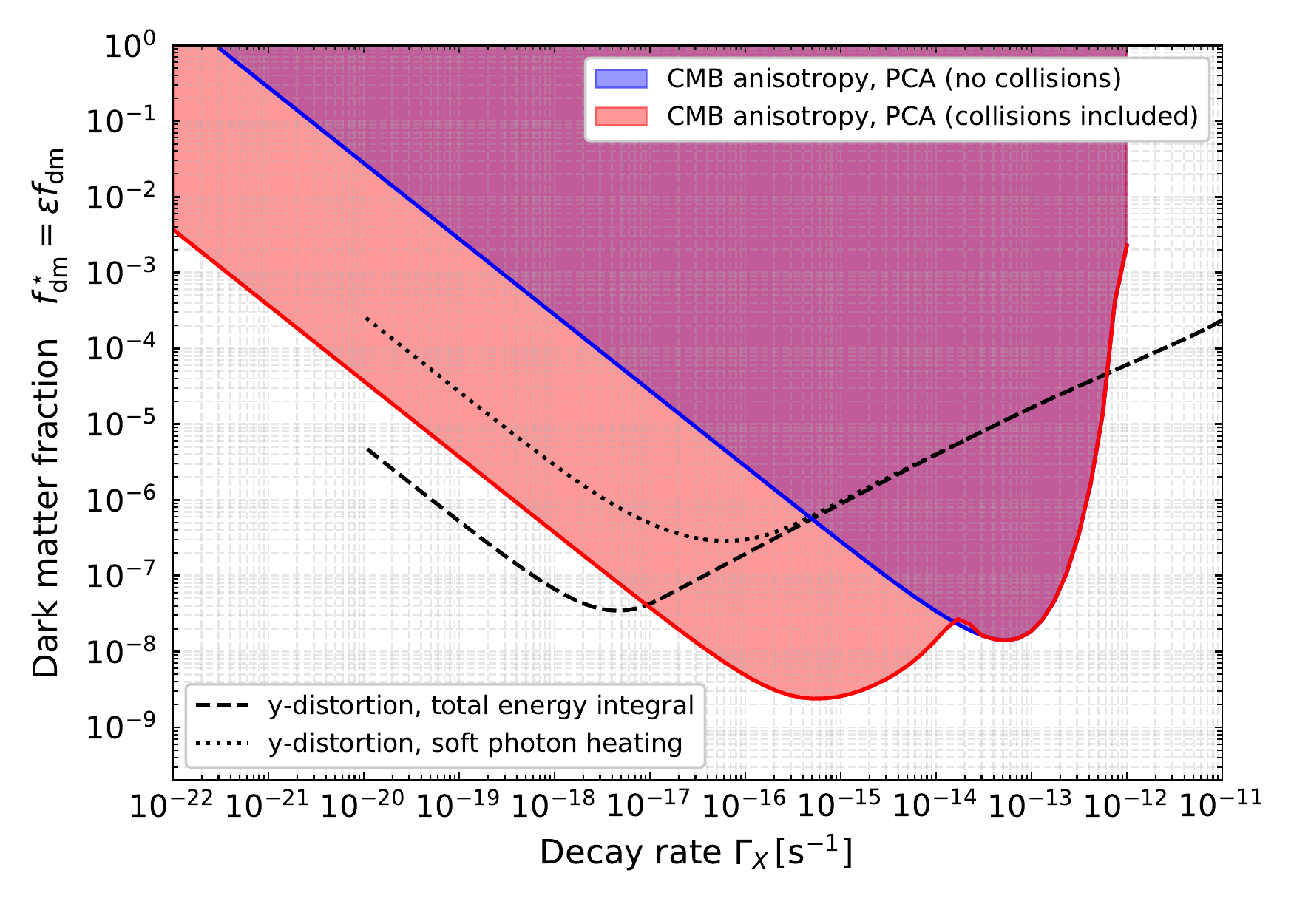}
\par\end{centering}
\vspace{-1mm}
 \caption{Derived 95\% CL dark matter fraction limit, $\fdms=\efdm$, for a range of lifetimes and $\xinjc \lesssim 10^{-8}$. The CMB anisotropy limits were obtained with the PCA projection method, while the other two are derived assuming $4y \lesssim \pot{6}{-5}$ (95\% CL) from \FIRAS. The dotted curve is obtained using Eq.~\eqref{eq:y_approx_defInt}, and the solid dashed line using the total energy integral with the $y$-era visibility function. For the CMB anisotropy constraints we considered a case where collisional ionisations are not included (blue region) and a case where these are included (red region). As can be seen, atomic collisions (see \ref{sssec:colls} and \ref{sec:collisions_effect}) are driving post-recombination decay constraints ($\Gamma_X\lesssim10^{-14}\,\mathrm{s}^{-1}$). We emphasize that these constraints are for very low mass decaying particles ($m_X \ll 1\mathrm{eV}$) where all the injected energy in the form of photons is converted into heat, in contrast with scenario studied in  \citep{Poulin2017, Lucca2020} where the energy is deposited and leads to direct ionisation.}
\label{fig:fdm_int}
\vspace{-2mm}
\end{figure}
%----------------------------------------------------------

Below we will estimate the CMB anisotropy constraints using Eq.~\eqref{eq:PCA}. However, for very low-frequency injection cases causing changes to $\Ne$ by pure heating (e.g., see Fig.~\ref{fig:soft_heating} for illustration of the ionisation history in these cases), the corresponding constraint can be directly computed using {\tt CosmoRec/Recfast++} and is illustrated in Fig.~\ref{fig:fdm_int}. For comparison we also show the limit derived from the corresponding $y$-distortion, which was computed consistently from the recombination output using 
%----------------------------------------------------------
\begin{align}
\label{eq:y_approx_defInt}
y\approx \int \frac{k[\Te-\Tg]}{\me c^2}\,\Ne \sigT c \id t
\end{align}
%----------------------------------------------------------
and then assuming $4y\lesssim \pot{6}{-5}$ (95\% CL) [see previous section]. For these low-frequency injection, where the energy is fully converted into heat, the constraint obtained from CMB anisotropy data supersedes the direct spectrometer constraint in all cases with $\Ginj \lesssim \pot{6}{-13}\,{\rm s}^{-1}$. The main reason is that, due to the large excess of photons over baryons, it is a lot easier to perturb the ionization history than to affect the energy spectrum of the CMB.
This illustrates the powerful complementarity of CMB anisotropy and spectrometer constraints.

To obtain the PC projection limit in Fig.~\ref{fig:fdm_int},  we computed the extra heating assuming all the injected photons are converted by free-free absorption. Then, for a given lifetime, we solve Eq.~\eqref{eq:PCA}. 

We would like to highlight that in the low-redshift Universe not all the injected energy creates a distortion signal. As indicated by Fig.~\ref{fig:fdm_int}, assuming that all the energy reaches the CMB spectrum to create a $y$-distortion yields significantly tighter limits for lifetimes $\Ginj \lesssim 10^{-15}\,{\rm s}^{-1}$ (dashed blue line). This is because the thermal coupling between electrons and photons reduces at $z\lesssim 200$ and a fraction of the energy is indeed used to raise the electron temperature well above the CMB temperature. 
We also mention that collisional-ionization turned out to be highly relevant. Without collisions, the changes to the ionization history were significantly weaker at $\Ginj \lesssim 10^{-14}\,{\rm s}^{-1}$. The difference indeed reached two to three orders of magnitude at $\Ginj \lesssim 10^{-16}\,{\rm s}^{-1}$ and highlights the importance of modeling modifications to the ionization history accurately in these non-standard scenarios (compare red and blue regions on Fig.~\ref{fig:fdm_int}).

We also note that with the same PC projection method we can estimate the constraints on high-mass particle decays. In this case, we again simply ran {\tt CosmoRec/Recfast++}, but including the description of \citet{Chen2004} for the branching ratios into heating, ionization and excitation from the decay products. The corresponding constraint on $\fdm$ was found to be in reasonable agreement with direct MCMC approaches \citep{Poulin2017, Lucca2020} given the differences in the computations of the energy branching ratios and deposition efficiencies. Therefore, we believe that the PC projection method provides an efficient way of estimating the CMB limit from \Planck. However, a more detailed comparison of the two approaches will be the subject of future work.

%----------------------------------------------------------
\begin{figure}
\includegraphics[width=\columnwidth]{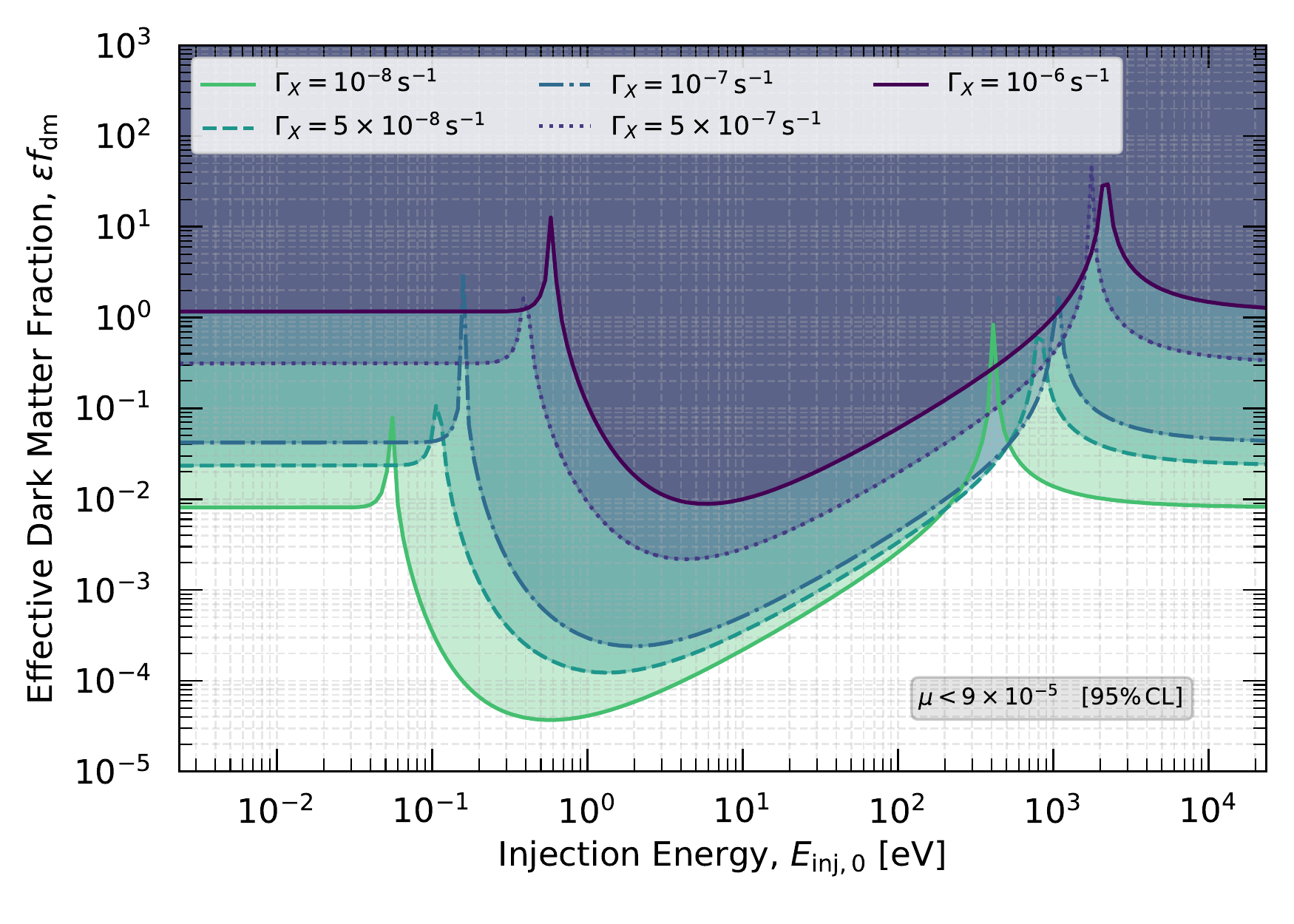}
\vspace{-6mm}
\caption{
\FIRAS constraints (95\% CL) on the effective DM fraction for decaying particle scenarios with short lifetimes. We use the analytic modeling in the $\mu$-era given in Eq.~\eqref{eq:sol_mu} to derive these limits. For the longest considered lifetime, this agrees very well with the full numerical result.
} 
\vspace{-3mm}
\label{fig:fdm_highz} 
\end{figure}
%----------------------------------------------------------

\vspace{-0mm}
\subsection{Model-independent constraints for various lifetimes}
\label{ss:mic}
%-------------------------------------------------------------------
We now have all the ingredients to compute constraints on various decaying particle scenarios. We start with model-independent limits and then use the results for the discussion of specific scenarios. 
Given the CMB spectrum data at $N$ frequencies, $\mathbf{I}^\mathrm{data}=(I_{\nu_1},..,I_{\nu_N})$, we minimize $\chi^2=\mathbf{\Delta}^{t}C^{-1}\mathbf{\Delta}$, where $C$ is the $N\times N$ covariance matrix and $\mathbf{\Delta}=\mathbf{I}^\mathrm{data}-\mathbf{I}^\mathrm{theory}$. For the theory model, we use
%--------------------------------------------
\begin{equation}
    I(\nu) = B(T_0) +\Delta T \left.\frac{\partial B}{\partial T}\right|_{T_0} + G_0 g(\nu) +  A _\mathrm{PI} \,\Delta I^\mathrm{PI}(\nu;x_\mathrm{inj,0},\Gamma_\mathrm{inj}),
\end{equation}
%--------------------------------------------
and fit for the CMB temperature $\Delta T=\TCMB-T_0$, the amplitude of galactic contamination $G_0$, and the amplitude of the photon injection spectral distortion $A _\mathrm{PI}$. The photon injection spectral distortion $\Delta I^\mathrm{PI}(\nu;\xinjc,\Ginj)$ is obtained at each frequency $\nu_{i=1..N}$ by interpolating the {\tt Cosmotherm} database (see section \ref{ss:database}).

The constraint on the parameter $A _\mathrm{PI}$ can then be translated into a constraint on the effective dark matter fraction $\fdms$, as follows. The spectra in the database are computed for $\Delta \rho/\rho\big|_\mathrm{inj}=3\times 10^{-5}$ which corresponds to a certain value $f_\mathrm{inj}^\mathrm{ct}$, via Eq.~\eqref{eq:sol_norma}. The value of  $f_\mathrm{inj}$ corresponding to $A _\mathrm{PI}$ is then given by $f_\mathrm{inj}=A _\mathrm{PI}f_\mathrm{inj}^\mathrm{ct}$ and the dark matter fraction is then simply obtained using Eq.~\eqref{eq:f_inj}. 

%----------------------------------------------------------
\begin{figure*}
\includegraphics[trim={0 0mm 0 0mm}, clip, width=1.7\columnwidth]{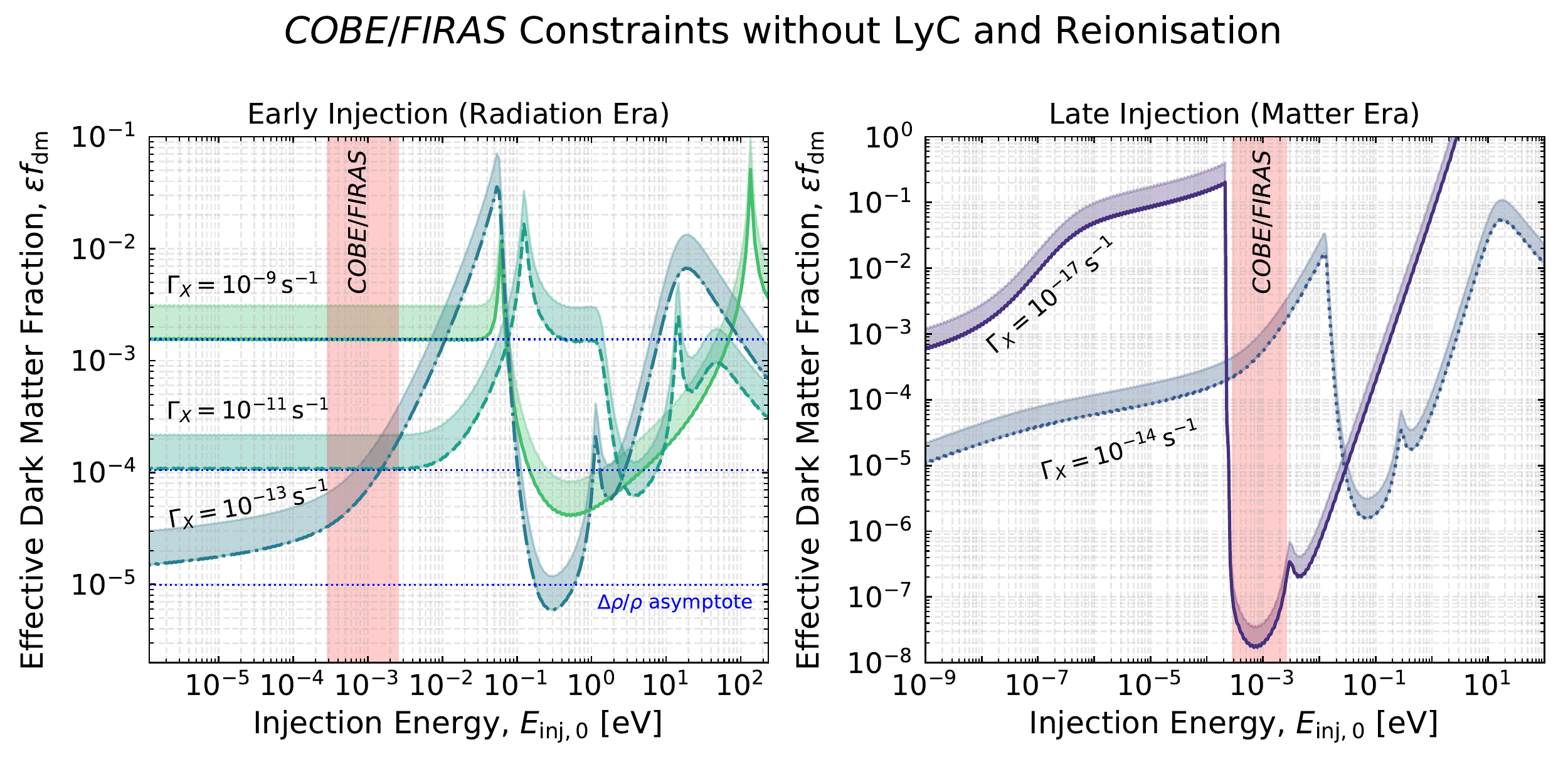}
\\
\includegraphics[trim={0 0 0mm 0mm}, clip, width=1.7\columnwidth]{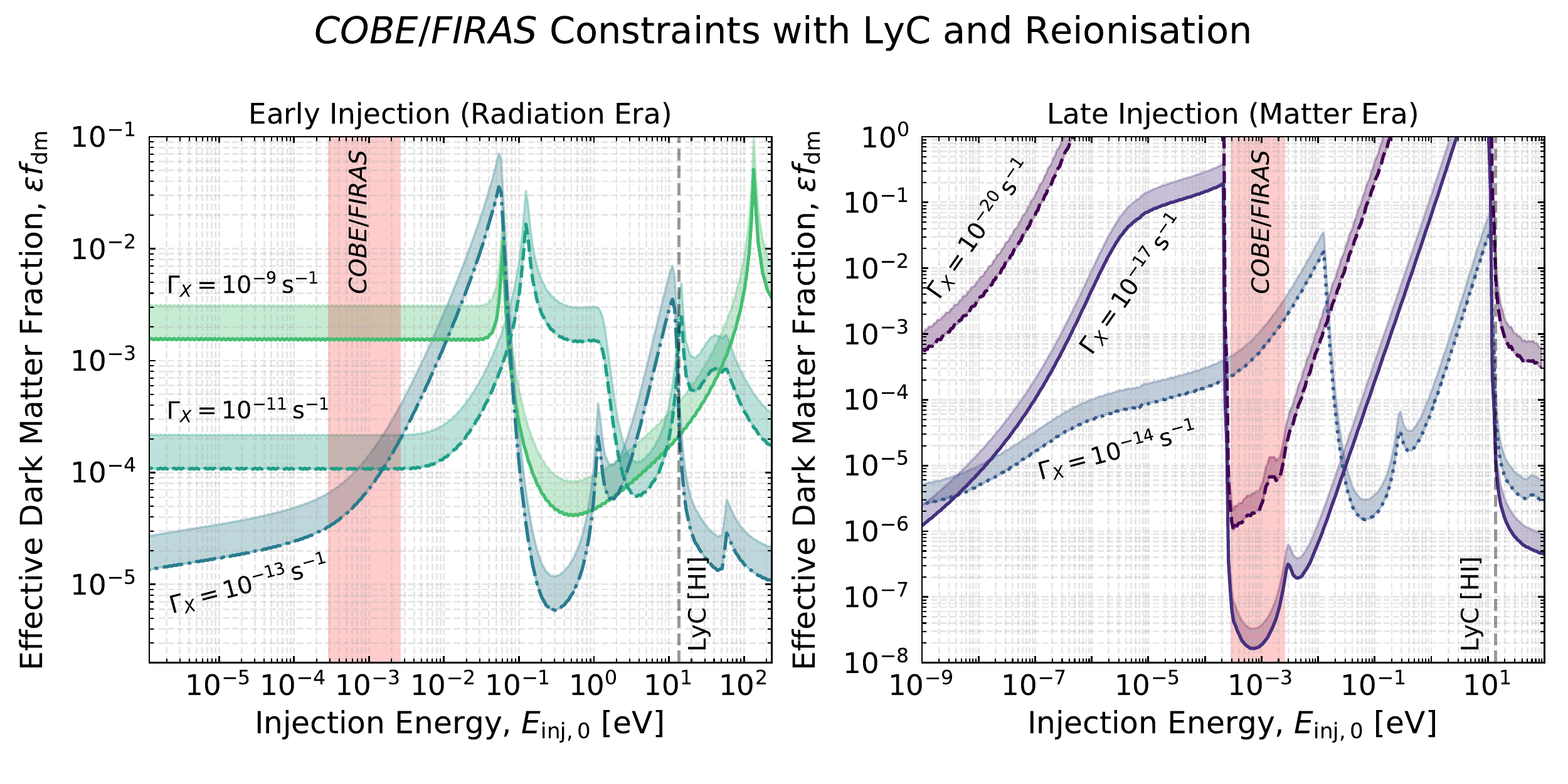}
\vspace{-4mm}
\caption{
\FIRAS constraints (95\% CL) on the effective DM fraction for decaying particle scenarios with decay rates $\Ginj \leq 10^{-9}\,{\rm s}^{-1}$. The upper panels show the limits without reionization and photo-ionization effects included, while for the lower panels these are switched on. See main text for discussion. 
} 
\vspace{-3mm}
\label{fig:fdm_no_edges}
\end{figure*}
%----------------------------------------------------------

\vspace{-0mm}
\subsubsection{CMB distortion limits: short lifetimes}
%-------------------------------------------------------------------
For particles that decay in the $\mu$-era, i.e., $\Ginj \gtrsim 10^{-8} \,\mathrm{s}^{-1}$, we can obtain constraints on $\fdms$ based on the \FIRAS limit on the chemical potential $\mu_\mathrm{COBE}$. As confirmed in Sect.~\ref{sec:mu_analytic}, for these particles, the distortion is a simple $\mu$ distortion whose amplitude is given by Eq. \eqref{eq:sol_mu} and is proportional to $f_\mathrm{inj}$. Hence, the value of $f_\mathrm{inj}$ that corresponds to $\mu_\mathrm{COBE}$ is given by $f_\mathrm{inj}=\mu_\mathrm{COBE}/\mu$, where $\mu$ is computed with Eq.~\eqref{eq:sol_mu} at $f_\mathrm{inj}=1$, and the dark matter fraction $\fdms=\efdm$ is then obtained using Eq.~\eqref{eq:f_inj}.  

Figure \ref{fig:fdm_highz} shows the constraints for several short lifetimes. The overall limit weakens for shorter lifetimes, as expected from the fact that the distortion visibility (and hence the distortion amplitude) drops. We can also observe the frequencies at which the final distortion has $\mu\simeq 0$ due to the balance between number and energy of the  injected photons. In the intermediate regimes between the nulls, the distortion is indeed negative, as explained in Sect.~\ref{sec:mu_analytic}. 
Both at very low and very high injection energies, the constraint is close to what would be expected from simple energy constraints on $\Delta \rho/\rho$, but generally this estimate fails.

We mention that our computations do not include corrections for the evolution of large distortions, which can become relevant to cases with $\Ginj>10^{-6}\,{\rm s}^{-1}$ \citep{Chluba2020large}. In addition, for the considered scenarios, the particles need not be DM particles and hence the bound $\efdm<1$ can principally be avoided, as long as the Universe remains radiation dominated. If considering DM as the source of the photons, another interpretation of the results is related to the excited state energy, with $\epsilon<1$, as we discuss below.

%----------------------------------------------------------
\begin{figure*}
\includegraphics[width=2\columnwidth]{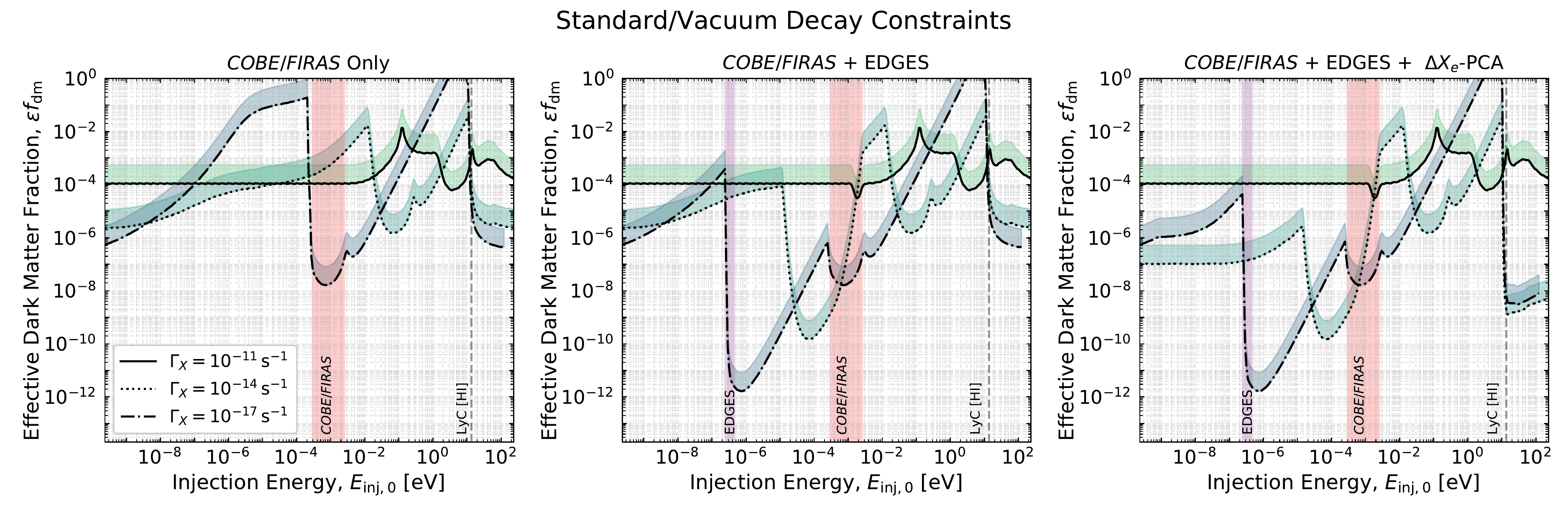}
\\[-0mm]
\includegraphics[width=2\columnwidth]{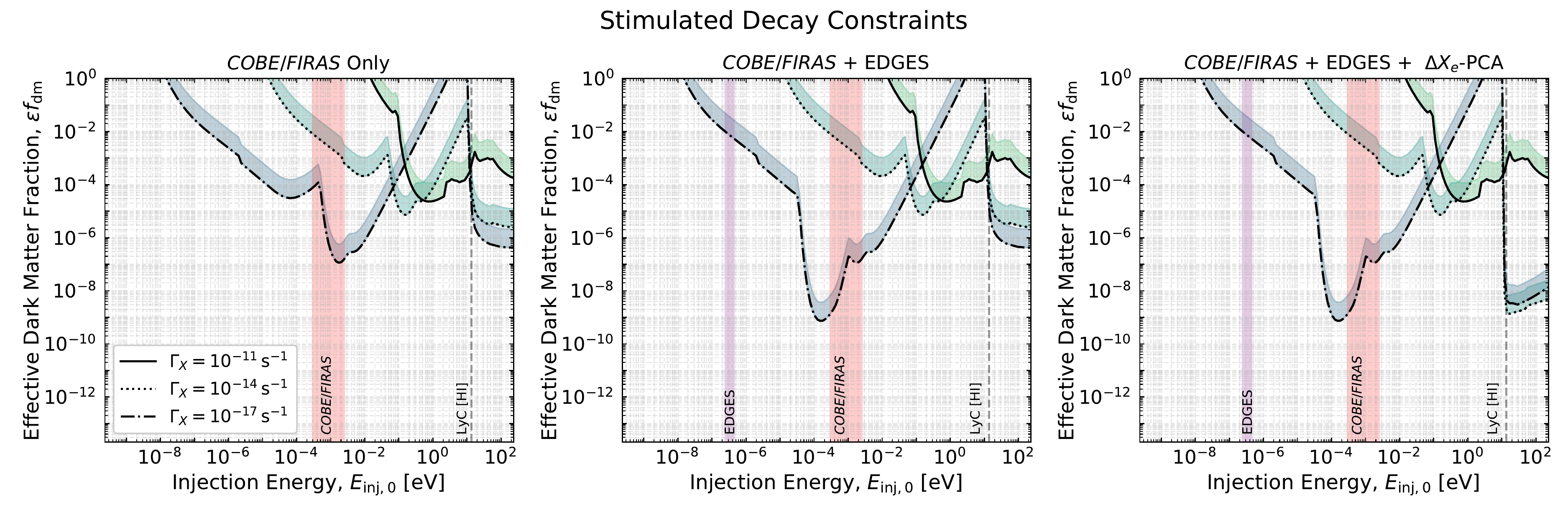}
\caption{
Constraints on the effective DM fraction for decaying particle scenarios (95\% CL) for various combinations of data from \FIRAS, \EDGES and \Planck.
The upper panels show the constraints for lifetimes $\Ginj \leq 10^{-11}\,\ratesec$ and various combinations of datasets when assuming decay to occur in vacuum. The lower panels show several examples when including stimulated decay. See main text for discussion.}
\vspace{-0mm}
\label{fig:fdm_with_edges}
\end{figure*}
%----------------------------------------------------------

\subsubsection{CMB distortion limits: late injection}
%-------------------------------------------------------------------
Next, we consider the \FIRAS only limits on decaying particle scenarios with $\Ginj \leq 10^{-9}\,{\rm s}^{-1}$. Our  results are summarized in Fig.~\ref{fig:fdm_no_edges}. 
We already understand from the discussion in Sect.~\ref{sec:distortion_results} that both reionization and photo-ionization effects modify the resultant distortion significantly. To illustrate the differences we compare the results for the spectra obtained with and without these effects included.
For $\Ginj \geq 10^{-11}\,{\rm s}^{-1}$, the corresponding constraints are not affected significantly by these processes, while for $\Ginj \leq 10^{-11}\,{\rm s}^{-1}$, the high-energy limits are tightened due to photo-ionizations. At $\Ginj \leq 10^{-13}\,{\rm s}^{-1}$, we furthermore find the low-energy constraint to tighten noticeably, simply because reionization affects the free-free transparency (see Fig.~\ref{fig:regimes}).

Considering the case with $\Ginj = 10^{-13}\,{\rm s}^{-1}$ (left panels of Fig.~\ref{fig:fdm_no_edges}), we see that the constraint shows significant structure around $\Einjc\simeq 0.05\,\eV$. This marks the domain in which the final distortion at $z=0$ mainly appears redward of the \FIRAS bands, but above the regime where free-free absorption is highly effective in converting the injected energy into heat. Remembering that the injection mostly occurs at $\Einj=\Einjc/(1+\zi)$ then also explains why the limits are strongest blueward of the \FIRAS bands at $z=0$. Moving to $\Einjc>0.05\,\eV$, we observe a complicated pattern that essentially reflects the variation in the shape of the distortion at $z=0$ across the \FIRAS bands. Finally, once the \HI Ly-c threshold is crossed (visible in the lower panel of Fig.~\ref{fig:fdm_no_edges}), the distortion limit tightens again, as most of the photons are converted into heat, yielding noticeable $y$-type contributions directly constrained by \FIRAS.

For $\Ginj \leq 10^{-13}\,{\rm s}^{-1}$ (right panels of Fig.~\ref{fig:fdm_no_edges}), we observe a fairly self-similar constraint curve that gradually shifts across $\Einjc$ with varying lifetime. For the longest lifetime, we can see that the significant drop in the limits at $\Einjc\simeq \pot{3}{-4}\,\eV$ and the feature at $\Einjc\simeq \pot{3}{-2}\,\eV$ directly coincide with the edges of the \FIRAS bands. The positions of these features move upward as the lifetime decreases, reflecting the $\Einj=\Einjc/(1+\zi)$ scaling for the maximal injection energy. Overall, in both the low- and high- energy limits only a very small value of $\fdms$ is allowed when photo-ionization and reionization are included. In particular on the high-energy end, the constraint is weakened by several orders of magnitude without photo-ionization effects. Significant holes in the limit from \FIRAS alone exist redward of the \FIRAS band and below the \HI Ly-c threshold energy. The former will be tightened once \EDGES data is included. 

We also stress that naive estimate based only on energetic arguments are generally inaccurate. When considering the low-energy limit of the constraint, we observe that the limit becomes independent of the injection energy. These asymptotes define the constraint obtained purely by considering $\Delta \rho/\rho$ and hence $\mu$ and $y$ distortions. Extrapolating these asymptotes to higher injection energies confirm our statement, which both vastly over- and underestimated SD constraints with respect to these asymptotes (e.g., see upper left panel of Fig.~\ref{fig:fdm_no_edges}). Thus, the detailed treatment of the thermalization problem presented here is required to obtain reliable limits.

\vspace{-0mm}
\subsubsection{CMB distortion limits: quasi-stable particles}
%-------------------------------------------------------------------
The constraints on $f_\mathrm{dm}$ for quasi-stable particles, i.e., for decay rates $\Gamma_X\lesssim \Gamma_X^\mathrm{QS}=10^{-20} \mathrm{s}^{-1}$, can be simply deduced using the scaling presented in Sect.~\ref{LLp}. If $f_\mathrm{dm}^\mathrm{QS}$ is the limit corresponding to $\Gamma_X^\mathrm{QS}$, then $f_\mathrm{dm}=(\Gamma_X^\mathrm{QS}/\Gamma_X)\times f_\mathrm{dm}^\mathrm{QS}$ is the constraint for $\Gamma_X\lesssim \Gamma_X^\mathrm{QS}$. In these cases, the bounds have the same interpretation as the ones from the longest lifetimes case in the bottom right panel of Figure \ref{fig:fdm_no_edges}. 

%----------------------------------------------------------
\begin{figure*}
\includegraphics[trim={0mm 2mm 8mm 0},clip, width=0.96\columnwidth]{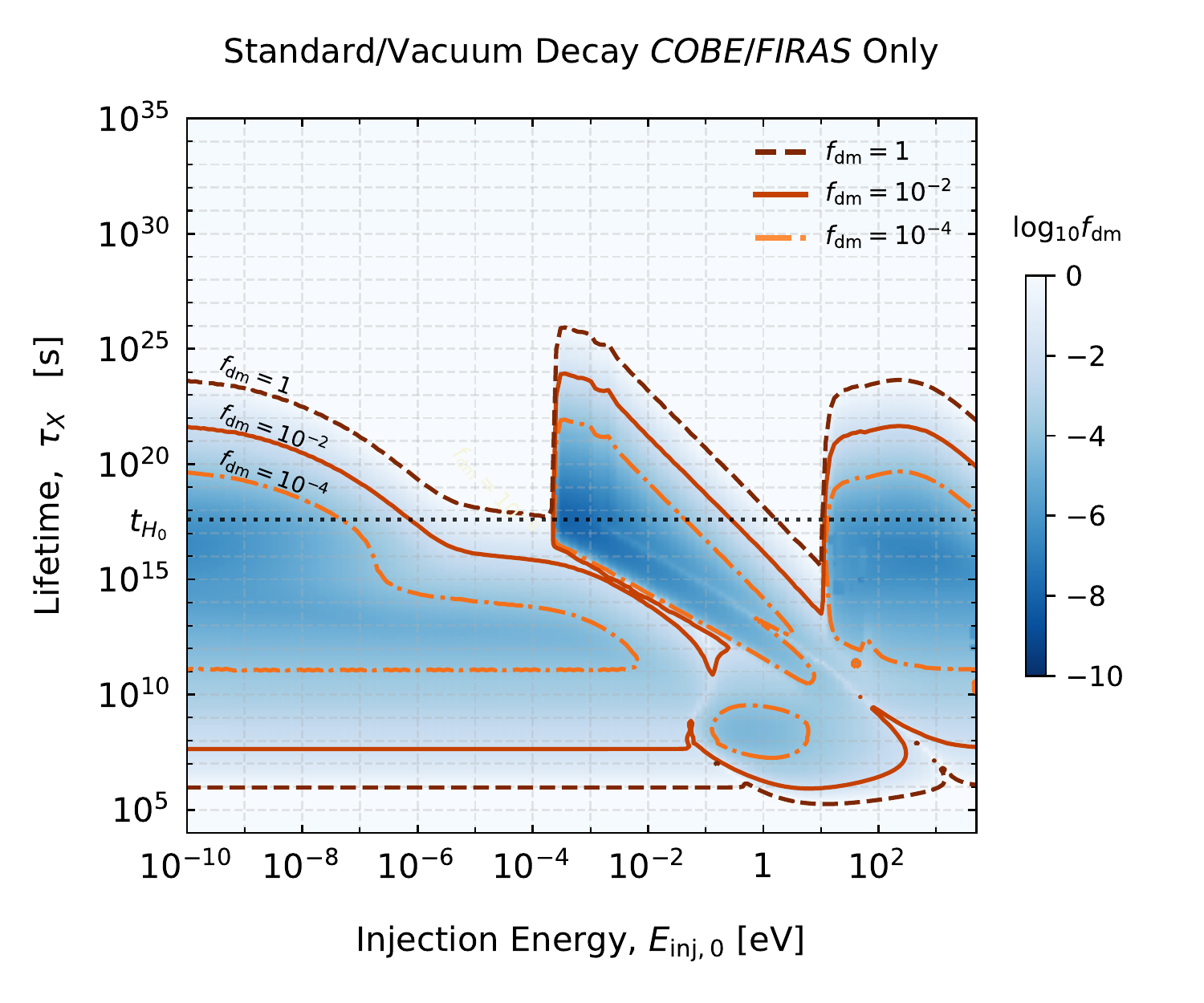}
\includegraphics[trim={0mm 2mm 8mm 0},clip, width=0.96\columnwidth]{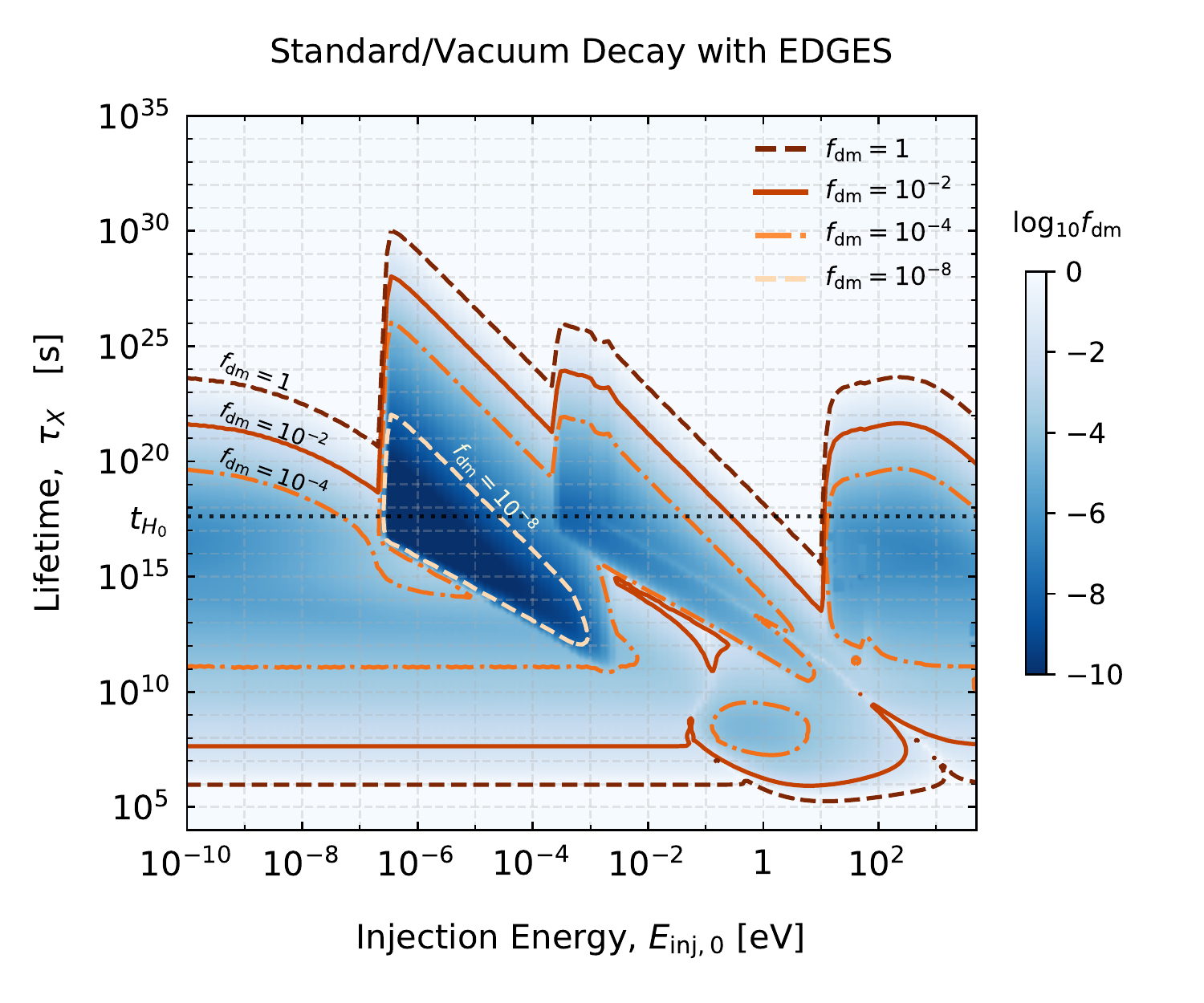}
\\[-2mm]
\includegraphics[trim={0mm 2mm 8mm 0},clip, width=0.96\columnwidth]{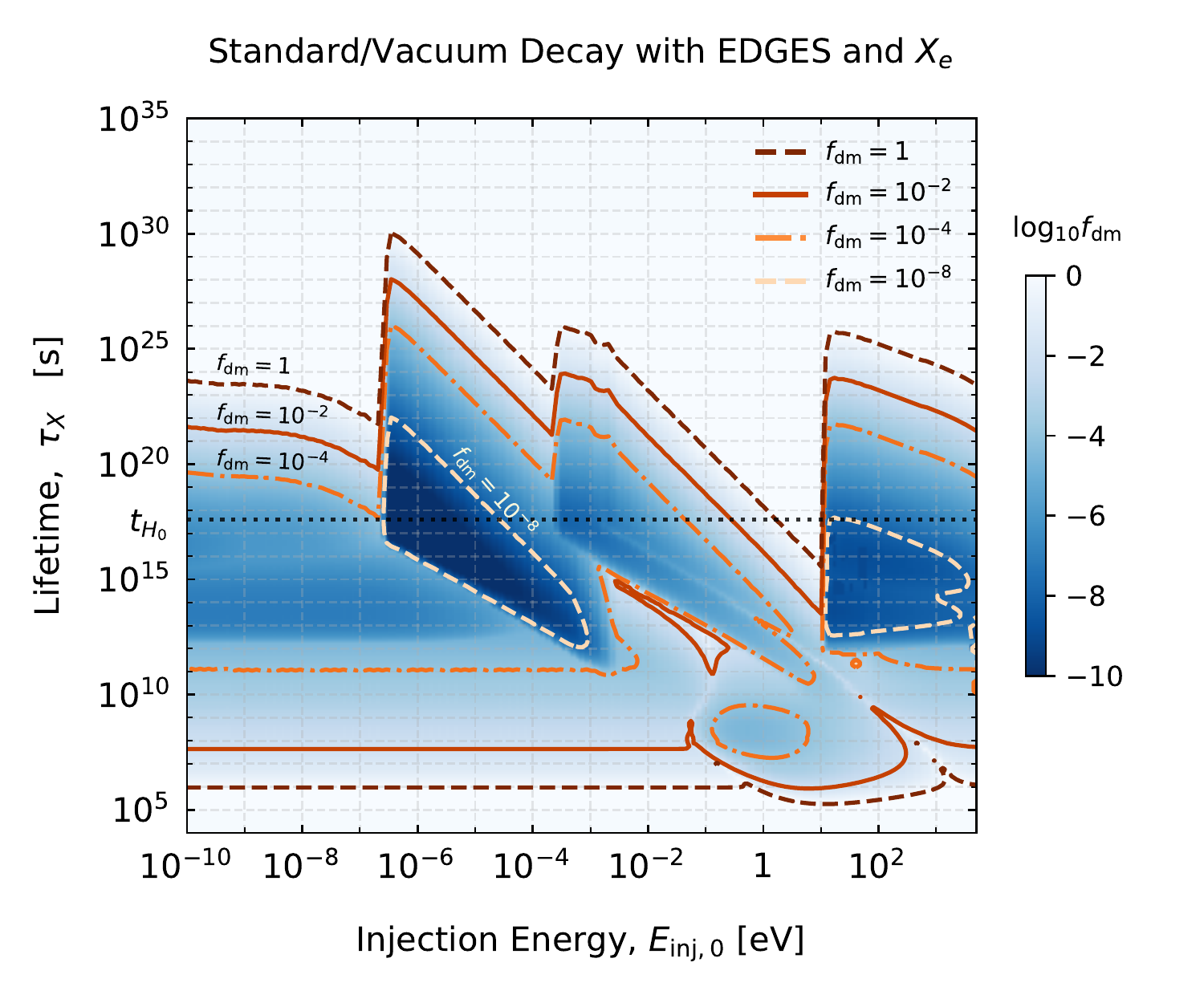}
\includegraphics[trim={0mm 2mm 8mm 0},clip, width=0.96\columnwidth]{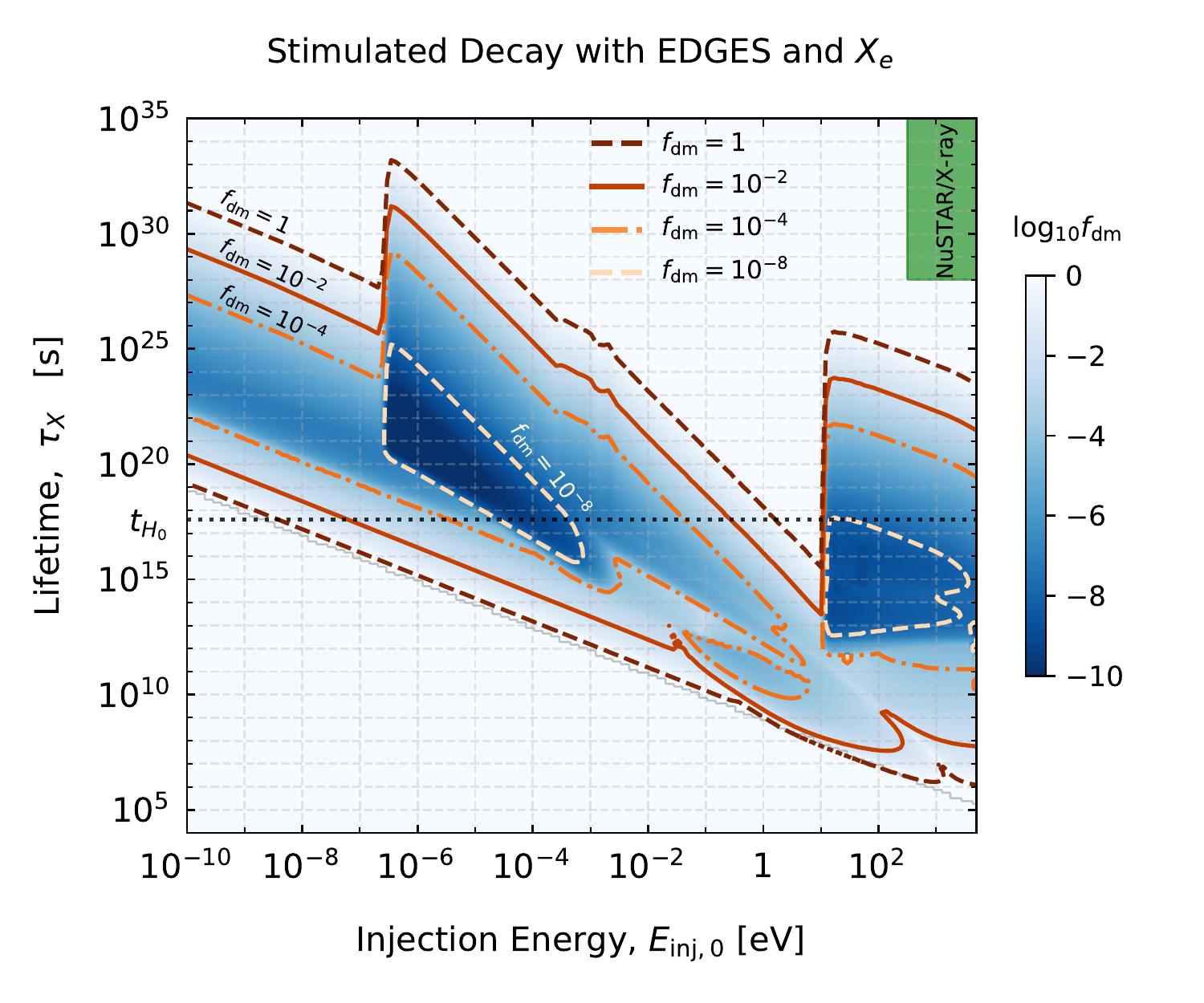}
\vspace{-3mm}
\caption{Constraints (95\% CL) on decaying particle models when subsequently including data from 
\FIRAS, \EDGES and \Planck. The lower right panel furthermore shows the cumulative constraint when including stimulated decay. The addition of \EDGES affects the constraints at $\Einj\simeq\pot{3}{-7}-\pot{3}{-4}\,\eV$, while the addition of \Planck has the largest effect at high energies above $\Einj\simeq 13.6\eV$. The dominant effect of simulated decay is to shift and compress the limits along the $\tau_X$ axis when approaching the low energies. We also mark the lifetime corresponding to the age of the Universe ($\tau_X \equiv t_{H_0}\simeq \pot{4.4}{17}\, \secs$). In the bottom right plot, for comparison, we have added the current level of constraints of sterile neutrino decay in the X-ray frequency range, namely the lifetime has to be longer than $10^{28}\,\mathrm{s}$ for masses between $5-50$keV \citep[green area labeled NuSTAR/X-ray, see Fig 5 of ][]{Roach_2020}.
} 
\label{fig:fdm_with_edges_contours}
\vspace{-2mm}
\end{figure*}
%----------------------------------------------------------

\vspace{-3mm}
\subsubsection{Adding \EDGES}
%-------------------------------------------------------------------
After having considered \FIRAS only constraints (i.e., Fig.~\ref{fig:fdm_highz} and \ref{fig:fdm_no_edges}), we now add the data point from \EDGES to the likelihood, as described at the end of Sect.~\ref{ss:edges}. For the low-energy limit, this is expected to significantly tighten the constraint for scenarios with most of the injection occurring in the post-recombination era, while it should leave the other cases mostly unaffected. 

In Fig.~\ref{fig:fdm_with_edges}, we present the comparison of scenarios with decay rates $\Ginj\lesssim 10^{-11}\,\ratesec$ with and without \EDGES included. For $\Ginj\lesssim 10^{-15}\,\ratesec$, the limits are significantly affected for $\Einjc$ between the \EDGES and \FIRAS bands ($\pot{3}{-7}\,\eV\,\lesssim \Einjc\lesssim \pot{3}{-4}\,\eV$), while for $\Ginj\simeq 10^{-11}\,\ratesec$, \EDGES clearly adds little to the constraint.
The lower sharp edge is defined by the fact that for these scenarios the direct distortion remains below the \EDGES band, such that the significantly weaker $y$-distortion contribution drives the limits. Similarly, for $\Ginj\simeq 10^{-11}\,\ratesec$ no noticeable low-frequency signal remains visible at $z=0$ and hence the limit again mainly comes from \FIRAS.

Overall, our analysis shows that adding \EDGES to \FIRAS rules out a significantly increased portion of parameter space. For the aforementioned energies, the limit improved by many orders of magnitude, strongly superseding the simple $\Delta\rho/\rho$ limits from \FIRAS. Therefore, we see that future measurements of the global 21cm signal in combination with CMB spectral distortions will provide a novel probe of particle physics. 
We also highlight that the limits at $\pot{3}{-7}\,\eV\,\lesssim \Einjc\lesssim \pot{3}{-4}\,\eV$ for the long-lifetime cases implies that scenarios with excited low-energy states in DM can be tightly constrained. Assuming $\fdm=1$, means that $\epsilon=\Einj/m_X c^2\ll 1$, implying cases with DM masses in the range of $m_X c^2\simeq $~eV to MeV can be probed. We will consider these cases in Sect.~\ref{sec:excited_states}.

\vspace{-3mm}
\subsubsection{Adding CMB anisotropy limits}
\label{sec:PCA_limit_discussion}
%-------------------------------------------------------------------
As explained in Section \ref{sec:xe} and \ref{xeconstraints}, photon injection at both low and high energies can modify the ionization history significantly. Therefore, we expect that the addition of CMB anisotropy data from \Planck \citep{Planck2015params} can further tighten the limits on decaying particle scenarios.
In the right panels of Fig.~\ref{fig:fdm_with_edges} we show the constraints on the effective DM fraction when we also use this information. As expected, for post-recombination injection ($\Gamma_X\lesssim 10^{-13}\,\mathrm{s}^{-1}$) we see a significant improvement of the constraint at low energy, where injected photons are quickly converted into heat that prevents electrons and protons from recombining as efficiently, and at high energy, where the injected photons can directly ionize hydrogen and helium.  At  intermediate energy, the plasma is optically thin to ionizations and heating (see Fig.~\ref{fig:regimes}), and the constraints are driven by the direct distortion signal. 

We mention that the PC projection constraints could be improved by performing a second iteration with the obtained constraint on $\efdm$ in the {\tt CosmoTherm} computation. As mentioned above, a key assumption is that the changes in $X_{\rm e}$ remain linear and small. However, for large injections, this is not always guaranteed even if $\Delta\rho/\rho_{\rm inj}=\pot{3}{-5}$ is warranted. 
A full exploration of this iterative approach is beyond the scope of this paper (and can become time-consuming). Nevertheless, we confirmed in a few cases that the results we obtained are not affected by more than a factor of $\simeq 2$. This part of our constraints, therefore, remains the least accurate within the current treatment. We generally expect our result to underestimate the limit, given the discussion surrounding Fig.~\ref{fig:fdm_int}.

%-------------------------------------------------------------------
\subsubsection{Effect of stimulated decay}
\label{sec:stimulated_decay}
%-------------------------------------------------------------------
In the lower panel of Fig.~\ref{fig:fdm_with_edges}, we repeat the same steps as for the vacuum decay scenarios, but this time including blackbody-induced stimulated decay (see Sect.~\ref{sec:stim}). For a given lifetime, the differences are mainly visible for low-energy injections (compare cases in upper and lower panels of Fig.~\ref{fig:fdm_with_edges}). For the examples we have chosen, the constraints {\it weaken} significantly when decreasing the injection energy. This stems from the fact that the effective lifetime with stimulated decays is always shortened with respect to the vacuum decay rate. For a given vacuum decay rate $\Ginj$, one can therefore always find a value of $\xinjc$ for which most of the injection happens at $z\simeq \pot{2}{6}$. Beyond that value, the constraint weakens exponentially due to the onset of efficient thermalization.
Conversely, longer lifetime cases (not shown in the figure here) start playing the role of shorter lifetime cases for the vacuum decay scenarios. This transformation is reflected in the final constraints presented in the next section.

%-------------------------------------------------------------------
\subsubsection{Final contours in the lifetime-energy plane}
\label{ss:contours_LE_plane}
%-------------------------------------------------------------------
We are now in the position to present the final model-independent constraints in the lifetime-energy plane. The results are presented in Fig.~\ref{fig:fdm_with_edges_contours} when subsequently including data from 
\FIRAS, \EDGES and \Planck, and also stimulated decays. Considering the vacuum decay cases (all panels but the lower right one), we find that \FIRAS alone provides extremely tight limits for decaying particles with lifetimes $10^5\,\secs\lesssim \tau_X\lesssim 10^{11}\,\secs$ independent of energy injection, essentially ruling out $\efdm=1$ inside the full domain except for some narrow region along the extension of the ${\mu\approx 0}$ line (visible when starting from the lower right corners in the $\tau_X-\Einj$ plane; see Sect.~\ref{sec:mu_analytic} for explanation). The long-lifetime edges have a complicated shapes that depends on the chosen $\fdm^*$ threshold and particle energy. The addition of \EDGES in particular adds constraining power for injections at $\Einj\simeq \pot{3}{-7}\,\eV - \pot{\rm few}{-4}\,\eV$ and long lifetimes $10^{13}\,\secs\lesssim \tau_X$, which ensures that a direct distortion remains visible in the \EDGES bands. 
Adding data from \Planck through the PCA projection tightens the limit mostly at $\Einj\gtrsim 13.6\eV$. Given the rich structure of the contours for various $\fdm^*$ levels, it is clear that the limits cannot be easily estimated using simple energetic arguments.

Turning to the case of stimulated decays, we observe that the main change is visible at $\Einj\lesssim 10\eV$, although some small modifications are also noticeable at higher energies for very short lifetimes. The tendency of stimulated decay is to tilt the constraints towards longer lifetimes. This is expected when realizing that for a given vacuum decay rate the effective lifetime is significantly {\it shortened} when including stimulated effects for injection at low energies. For particle lifetimes $\tau_X\lesssim 10^5\,\secs\, [\Einj/\keV]^{-1}$ (i.e., towards the lower left corner), CMB anisotropy measurements of $N_{\rm eff}$ (assuming BBN consistency relations) provide additional constraints on these models. However, we do not consider them in this work. Some related discussions can be found in \citet{Simha2008}, \citet{Baumann2016} and \citet{DeZotti2020}.

Additional constraints on long-lived decaying particles with injection energies above the typical nuclear dissociation thresholds ($\simeq 0.1 - 1\MeV$) can be obtained from measurements of light element abundances \citep[e.g.,][]{Kawasaki2005, Kawasaki2018, Keith2020, Kawasaki2020, Depta2020}. These cases reach into the regime of SDs from non-thermal particle cascades at early times \citep[e.g.,][]{Acharya2018, Acharya2020b}, which we specifically avoid here. The constraints provided here complement these works on the low-energy end, to which the former are largely insensitive.

We close by highlighting that once one allows small fractions of decaying particles to be present in the Universe, the limits weaken significantly (see Fig.~\ref{fig:fdm_with_edges_contours}). In these cases, future measurements of the CMB spectrum could provide extremely valuable improvements, which will be hard to match using other observables.
Around $\Einj\simeq 1-10\,\eV$, the current SD constraint cannot rule out $\efdm=1$ inside a small wedge of lifetimes slightly less than the age of the Universe, a hole that could potentially be closed using future SD data, to highlight just one of the opportunities.

%-------------------------------------------------------------------
\subsection{Model-dependent interpretation}
\label{ss:mdc}
%-------------------------------------------------------------------
Our model independent constraints on the effective dark matter fraction $\efdm$, decay rate  $\Gamma_X$, and injection energy $E_\mathrm{inj,0}$ (see previous section) can be readily mapped to specific models. In this section, we translate our constraints to ALPs and excited states of dark matter, highlighting two applications of our SD library.

\subsubsection{Axion-Like-Particles}
\label{sec:alps_limits}
%-------------------------------------------------------------------
Axion-like particles (ALPs) form a class of DM particle models motivated by  solutions to the strong CP problem \citep{Weinberg1978, PhysRevLett.40.279,PhysRevLett.38.1440} and string theory \citep{Arvanitaki:2009fg}. For an extensive review on the role of axions in cosmology and existing constraints we refer the reader to \cite{Marsh_2016}. 
Here, we are mainly interested in ALPs coupled to electromagnetism via the two-photon decay channel ($\epsilon=1$ and $f_\gamma=2$), as characterized by a coupling constant
%-------------------------------------------------------------------
\begin{align}
    g_{a\gamma\gamma}
    &=\left(\frac{64\pi\Gamma_a}{m_a^3}\right)^{1/2}
\approx 
\frac{\pot{3.63}{-2}}{\GeV}\left[\frac{\Gamma_a}{10^{-17}\mathrm{s}^{-1}}\right]^{1/2}\left[\frac{m_a c^2}{\meV}\right]^{-3/2},
    \label{eq:gagg}
\end{align}
%-------------------------------------------------------------------
where $\Gamma_a$ and $m_a$ are the vacuum decay rate and mass of the ALP. Estimates for SD constraints on these models were presented in \citet{Masso1997}, \cite{cadamuro2012cosmological}, \cite{Cadamuro_2012} and \cite{Millea_2015}. 
Here, we significantly extend these analyses by simultaneously solving the thermalization problem and accounting for modifications to the ionization history. The SD constraints derived here allow us to go beyond the simple $\mu$ and $y$ distortion description, which becomes inaccurate for long lifetimes (see discussion surrounding Fig.~\ref{fig:fdm_no_edges}).

\begin{figure}
\includegraphics[width=\columnwidth]{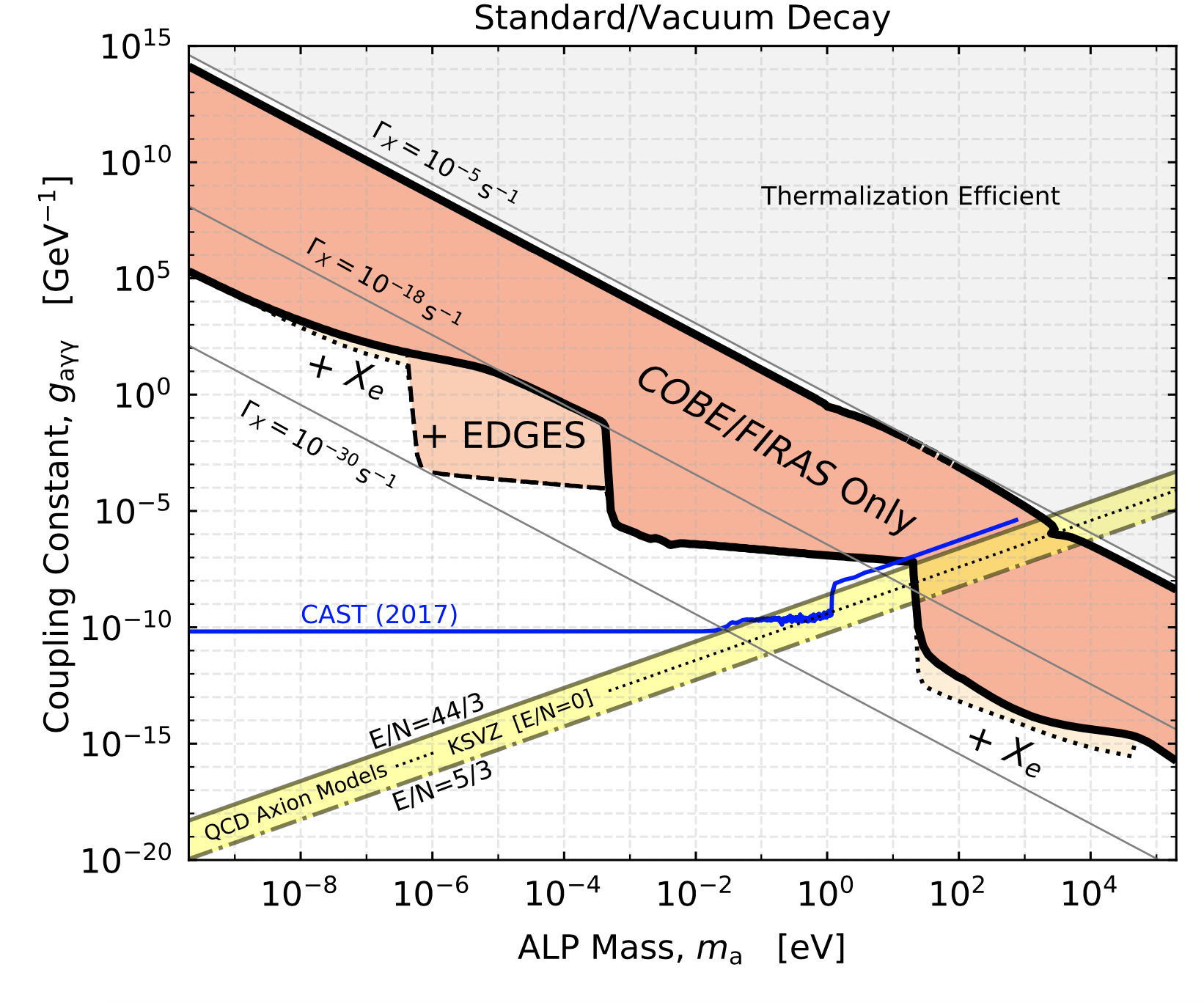}
\\[2mm]
\includegraphics[width=\columnwidth]{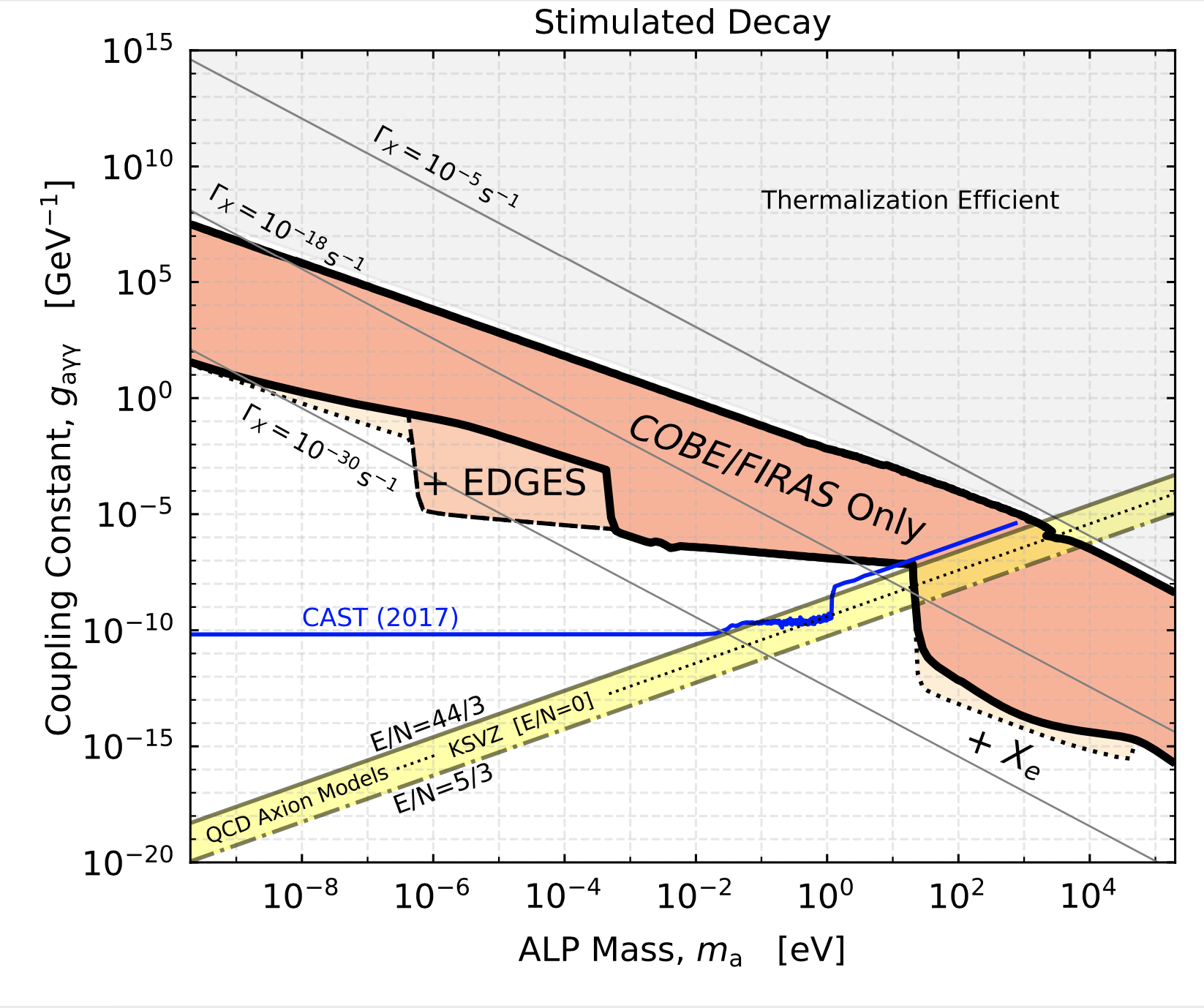}
\vspace{-2mm}
\caption{
Cumulative \FIRAS, \EDGES and \Planck constraints (95\% CL) on axion models assuming the relation Eq.~\eqref{eq:gagg} for the coupling constant and $\fdm=1$ (and $\epsilon=1$). The upper panel assumes vacuum decay, while the lower panel includes blackbody-induced stimulated decay.
For reference, we also show that constraint from CAST \citep{Anastassopoulos:2017ftl} and the physically-motivated region valid for the KSVZ axion model over a range of anomaly coefficients E/N (yellow region). SDs provide competitive constraints at axion masses of $m_a c^2\gtrsim 27\,\eV$.
} 
\label{fig:fdm_with_edges_contours_ALP}
\vspace{-2mm}
\end{figure}

By applying Eq.~\eqref{eq:gagg} for the assumptions of the previous sections, we can convert the constraints presented in Fig.~\ref{fig:fdm_with_edges_contours} into constraints on $g_{a\gamma\gamma}$ as a function of $m_a$. The results are shown in Fig.~\ref{fig:fdm_with_edges_contours_ALP} for both vacuum and stimulated decay scenarios and assuming\footnote{The considered case naturally requires $\epsilon=1$.} $\fdm=1$. 
For reference, we show the constraint from the CERN Axion Solar Telescope \citep{Anastassopoulos:2017ftl}. These are based on the non-detection of axion to photon conversion inside a strong laboratory magnetic field, assuming axions are produced by the Sun. The production mechanism only depends on stellar physics and the Primakoff process \citep{Primakoff1951}, i.e., $\gamma\rightarrow a$ scattering in the Coulomb field of the stellar plasma. Therefore, the derived limit on $g_{a\gamma\gamma}$ does not depend on the cosmological axion abundance.
In addition, we highlight the region in the $g_{a\gamma\gamma}-m_a$ plane expected for the QCD axions in yellow. This region corresponds to physically-motivated values for the anomaly coefficient E/N, with the benchmark KSVZ axion model having $\mathrm{E/N}=0$ (dotted line). We refer to \cite{Marsh_2016} for details about these models. 

%----------------------------------------------------------
%----------------------------------------------------------
%\newpage

As shown in Fig.~\ref{fig:fdm_with_edges_contours_ALP}, SDs alone already provide competitive constraints on axions with masses $m_a c^2\gtrsim 27\,\eV$. Including \EDGES tightens the limits in the range of $m_a c^2 \simeq \pot{3}{-7}-\pot{3}{-4}\,\eV$. \Planck data improves the limits slightly at $m_a c^2 \lesssim \pot{3}{-7}\,\eV$ and also noticeably for $m_a c^2 \gtrsim 27\,\eV$, superseding those from \FIRAS in that regime. However, it is worth stressing that the \FIRAS limits pre-date all of the other constraints shown in the figure.

For lower masses ($m_a c^2 \lesssim 27\,\eV$), the constraints presented here weaken, quickly exceeding the CAST limit by orders of magnitude. Including stimulated decays in the computation tightens the constraint significantly at $m_a c^2\lesssim 10^{-2}\,\eV$ at the lower boundary of the SD limit (i.e., towards lower values of $g_{a\gamma\gamma}$) and $m_a c^2\lesssim 10^{3}\,\eV$ for the upper. Inside the gray domain, thermalization processes are too efficient to allow any constraint from CMB SDs. Here complementary constraints from measurements of $N_{\rm eff}$ apply \citep[e.g.,][]{Millea_2015, millea2020new}, but are not presented.

We note that for $m_a c^2\gtrsim 10\,\keV$, the upper boundary of our SD domain (towards larger values of $g_{a\gamma\gamma}$), our constraints seem to be consistent with the those given in Fig.~3 of \citet{Millea_2015} after converting into the $g_{a\gamma\gamma}-m_a$ plane. However, at lower masses the distortion limit can become extremely weak [occasionally resulting in a chemical potential $\mu\simeq 0$ and thus no constraint (see Sect.~\ref{sec:mu_analytic} and Fig.~\ref{fig:fdm_highz})], an effect that apparently was not captured in previous analyses. 
We also mention that for $m_a c^2\gtrsim 10\,\keV$, non-thermal distortion contributions are expected to become relevant only if the distortion is primarily produced at redshift $z\lesssim \pot{3}{5}$ or with lifetime $\tau_a\gtrsim 10^{8}\,\secs$, thus leaving the shape of the upper boundary unaffected.

%----------------------------------------------------------
\begin{figure}
\begin{centering}
\includegraphics[width=\columnwidth]{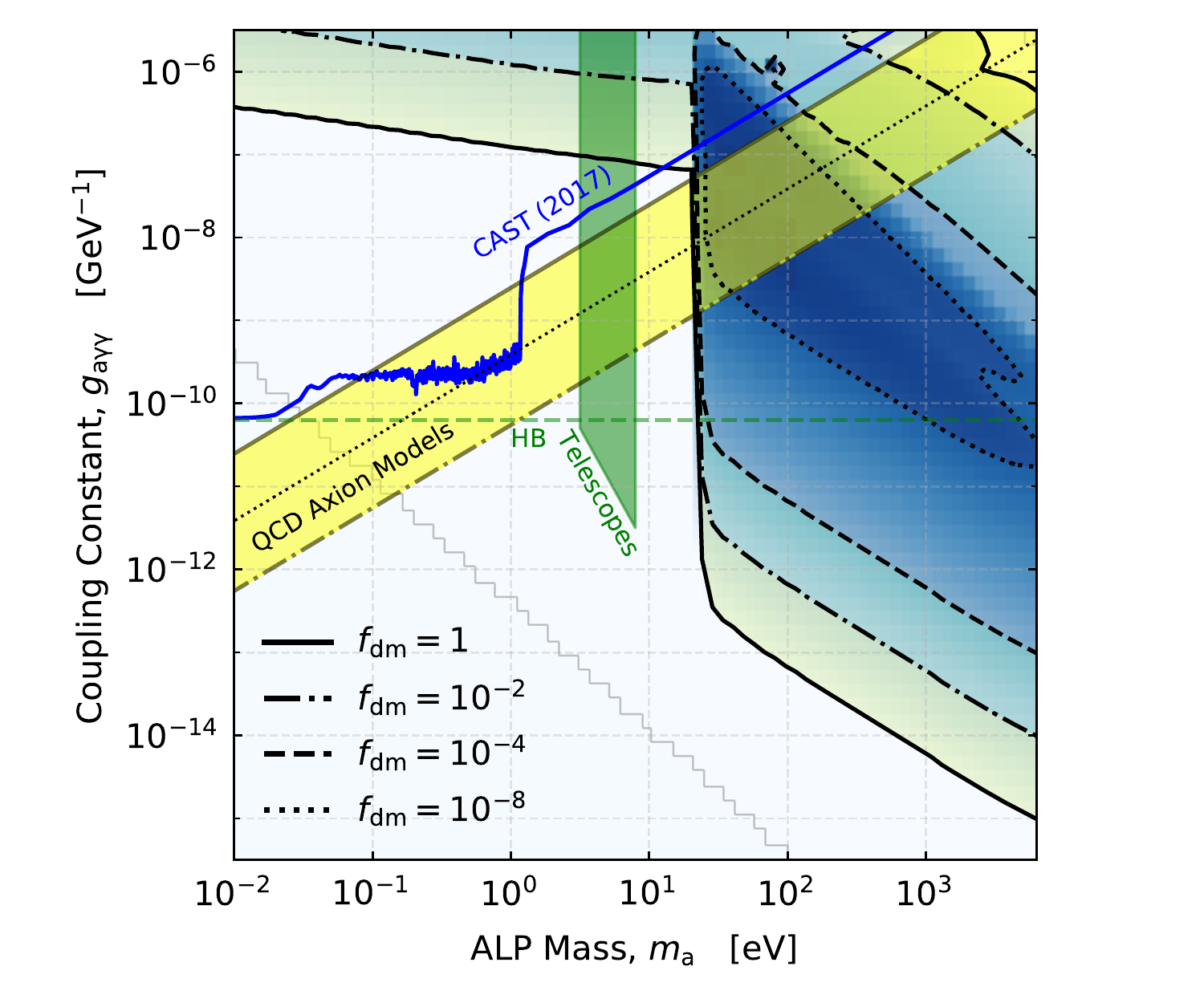}
\end{centering}
\vspace{-4mm}
\caption{
Same as Fig.~\ref{fig:fdm_with_edges_contours_ALP} for \FIRAS + \Planck zoomed in on the high-mass end and for various values of $\fdm$. Stimulated decay is unimportant in this domain.
Axions with masses $m_a c^2\gtrsim 27\eV$ are heavily constrained even if they only constitute a small fraction of the DM. 
For comparison we have added constraints from Horizontal Branch Stars (green dashed line labeled `HB') and optical telescopes searches focused on Abell clusters \citep[green shaded area,][]{PhysRevLett.66.1398,Grin_2007}. We refer to Fig 91.1 of \citet{PhysRevD.98.030001} for further details on these optical constraints and to \citet{Regis2021} for the most recent updates.}
\label{fig:fdm_with_edges_contours_ALP_zoomed}
\vspace{-3mm}
\end{figure}
%----------------------------------------------------------

For the lower boundary of the \FIRAS only limit (towards lower values of $g_{a\gamma\gamma}$) at $m_ac^2\gtrsim 27\,\eV$, our constraint is very similar to the one presented in \citet{Hektor2018}. The latter is purely derived from the rise in the electron gas temperature through the decaying axion heating process\footnote{Similar arguments for limiting the allowed amount of heating were earlier used to constrain DM decay \citep{Mitridate:2018iag} and  annihilation scenarios \citep{DAmico2018}.}, while here the constraint is alone driven by \FIRAS.
The SD constraint obtained here is about $\simeq 2-3$ orders of magnitude weaker than the one presented in \citet{Hektor2018}. Further including \Planck data through the effect on the ionization history improves our limit by a factor of $\simeq 20-50$ in the mass range we have considered, still remaining less stringent. However, we mention that the \EDGES result indeed requires $T_k$ at $z\simeq 17$ to be {\it lower} than the standard case to explain the measurement if the background photon spectrum is the undistorted CMB blackbody. In this case, a strong additional cooling process has to be present \citep[e.g.,][]{Barkana2018}, potentially offsetting any extra heating through high-energy injection. The constraint given here could be seen as slightly more robust and independent of the interpretation for the \EDGES measurement. Nevertheless it is clear that accurate measurements of the 21cm global signal (or more generally at low frequencies we where, for example, ARCADE2) provide a unique path for constraining energy release at late times.

%-------------------------------------------------------------------
\begin{figure*}
 \includegraphics[width=0.95\columnwidth]{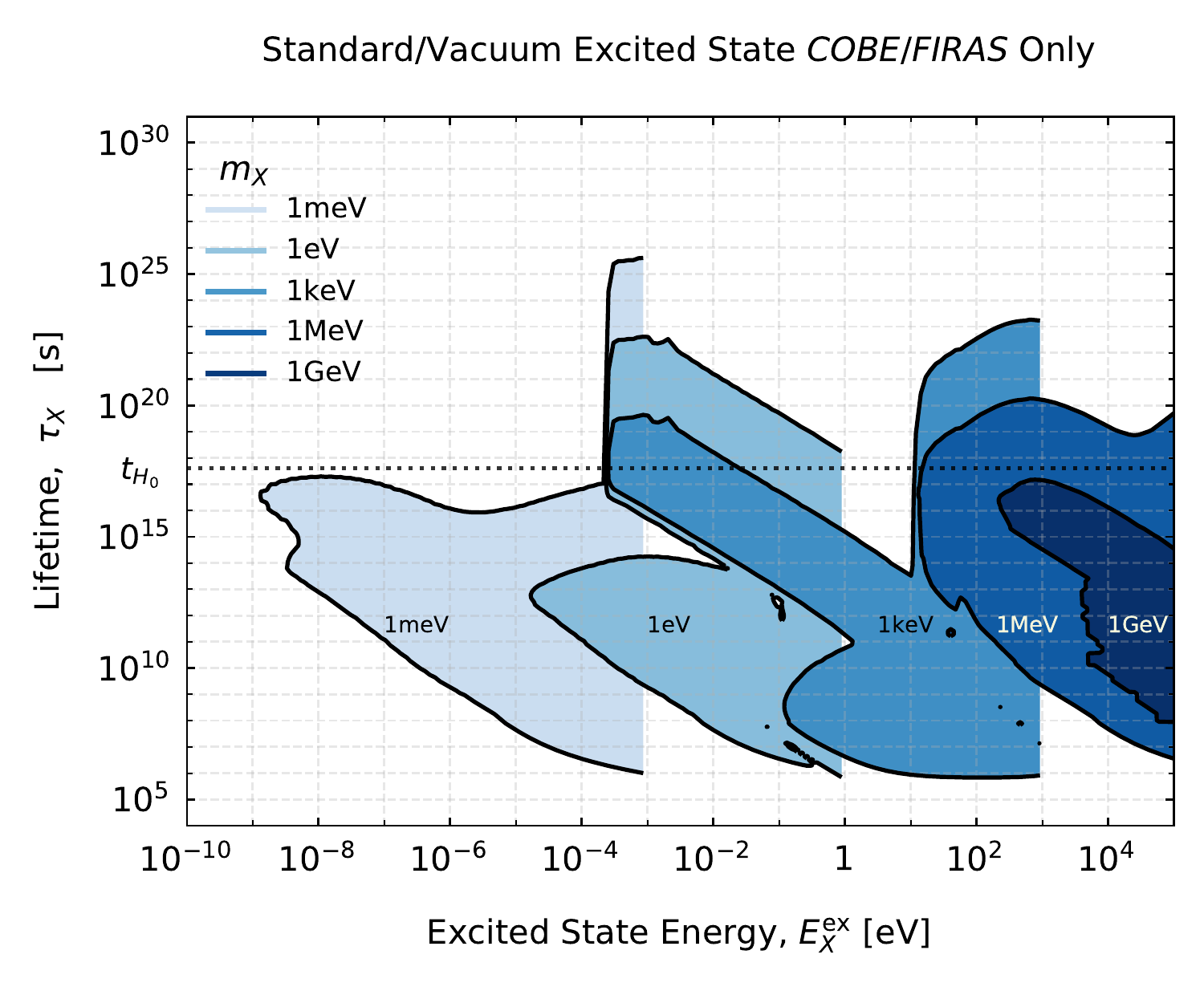}
 \hspace{3mm}
 \includegraphics[width=0.95\columnwidth]{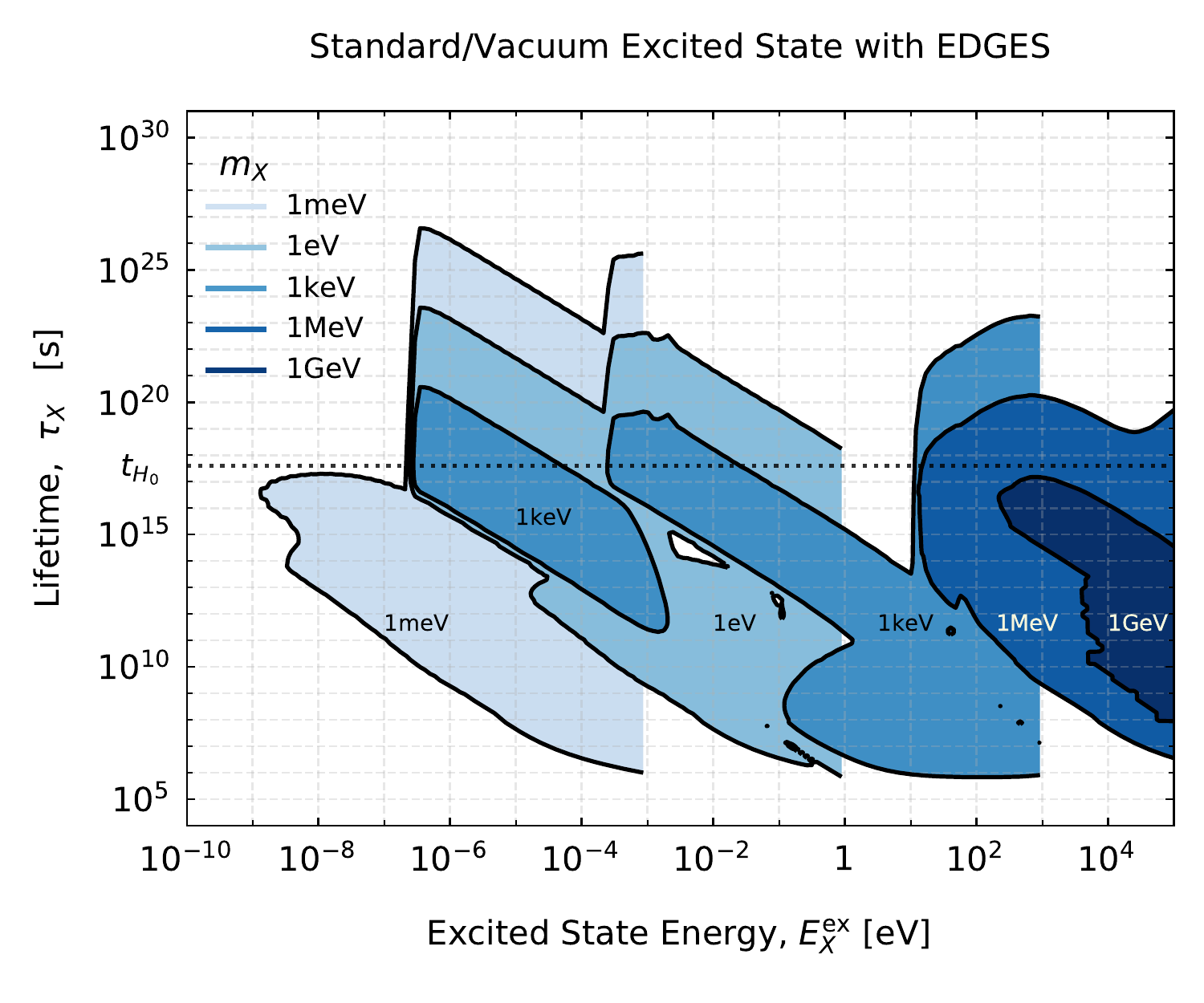}
 \\[-2mm]
 \includegraphics[width=0.95\columnwidth]{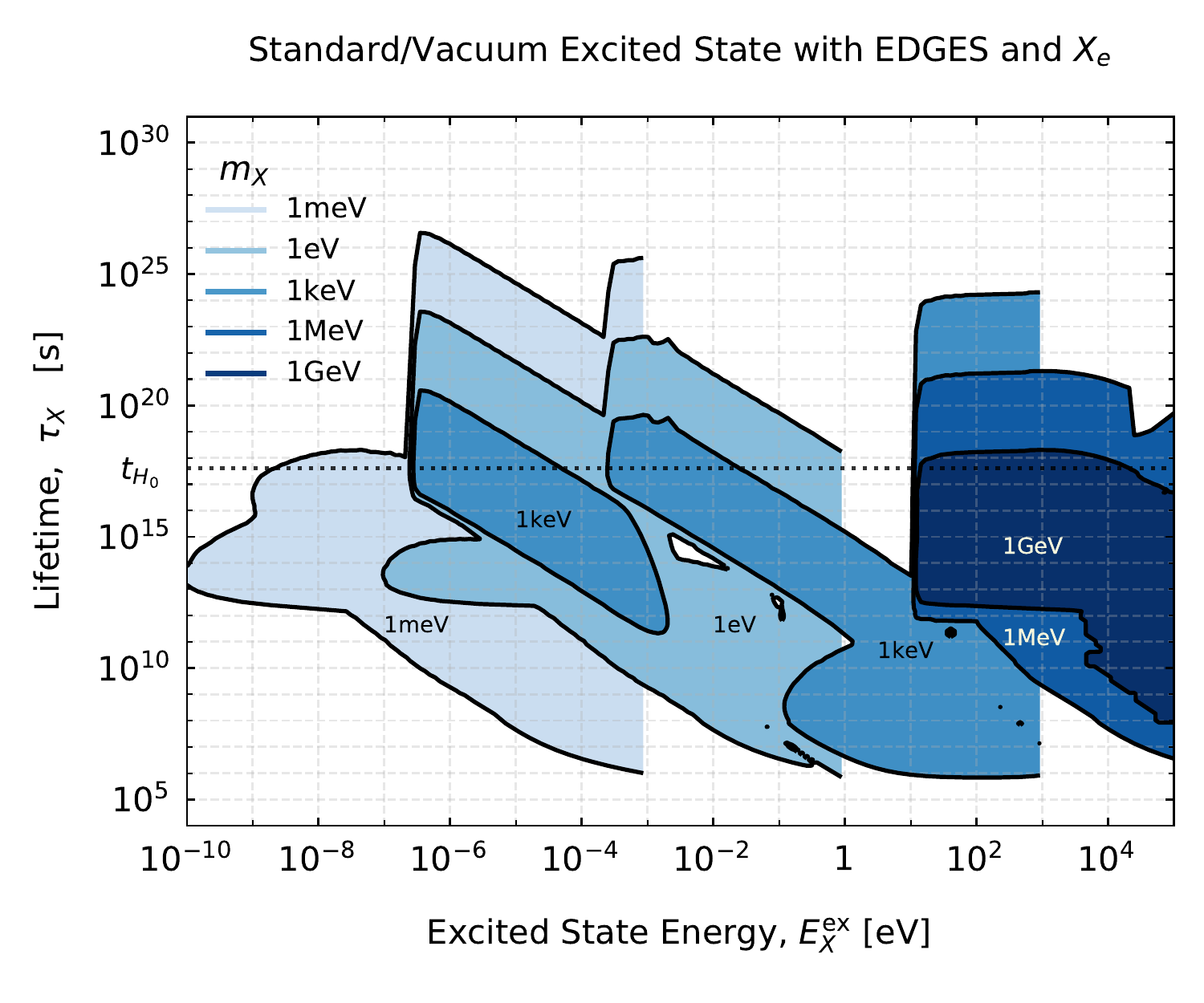}
 \hspace{3mm}
 \includegraphics[width=0.95\columnwidth]{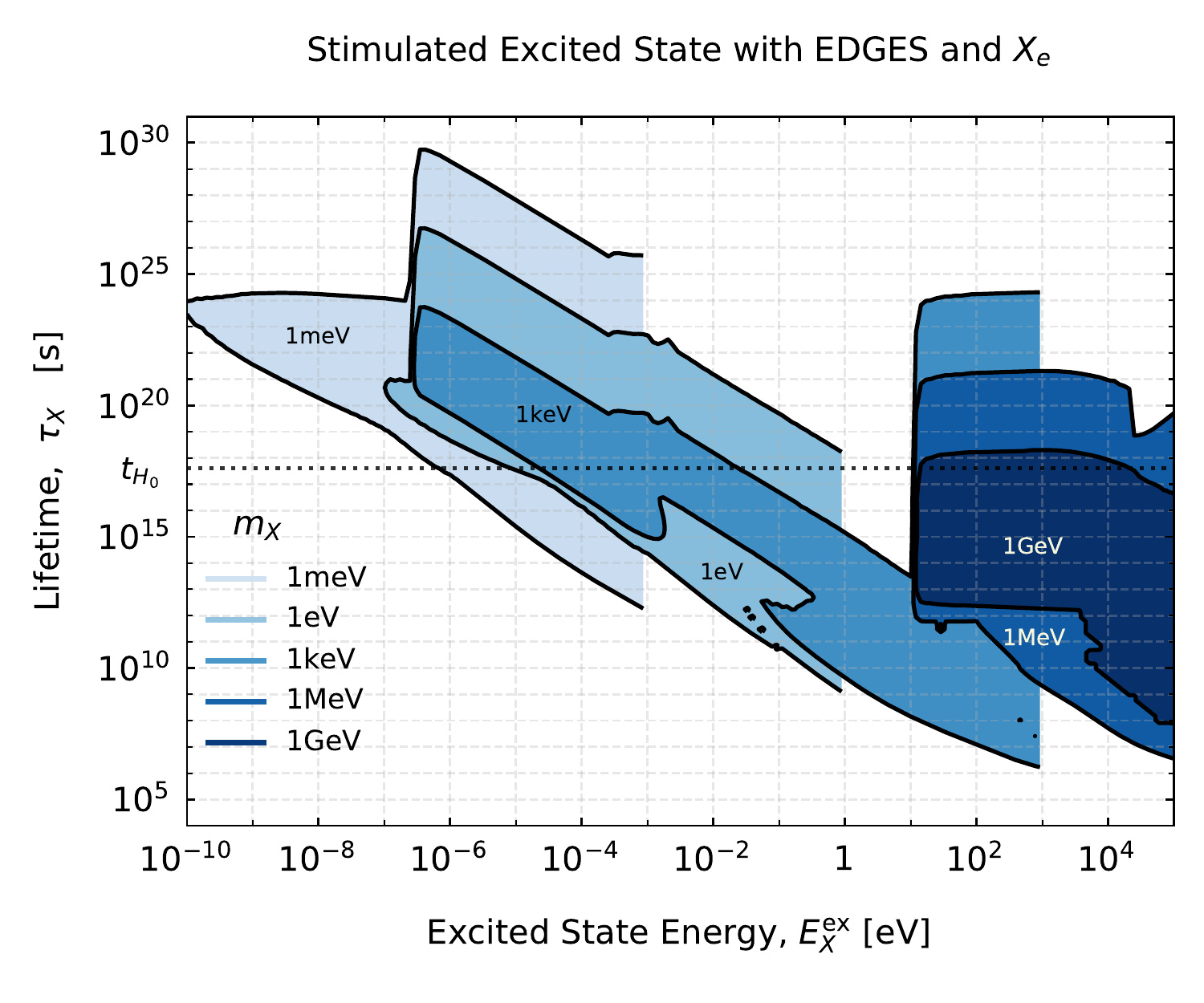}
 \caption{
Constraints (95\% CL) on excited states of DM for various combinations of \FIRAS, \EDGES and \Planck data as labeled. For each exclusion region, we varied the DM mass as a parameter. The differences between the constraints for the various data combinations follow from arguments that are similar to those relating to Fig.~\ref{fig:fdm_with_edges_contours}. 
For injection of photons in the $\simeq \keV$ range (of possible interest to the XENON1T result) very high DM masses (exceeding $m_X c^2\simeq 1\,\TeV$) are required to avoid any constraints. For $\MeV$- or $\GeV$-scale DM mass, the excited state either is extremely long-lived (longer than the age of the Universe) or has lifetimes $\tau_X\lesssim 10^{9}\,\secs$ for $m_X c^2=1\,\MeV$ or $\tau_X\lesssim 10^{12}\,\secs$ for $m_X=1\,\GeV$.
} 
\label{fig:minimal_mass_SD_final}
\end{figure*}
%-------------------------------------------------------------------

Our SD constraints can also limit ALPs if they are only a small fraction of the DM (see Fig.~\ref{fig:fdm_with_edges_contours_ALP_zoomed}) and with lifetimes shorter than the age of the Universe.
This case already puts significant pressure on $\keV$-mass axion explanations of the XENON1T anomaly \citep{aprile2020observation}, even if a very small fraction of the DM. Arguments against this possibility, but based on measurements of $N_{\rm eff}$, have recently been presented in \citet{millea2020new}. 
For short lifetimes, the CAST limit would furthermore weaken, as the expected number density of ALPs originating from the Sun would be depleted\footnote{This assumes the ALPs travel at non-relativistic speeds from the Sun.} by the time they reach the Earth, hence decreasing the chance of detection. 

We close our brief discussion of axions by mentioning that SD measurements can also be used to constraint resonant axion-photon conversion processes \citep{Tashiro2013, Moroi2018} and similarly dark radiation to photon couplings \citep[e.g.,][]{Bondarenko2020}. In these cases, the time-dependence of the injection process differs significantly from the decaying particle scenarios considered here. Our results are not directly applicable, but could, however, be extended using the appropriate source term. 
We noted that resonant photon conversion mediated by magnetic fields inside our galaxy and galaxy clusters can lead to anisotropic spectral distortions \citep[both polarized and unpolarized, e.g.,][]{Mukherjee2019a, Mukherjee2020}, providing exciting targets for future CMB imagers \citep{Delabrouille2019Voyage, Chluba2019Voyage}.
All these examples highlight the immense potential of future spectral distortion measurements as a probe of particle physics.

%-------------------------------------------------------------------
\subsubsection{Excited states of dark matter}
\label{sec:excited_states}
%-------------------------------------------------------------------
So far we have focused our discussion on the decay of a particle into two photons. Another possibility is the de-excitation of a meta-stable excited state of DM associated with the emission on one photon. Models of excited states in DM have been considered in connection with the $511\,\keV$ line observed by INTEGRAL and the DAMA anomaly \citep[e.g.,][]{Finkbeiner2007, Finkbeiner2009}. More recently, similar models were also used to explain the XENON1T result \citep{Baryakhtar2020}.

While typically the excitation energies (related to the mass-splitting) for the aforementioned models are large ($E^{\rm ex}_X>{\rm few}\times\keV$), with SDs we can probe cases with significantly lower energies. By setting $\fdm=1$ (i.e., assuming excited states in DM) and assuming that the excited state has been populated in the primordial Universe (or at least before $z\simeq \pot{2}{6}$), we can then convert the limit on $\epsilon = E^{\rm ex}_X/m_X c^2$ into a limit on the {\it minimally} required mass to not violate the observation. The results of this exercise are presented in Fig.~\ref{fig:minimal_mass_SD_final} in the $\tau_X-E^{\rm ex}_X$ plane for various threshold masses. 
The colored domains are excluded by the various dataset combinations. We also assume $E^{\rm ex}_X<m_X c^2$ by construction.

All the limits shown in Fig.~\ref{fig:minimal_mass_SD_final} naturally weaken for the longest and shortest lifetimes. Adding \EDGES and \Planck data to \FIRAS modifies the contours as expected from the discussion in Sect.~\ref{ss:contours_LE_plane} and of Fig.~\ref{fig:fdm_with_edges_contours}. Similarly, stimulated decays transform the contours for the respective vacuum decay case as expected. The size of the excluded domain generally shrinks when increasing the mass of the DM particle. This is simply because a larger DM mass means fewer particles that could inject photons, hence causing a lower signal.

One important conclusion from the lower panels of Fig.~\ref{fig:minimal_mass_SD_final} is that quasi-stable excited states of DM which inject energy in the $\keV$ range either need to originate from a DM particle with masses in excess of $\simeq 1\,\GeV$ or have lifetimes longer than the age of the Universe. For short lifetimes, a $\GeV$-scale DM mass can be accommodated for $\tau_X\lesssim 10^{12}\,\secs$. For $\MeV$-scale DM mass this drops to $\tau_X\lesssim 10^{9}\,\secs$.
However, to allow us to draw strong conclusions, e.g., relevant to the XENON1T anomaly, a more careful analysis that includes specific details of the production mechanism should be carried out. We furthermore mention that for meta-stable excited states, decay into multiple photons is a viable possibility \citep[see][for some discussion]{Baryakhtar2020}. In this case, a continuum of photons will be created from the ensemble of DM particles, violating the assumptions of our calculations. In certain regimes, de-excited DM particle can furthermore carry away some of the momentum further broadening the injection spectrum. This is in particular important when $E^{\rm ex}_X\simeq m_X c^2$. These cases both require a more extended treatment that is beyond the scope of this paper.
 
\section{Conclusions}
%-------------------------------------------------------------------
We have presented a detailed exploration of the CMB SD signals caused by photons injected due to decaying particle scenarios over a wide range of model parameters. This is the first comprehensive study of the distortion shapes using detailed consideration of the crucial thermalization processes and effects on the cosmic recombination history with {\tt CosmoTherm} (see Sect.~\ref{sec:PHO}, Sect.~\ref{sec:domains_ana} and Fig.~\ref{fig:regimes} for details). 
The main outcome of this study is a detailed SD database\footnote{The database also includes the detailed solutions for the associated ionization history.} that can be used to quickly approximate the expected distortion signal given the abundance of the particle, its lifetime and the photon injection energy. This allowed us to perform detailed analyses using \FIRAS, \EDGES and \Planck data without the need to compute the distortion on a case-by-case basis, overcoming one of the main computational challenges.

In Sect.~\ref{sec:num_sols}, we illustrated the range of solutions and their dependence on the various physical processes (e.g., see Fig.~\ref{fig:distortion_sols} to \ref{fig:distortion_sols_Lyc_reion}). For injection at $\Einj>13.6\,\eV$, the inclusion of photo-ionization processes is crucial for reliable SD predictions. Similarly, the detailed low-frequency SD depends directly on the ionization history, which defines the free-free opacity and emission of the cosmic plasma, and is explicitly modeled using {\tt CosmoTherm}. We also illustrated the effect of CMB blackbody-induced stimulated decay on the SD and its evolution (see Sect.~\ref{sec:stim} for details). This effect is relevant when the decay is directly into photons, with the main consequence that the effective lifetime can be significantly reduced over the vacuum decay case if most of the injection occurs in the Rayleigh-Jeans tail of the CMB blackbody. 

For significant injection of photons at low frequencies, we furthermore identified an intermittent recombination mode (see Fig.~\ref{fig:collisions-critical}). Due to rapid free-free absorption the effect of photon injection at $x\lesssim 10^{-8}$ can always be modeled as heating of the plasma. However, photon injection at higher frequencies, just passing into the regime that is optically-thin to the free-free process, can lead to runaway feedback cycles of absorption, heating and collisional ionization, causing the intermittent behavior. We find that the problem becomes non-linear for certain critical parameter combination, making the {\tt CosmoTherm} calculation challenging. A more detailed exploration at this transition is, therefore, left to future work. Nevertheless, we could identify the interplay between collisional ionization and the evolution of the photon field as the source of this feedback loop. For the SD constraints we omit these cases, given that they only occupy a small parameter volume.

In Sect.~\ref{sec:constraints}, we explain in detail how we use existing measurements from \FIRAS, \EDGES and \Planck to obtain the decaying particle constraints. Our main model-independent results are presented in Fig.~\ref{fig:fdm_with_edges_contours}, both for vacuum and stimulated decay scenarios. Photon-injection SDs are sensitive to decaying particles with injections taking place at $z\lesssim \pot{2}{6}$, explaining the shape of the contours at short lifetimes. We, however, highlight that for specific injection energies the limit differs significantly from the heating-only case (see Sect.~\ref{sec:mu_analytic} and Fig.~\ref{fig:fdm_highz}), under certain conditions yielding a vanishing SD \citep[see also][]{Chluba2015GreensII}.
On the long-lifetime end of the constraint, the contour has a complicated shape which directly reflects the rich phenomenology of the SD shapes obtained in the various cases. Overall, this is the first comprehensive study deriving explicit SD limits using full thermalization and recombination calculations combining the aforementioned datasets.
Naive estimates, based solely on energetic arguments, fail to describe the obtained structure of the constraints.

As one example, we convert our constrains into limits on the decay of ALPs. The main results are presented in Fig.~\ref{fig:fdm_with_edges_contours_ALP} and Fig.~\ref{fig:fdm_with_edges_contours_ALP_zoomed}, and are discussed in Sect.~\ref{sec:alps_limits}. The constraints we obtained are competitive with existing constraints at axion masses $m_a c^2>27\,\eV$. Our detailed treatment overcomes some of the limitations of previous works, delivering reliable constraints using \FIRAS. The limits derived when accounting for changes to the ionization history are slightly tighter but also bear additional uncertainty due to the more approximate nature of our treatment (see Sect.~\ref{xeconstraints} and \ref{sec:PCA_limit_discussion} for discussion). Improvements of the latter will be further assessed in the future.

As an additional example, we considered the constraints on excited states of DM. These received renewed interests due to the recent reporting of the XENON1T anomaly \citep[e.g.,][]{Baryakhtar2020, Boehm2020}. 
We find that for photon injection in the $\keV$-range, of possible interest to the XENON1T measurement, our results exclude $\GeV$-scale DM masses unless the lifetime of the excited state exceed the age of the Universe or is shorter than $\tau_X\simeq 10^{12}\,\secs$. For $\GeV$-scale DM masses, lifetimes outside the region $10^{9}\,\secs\lesssim \tau_X\lesssim 10^{20}\,\secs$ are required (see Fig.~\ref{fig:minimal_mass_SD_final} for more cases and details).
However, a more detailed treatment of the production mechanism for the excited state is required to reach unambiguous conclusions.

These are just two applications of our SD constraints in particle physics. Additional scenarios directly mapping into our calculations are due to (sterile) neutrino decay \citep[e.g.,][]{Aalberts2018, Chianese2019} and gravitino decay \citep[e.g.,][]{Dolgov2013, Dimastrogiovanni2016}. The former could also play a role for the $\simeq 3.55\,\keV$ line seen in X-ray observations of galaxy clusters \citep{Bulbul2014}; however, usually these scenarios involve high-energy photons (and other particles), such that a more detailed consideration of non-thermal processes may become important (see Sect.~\ref{sec:energy_exchange}).

Going beyond the photon injection scenarios covered here, a wider range of ALP models with varying relation between mass, lifetime and coupling constants \citep[e.g.,][]{DiLuzio2020} could be considered. In this respect our model-independent constraints are extremely useful, as the corresponding relations for the coupling constant can be directly used in Fig.~\ref{fig:fdm_with_edges_contours}. 
In addition resonant axion-photon conversion or dark radiation to photon coupling come to mind, as briefly discussed at the end of Sect.~\ref{sec:alps_limits}. The crucial point to make is that in these cases the time-dependence of the injection process can differ significantly from the quasi-exponential lifetime-driven decay considered here. This implies that a more detailed case-by-case treatment is needed using {\tt CosmoTherm} to derive reliable SD constraints. In addition, a treatment incorporating all the details of the particle production and destruction process in the computation would be extremely important. This could also allow us to more carefully treat the differences between non-relativistic and relativistic particle scenarios. 
The broadening of the photon decay spectrum (which here was assumed to be a narrow line) by multiple-photon decays or through recoil effects in excited states of DM should also be incorporated and we leave them for future work.

While here we used \EDGES data to place limits on the decaying particle scenarios, our results can also be used to identify viable scenarios to explain the 21cm signal. However, we refrained from exploring this in more detail, as in this case also the evolution of the spin temperature should be considered. Given that both the low- and high-frequency spectrum can be modified in the thermalization calculation, this step requires a more detailed calculation of the corresponding transfer effects. Naively, we find that injections at low energies in the post-recombination era are most promising. For these considerations, also the interplay with free-free absorption in the thermal (possibly heated) plasma and the low-frequency emission of energetic non-thermal particles has to be considered more carefully, and, hence, this is left for future work. 

We close by remarking that future improved CMB spectrometers such as \PIXIE \citep{Kogut2011PIXIE} could tighten the SD limits on the considered scenarios considerably (possibly by 3-4 orders of magnitude). While standard $\mu$ and $y$-type distortion constraints are noticeably hampered by foreground marginalization \citep[e.g.,][]{Abitbol2017, Chluba2019Voyage, Rotti2020}, more narrow spectral features introduced by photon injection at $z\lesssim \pot{3}{5}$ should be less affected. This is particularly expected if most of the injection occurs close to the standard CMB bands directly constrainable by the CMB spectrometer ($x\simeq 0.1-10$ or $\nu\simeq 30-1000\,\GHz$). Together with direct measurements of the global 21cm signal (or generally measurements at low-frequencies outside the classical CMB bands, i.e., $\nu\lesssim 10\,\GHz$), this could provide a competitive probe of cosmo-particle physics, possibly being able to shed light on the nature of DM. We can only look forward to the advent of these novel experimental possibilities.

%-------------------------------------------------------------------
\section*{Acknowledgments}
We thank Sandeep Acharya for comments on the manuscript, Anthony Holloway and Sotiris Sanidas for help with the computing facility at JBCA. JC thanks Abir Sarkar for useful discussions of Coulomb scattering energy exchange rates at high energies and BB thanks Alessandro Podo, Ken Van Tilburg and Jay Wadekar for references and discussions about Dark Matter decay and ALPs constraints. We also thank a referee for useful comments on the manuscript.
This work was supported by the ERC Consolidator Grant {\it CMBSPEC} (No.~725456) as part of the European Union's Horizon 2020 research and innovation program.
JC was also supported by the Royal Society as a Royal Society URF (No. URF{\textbackslash}R{\textbackslash}191023) at the University of Manchester.
%-------------------------------------------------------------------

\section*{Data availability} 
The photon injection database will be made publicly available upon publication of the manuscript. 

\bibliographystyle{mnras} 
\bibliography{refs,Lit}
%----------------------------------------------------------------
\begin{appendix}
%----------------------------------------------------------------

\section{Results for \FIRAS}
\label{a:tab}
%----------------------------------------------------------------
Table~\ref{tab:chi2-firas} summarizes the results on $\mu$ and $y$ obtained with our analysis of \FIRAS data. 
%----------------------------------------------------------------
\setcellgapes{2pt}\makegapedcells
\begin{table*}
\begin{centering}
\begin{tabular}{|c|c|c|c|c|c|}
\cline{3-6} \cline{4-6} \cline{5-6} \cline{6-6} 
\multicolumn{1}{c}{} &  & $\TCMB$ & $\mu$ & $y$ & $G_{0}$\tabularnewline
\hline 
\multirow{3}{*}{} & \multirow{3}{*}{Fixsen et al. (1996)} & $2.725\mathrm{K}\pm10\mu\mathrm{K}$ & -- & -- & N. A.\tabularnewline
\cline{3-6} \cline{4-6} \cline{5-6} \cline{6-6} 
 &  & N. A. & -- & $\left(-1\pm6\right)\times10^{-6}$ & N. A.\tabularnewline
\cline{3-6} \cline{4-6} \cline{5-6} \cline{6-6} 
 &  & N. A. & $\left(-1\pm4\right)\times10^{-5}$ & -- & N. A.\tabularnewline
\hline 
\hline 
\multirow{2}{*}{$G_{0}$} & curvefit & $2.725020\mathrm{K}\pm10\mu\mathrm{K}$ & -- & -- & $0.002\pm0.030$\tabularnewline
\cline{2-6} \cline{3-6} \cline{4-6} \cline{5-6} \cline{6-6} 
 & MCMC & $2.725019\mathrm{K}\pm10\mu\mathrm{K}$ & -- & -- & $0.0016\pm0.030$\tabularnewline
\hline 
\hline 
\multirow{5}{*}{$\mu$} & curvefit & $2.725012\mathrm{K}\pm28\mu\mathrm{K}$ & $\left(-1.0\pm3.7\right)\times10^{-5}$ & -- & $0.005\pm0.031$\tabularnewline
\cline{2-6} \cline{3-6} \cline{4-6} \cline{5-6} \cline{6-6} 
 & MCMC & $2.725014\mathrm{K}\pm28\mu\mathrm{K}$ & $\left(-0.9\pm3.7\right)\times10^{-5}$ & -- & $0.004\pm0.031$\tabularnewline
\cline{2-6} \cline{3-6} \cline{4-6} \cline{5-6} \cline{6-6} 
 & curvefit $\left[T_{0}=2.725\,\mathrm{K}\right]$  & $2.725006\mathrm{K}\pm28\mu\mathrm{K}$ & $\left(-1.5\pm3.7\right)\times10^{-5}$ & -- & $0.002\pm0.031$\tabularnewline
\cline{2-6} \cline{3-6} \cline{4-6} \cline{5-6} \cline{6-6} 
 & curvefit $\left[\mu M\left(x\right)\right]$ & $2.725025\mathrm{K}\pm22\mu\mathrm{K}$ & $\left(-1.0\pm3.7\right)\times10^{-5}$ & -- & $0.005\pm0.031$\tabularnewline
\cline{2-6} \cline{3-6} \cline{4-6} \cline{5-6} \cline{6-6} 
 & non-linear curvefit $\left[T_{0}=2.725\,\mathrm{K}\right]$  & $2.725006\mathrm{K}\pm28\mu\mathrm{K}$ & $\left(-1.5\pm3.7\right)\times10^{-5}$ & -- & $0.002\pm0.037$\tabularnewline
\hline 
\hline 
\multirow{4}{*}{$y$} & curvefit & $2.725020\mathrm{K}\pm10\mu\mathrm{K}$ & -- & $\left(0.2\pm3.9\right)\times10^{-6}$ & $0.001\pm0.037$\tabularnewline
\cline{2-6} \cline{3-6} \cline{4-6} \cline{5-6} \cline{6-6} 
 & MCMC & $2.725020\mathrm{K}\pm10\mu\mathrm{K}$ & -- & $\left(0.2\pm3.9\right)\times10^{-6}$ & $0.003\pm0.037$\tabularnewline
\cline{2-6} \cline{3-6} \cline{4-6} \cline{5-6} \cline{6-6} 
 & curvefit $\left(T_{0}=2.725\,\mathrm{K}\right)$  & $2.725016\mathrm{K}\pm10\mu\mathrm{K}$ & -- & $\left(-0.4\pm3.9\right)\times10^{-6}$ & $0.001\pm0.037$\tabularnewline
\cline{2-6} \cline{3-6} \cline{4-6} \cline{5-6} \cline{6-6} 
 & non-linear curvefit $\left[T_{0}=2.725\,\mathrm{K}\right]$ & $2.725016\mathrm{K}\pm10\mu\mathrm{K}$ & -- & $\left(-0.5\pm3.9\right)\times10^{-6}$ & $0.002\pm0.037$\tabularnewline
\hline 
\hline 
\multirow{2}{*}{$\mu+y$} & curvefit & $2.724992\mathrm{K}\pm49\mu\mathrm{K}$ & $\left(-3.7\pm6.5\right)\times10^{-5}$ & $\left(3.4\pm6.9\right)\times10^{-6}$ & $-0.008\pm0.040$\tabularnewline
\cline{2-6} \cline{3-6} \cline{4-6} \cline{5-6} \cline{6-6} 
 & MCMC & $2.724993\mathrm{K}\pm49\mu\mathrm{K}$ & $\left(-3.6\pm6.5\right)\times10^{-5}$ & $\left(3.3\pm6.9\right)\times10^{-6}$ & $-0.007\pm0.040$\tabularnewline
\hline 
\end{tabular}
\par\end{centering}
\caption{Results of $\chi^{2}$ fit to the \FIRAS monopole measurements. The quoted values only give the {\it statistical uncertainties} at 68\% CL In all cases, `N.A.' means the number was not given while '--' indicates the variable was not varied. For our analysis, we had to chose a CMB monopole pivot, $T_0=2.728\,\Kel$, unless stated otherwise is stated. This has a small effect on the obtained central values.}
\label{tab:chi2-firas}
\end{table*}
%----------------------------------------------------------------

%----------------------------------------------------------------
\end{appendix}
%----------------------------------------------------------------

\small

\end{document}